\documentclass[11pt,a4paper]{article}
\usepackage{amsmath,amssymb,bm,epsfig,color,graphicx,cite,braket,mathrsfs,slashed,multirow,hyperref}

\renewcommand{\thefootnote}{\fnsymbol{footnote}}

\usepackage{hyperref,xcolor,doi}
\hypersetup{
    colorlinks=true,
    linktocpage=true,
    linkcolor=blue,
    filecolor=blue,      
    urlcolor=blue,
    citecolor=blue,
    }

\begin{document}

\title{
\begin{flushright}
\begin{minipage}{0.2\linewidth}
\normalsize
CTPU-PTC-23-55 \\*[50pt]
\end{minipage}
\end{flushright}
{\Large \bf 
Limits on heavy neutral leptons, $Z'$ bosons and majorons\\
from high-energy supernova neutrinos
\\*[20pt]}}

\author{
Kensuke Akita$^{1}$\footnote{
\href{mailto:kensuke8a1@ibs.re.kr}{kensuke8a1@ibs.re.kr}}
,\ \  Sang Hui Im$^{1}$\footnote{\href{mailto:imsanghui@ibs.re.kr}{imsanghui@ibs.re.kr}}
,\ \ Mehedi Masud$^{1, 2}$\footnote{\href{mailto:masud@cau.ac.kr}{masud@cau.ac.kr}}
\ \ and\ \ Seokhoon Yun$^{1}$\footnote{\href{mailto:seokhoon.yun@ibs.re.kr}{seokhoon.yun@ibs.re.kr}}\\*[20pt]
{\it 
$^1$Particle Theory and Cosmology Group, Center for Theoretical Physics
of the Universe,} \\
{\it Institute for Basic Science (IBS),
Daejeon, 34126, Korea} \\*[5pt]
{\it 
$^2$High Energy Physics Center, Chung-Ang University,}\\
{\it Dongjak-gu, Seoul 06974, Korea} \\*[50pt]
}

\date{
\centerline{\small \bf Abstract}
\begin{minipage}{0.9\linewidth}
\medskip \medskip \small 
Light hypothetical particles with masses up to $\mathcal{O}(100)\ {\rm MeV}$ can be produced in the core of supernovae. 
Their subsequent decays to neutrinos can produce a flux component with higher energies than the standard flux.
We study the impact of heavy neutral leptons, $Z'$ bosons, in particular ${\rm U(1)}_{L_\mu-L_\tau}$ and ${\rm U(1)}_{B-L}$ gauge bosons, and majorons coupled to neutrinos flavor-dependently.
We obtain new strong limits on these particles from no events of high-energy SN 1987A neutrinos and their future sensitivities from observations of galactic supernova neutrinos.
\end{minipage}
}

\maketitle{}
\thispagestyle{empty}
\addtocounter{page}{-1}
\clearpage
\noindent\hrule
\tableofcontents
\noindent\hrulefill

\renewcommand{\thefootnote}{\arabic{footnote}}
\setcounter{footnote}{0}

\section{Introduction}
\label{sec1}
The hot $(T\sim 30\ {\rm MeV})$ and dense $(\rho \sim 3\times 10^{14}\ {\rm g/cm^3})$ cores in supernovae (SNe) provide an environment to test feebly interacting particles (FIPs) \cite{Raffelt:1996wa}. 
In the core, particles with masses up to $m\sim 100\ {\rm MeV}$ can be produced.
If their interactions are considerably weak, they constitute a channel of energy loss, shortening the duration of neutrino burst from supernovae \cite{Raffelt:2012kt}. Using this argument, the observations of SN 1987A neutrino burst constrain various FIP scenarios (e.g., axions and axion-like particle~\cite{PhysRevLett.60.1793,Keil:1996ju,Chang:2018rso,Carenza:2019pxu,Carenza:2020cis,Bollig:2020xdr,Croon:2020lrf,Caputo:2021rux,Choi:2021ign}, heavy neutral leptons (HNLs)~\cite{Mastrototaro:2019vug,Carenza:2023old}, dark photons and gauge bosons~\cite{Chang:2016ntp,Chang:2018rso,Knapen:2017xzo,Croon:2020lrf,Shin:2021bvz,Shin:2022ulh,Cerdeno:2023kqo}, majorons~\cite{Choi:1987sd,BEREZHIANI1989279,Choi:1989hi,Chang:1993yp,Kachelriess:2000qc,Tomas:2001dh,Hannestad:2002ff,Farzan:2002wx,Heurtier:2016otg,Brune:2018sab}).

Besides the search by the energy loss argument, one can search for FIPs by studying secondary peculiar photon and neutrino fluxes due to their secondary interactions.
If FIPs with a sufficiently long lifetime escape from the SN core and decay into photons or neutrinos, the produced particles are not thermalized in the SN medium,
giving a secondary flux.
The typical energy of the secondary neutrino flux is $\gtrsim 100$ MeV, corresponding to the FIP energy related to the core temperature (or their mass, etc.), $E\gtrsim 3T$.
In addition, if FIPs decay into electron-positron pairs, the annihilation of the produced positron with electrons in the Galaxy can contribute a galactic $511\ {\rm keV}$ $\gamma$-ray signal. 
The search for the high-energy secondary fluxes from SNe in $\gamma$-ray and neutrino telescopes gives improved constraints on axion-like particles \cite{Grifols:1996id,Brockway:1996yr,Payez:2014xsa,Meyer:2016wrm,Jaeckel:2017tud,Calore:2020tjw,Ferreira:2022xlw,Diamond:2023scc}, dark photons \cite{Chang:2016ntp,DeRocco:2019njg,Calore:2021lih,Shin:2022ulh}, majorons \cite{Akita:2022etk,Fiorillo:2022cdq}, and HNLs \cite{Calore:2021lih,Mastrototaro:2019vug,Syvolap:2023trc}.
One can also search for light particles by using the excessive energy deposition in the stellar envelope by FIP decays over the observed energy in Type IIP SNe \cite{Falk:1978kf, Sung:2019xie,Calore:2021lih,Caputo:2021rux,Caputo:2022mah,Shin:2022ulh,Chauhan:2023sci}, the change of duration of SN 1987A neutrino burst \cite{Chang:2022aas,Fiorillo:2023cas,Fiorillo:2023ytr} and the observations of the secondary fluxes in other astrophysical phenomena \cite{Diamond:2023cto}.  

In this work, building upon the previous research~\cite{Mastrototaro:2019vug,Akita:2022etk,Fiorillo:2022cdq,Syvolap:2023trc}, we derive new supernova constraints on light particles interacting with neutrinos by estimating observable secondary neutrino fluxes due to their decay. As an illustrative example, we explore the implications for heavy neutral leptons (see e.g., ref.~\cite{Abdullahi:2022jlv} for review and motivation of HNLs), $Z'$ boson, particularly ${\rm U(1)}_{L_\mu-L_\tau}$~\cite{Foot:1990mn,He:1990pn,He:1991qd} and ${\rm U(1)}_{B-L}$~\cite{Davidson:1978pm,Mohapatra:1980qe,Wetterich:1981bx,Buchmuller:1991ce} gauge bosons. Additionally, we examine (pseudo-)scalar bosons coupled to neutrinos in a flavor-dependent way, referred to as flavored majorons~\cite{Chikashige:1980ui,Gelmini:1980re,Schechter:1981cv}. These particles can potentially address some experimental anomalies \cite{Asaadi:2017bhx,Chauhan:2018dkd,Smirnov:2021zgn,Dentler:2019dhz,deGouvea:2019qre,Jeong:2018yts,Abdallah:2022grs,Muong-2:2006rrc,Araki:2015mya,Borsanyi:2020mff,Muong-2:2021ojo,Carpio:2021jhu} and/or cosmological tensions \cite{vandenAarssen:2012vpm,Cyr-Racine:2013jua,Cherry:2014xra,Chu:2015ipa,Lancaster:2017ksf,Chu:2018gxk,Kreisch:2019yzn,Escudero:2019gzq,Grohs:2020xxd,RoyChoudhury:2020dmd,Araki:2021xdk,Abdullahi:2022jlv,Venzor:2023aka,Esseili:2023ldf,Asai:2023ajh}.
Secondary neutrino fluxes are produced by their decay in the energy region of $\mathcal{O}(100)$ MeV, higher than the standard neutrino burst with energy $E_\nu \simeq 15\ {\rm MeV}$ which emerges in the neutrino decoupling region outside the core (called the neutrino sphere). 
Furthermore, the detectability of SN neutrinos in neutrino telescopes is more sensitive to higher energies due to the scattering cross section, which scales roughly as $\sigma_\nu \propto E_\nu^2$ \cite{Strumia:2003zx,Formaggio:2012cpf,Kolbe:2002gk,Marteau:1999zp}.
The search for such high-energy SN neutrinos would significantly improve the supernova limits on these particles.
We derive limits on these light particles from the SN 1987A observations in Kamiokande-II \cite{Kamiokande-II:1987idp} and Irvine-Michigan-Brookhaven (IMB) \cite{Bionta:1987qt} water Cherenkov detectors.
The SN 1987A neutrinos are also observed in Baksan Underground Scintillation Telescope (BUST) \cite{Alekseev:1987ej} but we do not study it because of its small volume.
 We also discuss the projected sensitivities from future observations of a galactic supernova, which occurs a few times per century \cite{Rozwadowska:2020nab}, using Hyper-Kamiokande (HK) \cite{Hyper-Kamiokande:2018ofw}.

We comment on the differences between the previous works~\cite{Mastrototaro:2019vug,Akita:2022etk,Fiorillo:2022cdq,Syvolap:2023trc} and this work. 
While refs.~\cite{Akita:2022etk,Fiorillo:2022cdq} took into account secondary neutrino fluxes through flavor-universal decays of light bosons, we investigate a broader spectrum of particle physics models and flavor-dependent decays of light particles to neutrinos. Our analysis extends to estimating the secondary fluxes at neutrino detectors, accounting for the effects of neutrino oscillations during their journey to Earth.
Refs.~\cite{Mastrototaro:2019vug,Syvolap:2023trc} focused on secondary fluxes resulting from decays of HNLs and the future sensitivity from observations of galactic supernovae. In this work, we improve various calculations, including the production rate of HNLs in the SN core, neutrino oscillations, and the event rates at the detector. Furthermore, we obtain the current limits on HNLs from SN 1987A observations, building upon the above argument.

The paper is organized as follows: In section~\ref{sec2}, we present our benchmark model of the SN core. In section~\ref{sec3}, we discuss the effects of neutrino oscillations on the secondary neutrino fluxes by light particle decays.
In section~\ref{sec4}, we briefly characterize the secondary neutrino flux and fluence (time-integrated flux) by the light particle decays.
In section~\ref{sec5}, we show the new limits on HNLs, $Z'$ bosons, and flavored majorons, including the event rates in neutrino telescopes and the methods of our statistical analysis. Section~\ref{sec6} summarizes the paper. Several examples of neutrino and anti-neutrino spectra (fluences) and their $e^\pm$ signal spectra are presented in appendix~\ref{appa}.

\section{Reference SN core model}
\label{sec2}

For our estimations of light particle emissions from the collapsed core, we use the SN core model SFHo-18.8 \cite{Garching,Bollig:2020xdr,Caputo:2021rux,Fiorillo:2022cdq} in spherical symmetry from the Garching group.
For the actual values of physical quantities (temperature, the number densities, and the chemical potentials of the Standard Model (SM) particles) in the SN core, we take them from figures 1 and 5 in ref.~\cite{Caputo:2021rux} and figure S4 in ref.~\cite{Fiorillo:2022cdq}.
For the proton and electron number density $n_{p,e}$, we always take $Y_p\simeq Y_e \equiv n_{p,e}/n_b \simeq 0.2$ with the charge neutrality condition~\cite{Caputo:2021rux,Fischer:2021jfm}, where $n_b$ is the baryon number density. The number density for pions is computed using the relativistic virial expansion and other physical quantities in SFHo-18.8, based on ref.~\cite{Fore:2019wib}. 
The SN model SFHo-18.8 is a conservative model with low temperature in the core, where the final neutron star mass is at the lower edge of the allowed range for SN 1987A~\cite{Page:2020gsx}.
We neglect feedback onto the dynamics of the collapsed core due to the emission of hypothesis light particles.
This assumption would be correct, at least when the luminosity for light particles is smaller than that for neutrinos. However, even when the luminosity for light particles is comparable with or larger than that for neutrinos, we neglect the feedback because of lacking SN simulations including the light particles, as the rest of the literature does.

In a dense medium, nucleon masses are reduced due to nucleon forces \cite{Hempel:2014ssa}. We take the ratio of effective nucleon mass with respect to the vacuum one as a function of the mass density, based on the relativistic mean-field treatment, from ref.~\cite{Carenza:2019pxu}. In SN physics, the mass density is usually defined as $\rho\equiv m_u n_b$, where $n_b$ is the baryon number density and $m_u=931.494\ {\rm MeV}$ is the atomic mass. From the nucleon number densities and effective nucleon masses, we calculate the effective nucleon chemical potentials.

We summarize the typical values of physical quantities in the core required for our calculations. Inside the core radius of $r\sim 10\ {\rm km}$, where the SM particles are in thermal equilibrium, the temperature is $T\sim 30\ {\rm MeV}$ for the duration of $t=5\ {\rm s}$ after the core bounce. After $t=5\ {\rm s}$, the temperature decreases due to neutrino emissions. Inside the core, the baryon density is $\rho\sim 3\times 10^{14}\ {\rm g/cm^3}$, corresponding to the nuclear saturation density, with the effective nucleon masses of $m_{n,p}\sim \mathcal{O}(500)\ {\rm MeV}$ and the effective chemical potential of $\mu_{n}\sim \mathcal{O}(600)\ {\rm MeV}$ and $\mu_{p}\sim \mathcal{O}(500)\ {\rm MeV}$ in a neutron-rich star. Such a sizable difference $\mu_n-\mu_p\sim \mathcal{O} (100)\,{\rm MeV}$ implies that negatively charged particles such as electrons can be largely populated in the core.
Muons are indeed populated in the core because highly degenerated electrons with $\mu_e \sim \mu_n-\mu_p  \gtrsim m_\mu$, thermally distributed photons, and neutrinos with $T\sim 30\ {\rm MeV}$ can easily be converted into muons. 
Radiating a slight excess of $\nu_\mu$ compared to $\bar{\nu}_\mu$ create a net muon number~\cite{Bollig:2017lki}, corresponding to $Y_\mu \equiv n_\mu/n_b \sim 0.02$ with the muon number density $n_\mu$.
Negatively charged pions $\pi^-$ would also be populated in thermal and chemical equilibrium due to strong interactions with $Y_{\pi^-}\equiv n_{\pi^-}/n_b\sim 0.01$ \cite{Fore:2019wib,Fischer:2021jfm}.
The chemical potentials for $\nu_e$ and $\bar{\nu}_e$ are $\mu_{\nu_e}=-\mu_{\bar{\nu}_e}\sim 100\ {\rm MeV}$ for duration of $2\ {\rm s}$ \cite{Fiorillo:2022cdq} while those for the other flavors are approximately zeros. After 2 s, the chemical potential for $\nu_e$ becomes also zero by the depleptonization due to a large emission of $\nu_e$.

The standard SN neutrino fluxes follow a quasi-thermal spectrum \cite{Keil:2002in,Tamborra:2012ac}. We write the time-integrated SN neutrino flux as
\begin{align}
\frac{dN_{\nu}}{dE_\nu}=\frac{E_{\rm tot}}{6E_0^2}\frac{(1+\alpha)^{1+\alpha}}{\Gamma(1+\alpha)}\left(\frac{E_\nu}{E_0} \right)^\alpha e^{-(1+\alpha)E_\nu/E_0},
\label{STSN}
\end{align}
where $E_{\rm tot}$ is the total energy emitted from a supernova, $E_0$ is the average energy for a neutrino species and $\alpha$ is a "pinching" parameter that is 2 for the Maxwell-Boltzmann distribution and not 2 when the spectrum is pinched from the distribution. $\Gamma$ is the Gamma function.
$E_{\rm tot}/6$ is the total emitted energy for one neutrino species.
In the SN model SFHo-18.8, the total energy is $E_{\rm tot}=1.98\times 10^{53}\ {\rm erg}$ corresponding to the released gravitational binding energy. The exact effects of neutrino oscillations in the SN core are still under investigation. After averaging over all three $\bar{\nu}$, whose spectra produced in the core are almost the same \cite{Tamborra:2012ac}, the other parameters are $E_0=12.7\ {\rm MeV}$ and $\alpha=2.39$ \cite{Fiorillo:2022cdq}.

\section{Neutrino oscillations in supernovae and secondary neutrino fluxes on Earth}
\label{sec3}

To estimate the SN neutrino spectrum on Earth, we have to consider the effects of neutrino oscillations en route to Earth.  In the medium of the core, electron neutrinos are produced through both the neutral and charged current interactions while non-electron neutrinos ($\nu_x=\nu_\mu, \nu_\tau, \bar{\nu}_\mu, \bar{\nu}_\tau$) are produced through the approximately same neutral current interactions.
Then the standard initial SN fluxes for $\nu_x$ would be equal. However, in general, the secondary neutrino fluxes produced by the decays of light particles are different even for each non-electron neutrino flavor.
In this section, we extend the conventional discussion of neutrino oscillations in supernovae to its application to the secondary neutrino fluxes, following ref.~\cite{Dighe:1999bi}.

Effects of neutrino oscillations on the fluxes depend on the position of light particle decays and the density of the SN medium.
In the very high-density region, neutrino self-interactions are not negligible. 
Neutrino oscillations by the self-interactions are still under investigation \cite{Duan:2010bg,Mirizzi:2015eza,Chakraborty:2016yeg,Tamborra:2020cul}.
However, the secondary neutrino fluxes distinguishable from the standard one are produced outside the neutrino sphere, where neutrino self-interactions might decouple. In addition, the secondary neutrino flux in high-density region is produced by decays of light paricles with short lifetime, i.e., their strong coupling to neutrinos (see next section \ref{sec4} for details on secondary neutrino fluxes).
In this parameter region, a huge number of light particles and secondary neutrinos might be produced in SNe so that neutrino oscillations might have less effect on contours of our limits. This parameter region will also be excluded by the energy loss argument.
Then we will neglect neutrino oscillations by neutrino self-interaction in the high-density region.

Neutrinos also interact with the electron background in the SN envelope, which induces neutrino oscillations in matter (see e.g., refs.~\cite{Kuo:1989qe,Blennow:2013rca} for a review).
For a constant electron background, neutrino states in the matter basis, which diagonalize the Hamiltonian including neutrino masses and the electron potential, do not oscillate. For varying electron density, the matter eigenstates change while neutrinos propagate in matter. Then the eigenstates in the matter basis with a fixed density oscillate. Even for varying density, if the change is slow (i.e., adiabatic), the component of the prompt Hamiltonian eigenstates will retain that of the respective initial states during their propagation~\cite{Kuo:1989qe}. This approximation is called the adiabatic approximation, which is valid for \cite{Mikheyev:1985zog}
\begin{align}
\gamma \equiv \frac{\Delta m^2}{2E_\nu}\frac{\sin^2 2\theta}{\cos2\theta}\frac{1}{(1/n_e)(dn_e/dr)} \gg 1,
\end{align}
where $\Delta m^2$ and $\theta \in [0, \pi/2]$ are the neutrino mass squared differences and the mixing angle for flavor neutrinos in vacuum, respectively, and $n_e$ is the electron number density.
Related with the electron potential, there are two parameter combinations, $(\Delta m^2, \theta)=(\Delta m_{31}^2, \theta_{13})$ and $(\Delta m_{21}^2, \theta_{12})$ with $\Delta m_{ij}^2=m_i^2-m_j^2$. 
In the region of SN with densities of $\gtrsim 1\ {\rm g/cm^{3}}$, the electron fraction in the SN envelope is $Y_e \equiv n_e/n_b = n_em_u/\rho  \simeq 0.5$ \cite{Dighe:1999bi,Mirizzi:2015eza}. We assume the density profile in the SN envelope approximately as \cite{Woosley:1995ip,Nakazato:2012qf}
\begin{align}
\rho Y_e \sim  10^5\ {\rm g/cm^{3}}\left(\frac{r}{10^9\ {\rm cm}} \right)^{-3}.
\end{align}

In the SN envelope, the adiabatic approximation would be valid well for neutrinos with energy of $E_\nu \lesssim 500\ {\rm MeV}$ \cite{Dighe:1999bi}. For light particles decaying to two neutrinos, this approximation would be valid for their mass of $m \lesssim 1\ {\rm GeV}$, which is enough larger than masses that can be produced in the core. In the following, we adopt the adiabatic approximation to the effects of neutrino oscillations on the secondary fluxes.

The oscillation probability of flavor neutrinos from a production point of neutrinos with a given electron density $n_e$ to Earth is given by, averaging over time due to their long propagation,
\begin{align}
P_{\nu_\alpha \rightarrow \nu_\beta}=\sum_i\left|U_{\alpha i}^M\right|^2\left|U_{\beta i}\right|^2,
\label{ProbabilityM}
\end{align}
and the same expression holds for antineutrinos.
$U_{\beta i}$ is the Pontecorvo–Maki–Nakagawa–Sakata (PMNS) matrix and $U^M_{\alpha i}$ is the effective PMNS matrix at the production point approximately given by \cite{Blennow:2013rca}
\begin{align}
U^M&=
\begin{pmatrix}
1 & 0 & 0 \\
0 &  c_{23} & s_{23}\\
0 &  -s_{23} & c_{23}
\end{pmatrix}
\begin{pmatrix}
1 & 0 & 0 \\
0 &  1 & 0\\
0 &  0 & e^{i\delta_{\rm CP}}
\end{pmatrix}
\begin{pmatrix}
c^M_{13} & 0 & s^M_{13} \\
0 &  1 & 0\\
-s^M_{13} & 0 & c^M_{13}
\end{pmatrix}
\begin{pmatrix}
c_{12}^M & s_{12}^M & 0 \\
-s_{12}^M &  c_{12}^M & 0\\
0 & 0 & 1
\end{pmatrix},
\end{align}
where $c_{ij}=\cos\theta_{ij}$, $s_{ij}=\sin\theta_{ij}$, $\theta_{ij}$ and $\delta_{\rm CP}$ are the mixing angle and CP-violating phase in vacuum, respectively. $c^M_{ij}=\cos\theta^M_{ij}$ and $s^M_{ij}=\sin\theta_{ij}^M$ satisfy approximately \cite{Blennow:2013rca}, 
\begin{align}
\tan2\theta^M_{13}&=\frac{\tan2\theta_{13}}{1-\frac{2E_\nu V_e}{\Delta m_{31}^2\cos2\theta_{13}}}, \\
\tan2\theta^M_{12}&=\frac{\tan2\theta_{12}\cos\theta^M_{13}}{1-h},
\end{align}
with
\begin{align}
h=
\left\{
\begin{array}{ll}
\frac{\Delta m_{31}^2}{\Delta m_{21}^2}-\frac{\sin^2\theta_{12}}{\cos2\theta_{12}} & {\rm for}\ \ 2E_\nu V_e/\Delta m_{31}^2\cos2\theta_{13}  > 1 \\
\frac{2E_\nu V_e}{\Delta m_{21}^2\cos2\theta_{12}} & {\rm for}\ \ 2E_\nu V_e/\Delta m_{31}^2\cos2\theta_{13} < 1 
\end{array}
\right.
\end{align}
Here $\Delta m_{ij}\equiv m_i^2-m_j^2$, $V_e=\sqrt{2}G_F n_e$ for neutrinos, $V_e=-\sqrt{2}G_F n_e$ for antineutrinos and $G_F$ is the Fermi constant.

The secondary neutrino fluxes produced by light particle decays on Earth are given by
\begin{align}
F_{\nu_\beta}=\sum_{\alpha=e,\mu,\tau}P_{\nu_\alpha \rightarrow \nu_\beta} F_{\nu_\alpha}^0,
\end{align}
where $F_{\nu_\alpha}^0$ is the initial neutrino flux produced by their decays.
To reduce computational complexity, we approximate that light particle decays occur instantaneously when we estimate this effect, 
\begin{align}
P_{\nu_\alpha \rightarrow \nu_\beta}=P_{\nu_\alpha \rightarrow \nu_\beta}|_{r=r_{\rm D}}, \ \ \ \ r_{\rm D}=\tau \gamma \beta,
\label{Papp}
\end{align}
where $\tau$, $\gamma=E/m$, and $\beta=p/E$ are the lifetime, Lorentz boost factor, and velocity for the light particle, respectively.
For $\rho \rightarrow 0$, $P_{\nu_\alpha \rightarrow \nu_\beta}$ in eq.~(\ref{ProbabilityM}) reduces to the well-known time-averaged oscillation probability in vacuum. 

\begin{figure}[htbp]
\hspace{-6mm}
\begin{minipage}{0.5\hsize}
   \begin{center}
     \includegraphics[clip,width=9.4cm]{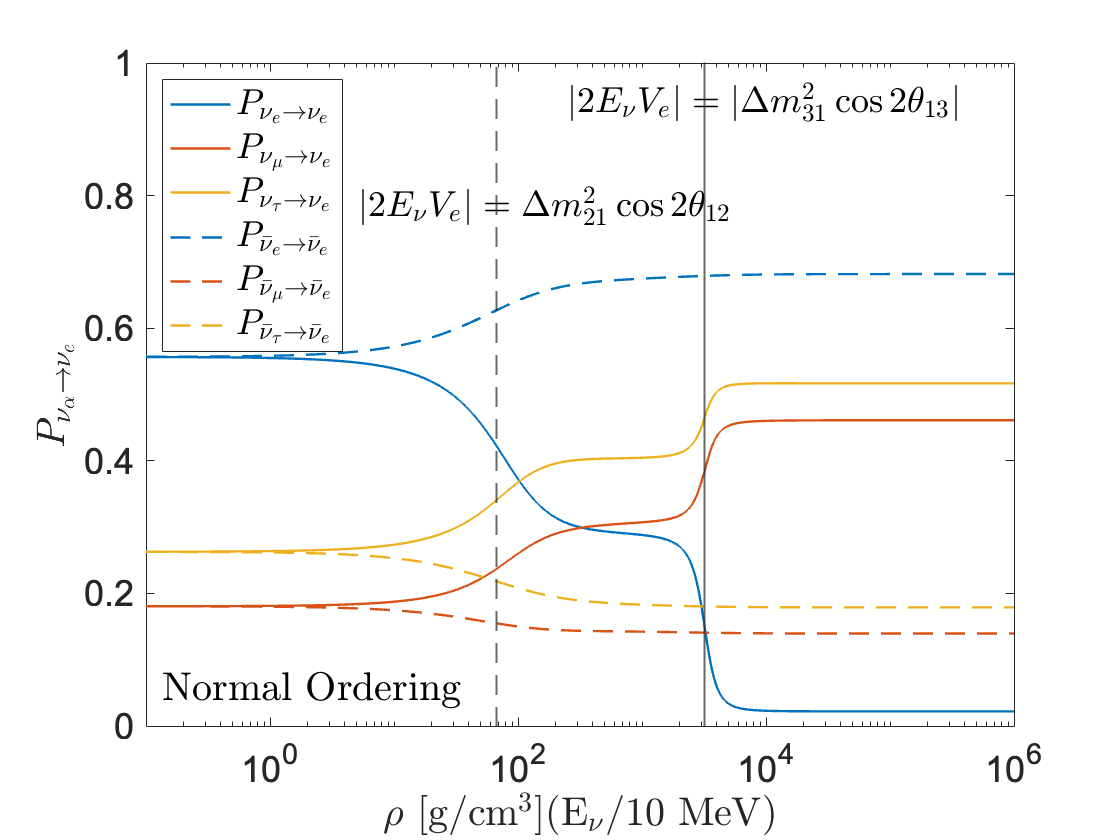}
    \end{center}
     \end{minipage} \hspace{1mm}
      \begin{minipage}{0.5\hsize}
      \begin{center}
     \includegraphics[clip,width=9.4cm]{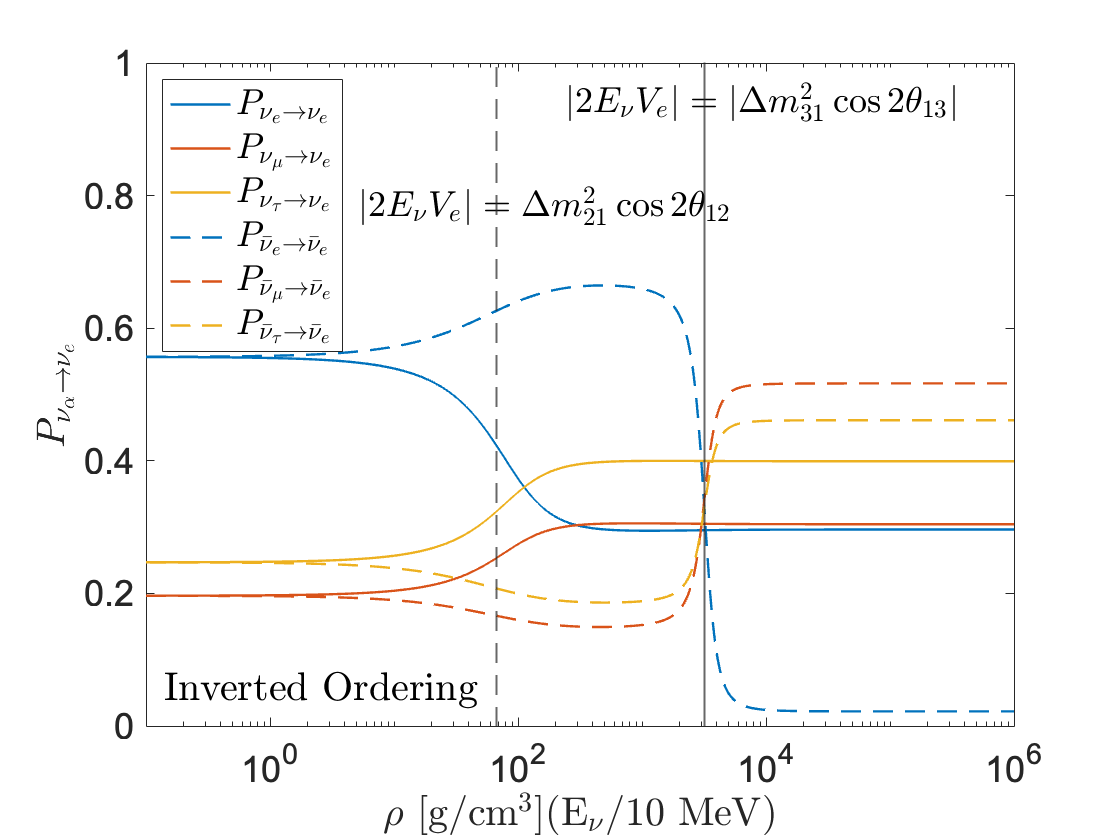}
    \end{center}
     \end{minipage}
    \vspace{-1mm}
 \caption{\small{The oscillation probabilities at Earth for $\nu_\alpha\rightarrow \nu_e$ and $\bar{\nu}_\alpha\rightarrow \bar{\nu}_e$, $P_{\nu_\alpha \rightarrow \nu_e}$, ($\alpha=e,\mu,\tau$) as a function of the initial matter density $\rho$ with $E_\nu$ in unit of $10\ {\rm MeV}$.  The vertical solid and dashed lines are the two resonance values, $|2E_\nu V_e|=|\Delta_{31}^2\cos2\theta_{13}|$ and $|2E_\nu V_e|=\Delta m_{21}^2\cos2\theta_{12}$, with $|V_e|=\sqrt{2}G_Fn_e$ ($n_e$: electron density). {\it Left}: Normal Ordering of neutrino masses, $\Delta m^2_{31}>0$. {\it Right}: Inverted Ordering of neutrino masses, $\Delta m^2_{31}<0$.} }
  \label{fig:Probability}
 \end{figure}

In figure~\ref{fig:Probability}, we show the oscillation probabilities at Earth, eq.~(\ref{Papp}), as a function of the initial matter density, $\rho$, with neutrino energy of $E_\nu=10\ {\rm MeV}$.
We only consider $P_{\nu_\alpha\rightarrow \nu_e}$ and $P_{\bar{\nu}_\alpha\rightarrow \bar{\nu}_e}$ $(\alpha=e,\mu,\tau)$ because water Cherenkov detectors mainly detect $\nu_e$ and $\bar{\nu}_e$ in the 10-100 MeV-energy region.
For neutrino mass-squared differences and mixing parameters in vacuum, we use their best-fit values \cite{Esteban:2020cvm} \footnote{\href{http://www.nu-fit.org}{NuFIT 5.2 (2022), www.nu-fit.org}} (see also ref.~\cite{deSalas:2020pgw}).
The left figure denotes the case of the normal ordering (NO) of neutrino masses, $\Delta m^2_{31}>0$, while the right figure denotes the inverted ordering (IO) of neutrino masses, $\Delta m^2_{31}<0$. From this figure, we can expect the following general results of neutrino oscillations on the secondary flux:

\begin{itemize}
  \setlength{\parskip}{0mm}
  \setlength{\itemsep}{2mm}
 
\item
$P_{\nu_\mu\rightarrow \nu_e}$ and $P_{\nu_\tau\rightarrow \nu_e}$ are very similar as well as $P_{\bar{\nu}_\mu\rightarrow \bar{\nu}_e}$ and $P_{\bar{\nu}_\tau\rightarrow \bar{\nu}_e}$. This indicates that the ratio between $\nu_\mu$ and $\nu_\tau$, and their anti particles, produced by light particle decays in the SN envelope does not affect the signal at Earth approximately. Collectively, we would still denote $\nu_\mu$ and $\nu_\tau$ as $\nu_x$ ($x=\mu,\tau$) as in the standard SN neutrinos.

\item
The difference between the total signal at Earth in the NO and IO cases would not be very significant because water Cherenkov detectors detect both $\nu_e$ and $\bar{\nu}_e$ (see e.g., FIG.~S1 in ref.~\cite{Fiorillo:2022cdq} for the detection cross sections in a water Cherenkov detector).
In the high density region, $P_{\nu_e\rightarrow \nu_e}$ and $P_{\bar{\nu}_e\rightarrow \bar{\nu}_e}$ are very different. However, if light particles decays to both $\nu_e$ and $\bar{\nu}_e$, this difference cancels out due to the difference between the detection rates for $\nu_e$ and $\bar{\nu}_e$ in the 100~MeV energy region. One exception would be the case light particles decay into only $\nu_e$ or $\bar{\nu}_e$ in the high density region.

\end{itemize}

\section{Secondary neutrino fluxes by decays of light particles}
\label{sec4}

\subsection{General description}
\label{sec4.1}

First, we introduce the secondary neutrino flux produced by the decays of hypothetical light particles generally and schematically. We denote light particles as $\chi$. 
To calculate the light particle spectrum produced in the core, we solve the Boltzmann equation inside the SN core, governing their production in region small enough to be constant density and large enough for particle interactions \cite{Sigl:1993ctk,Akita:2022hlx},
\begin{align}
\frac{\partial f_\chi}{\partial t}=\mathcal{C}[f],
\label{BEchi}
\end{align}
where $f_\chi$ is the distribution of light particles and $\mathcal{C}$ is the collision term summed over all possible collisional interactions.
The emitted spectrum of light particles is given by,
\begin{align}
\frac{dN_\chi}{dE_\chi}=\int dt \int dV e^{-\Gamma_\chi r_\nu/(\gamma \beta)}\int d\Omega\frac{E_\chi p_\chi}{(2\pi)^3}\frac{\partial f_\chi}{\partial t}
=\int dt \int dV e^{-\Gamma_\chi r_\nu/(\gamma \beta)}\frac{E_\chi p_\chi}{2\pi^2}\frac{\partial f_\chi}{\partial t},
\label{Spectrumchi}
\end{align}
where $t$ and $V$ denote the emitting time and the core volume, respectively, and $\Gamma_\chi$ is the total decay rate of light particles in the rest frame.
When light particles decay inside the neutrino sphere of $r_\nu\sim 20\ {\rm km}$ \cite{Fischer:2016cyd} (outside of which neutrinos decouple with the medium), the secondary neutrino flux would be thermalized, being the same with the standard flux \footnote{In general, neutrinos with higher energy decouple from the medium at lower medium density. Around the standard neutrino sphere, $r_\nu\sim 20\ {\rm km}$, the matter density rapidly decreases \cite{Fischer:2016cyd}. The neutrino sphere for high energy neutrinos, which is produced by the light particle decays, would be the same with the standard one.}. We account for the secondary flux produced only outside the neutrino sphere, plugging the exponential factor $e^{-\Gamma_\chi r_\nu/(\gamma \beta)}$.

If light particles interact with the SN medium feebly, they freely escape from the SN (called the free-streaming case). On the other hand, if they interact with the SN medium strongly, they are absorbed and/or rescattered (called the trapping case). In the trapping case, light particles would be less likely to both carry energy out of the SN and produce the secondary neutrino flux. This effect for their decays is already included as $e^{-\Gamma_\chi r_\nu/\gamma\beta}$ in eq.~(\ref{Spectrumchi}).
The effects for absorption and rescattering are characterized by the optical depth, which is given by \cite{Croon:2020lrf,Caputo:2021rux},
\begin{align}
\tau_{\rm abs}(E,r)=\int_r^\infty dr' \Gamma(E,r')/\beta,
\label{absrate}
\end{align}
where $\Gamma$ is the effective absorption (or rescattering) rate and $r$ is the production point for light particles. The effective absorption rate is defined by $\Gamma\equiv\Gamma_{\rm abs}\mp\Gamma_{\rm prod}$ for bosonic (fermionic) $\chi$ with $-\ (+)$ sign. $\Gamma_{\rm abs}$ and $\Gamma_{\rm prod}$ are the absorption and production rates, respectively, which are defined as $\mathcal{C}[f]=\Gamma_{\rm prod}(1\pm f_\chi)-\Gamma_{\rm abs}f_\chi -\Gamma_\chi f_\chi$ in eq.~\eqref{BEchi}. If the medium is in thermal equilibrium, the absorption and production rates are related as
\begin{align}
\Gamma_{\rm prod}= \Gamma_{\rm abs}e^{(-E_\chi-\mu_i-\mu_j-...+\mu_\alpha+\mu_\beta+...)/T}
\end{align}
by the principle of the detailed balance for a process $i+j+... \leftrightarrow \chi + \alpha + \beta+...$ with the corresponding chemical potential $\mu$. We use this relation when evaluating eq.~(\ref{absrate}). The emitted spectrum from SNe is given by
\begin{align}
\frac{dN_\chi}{dE_\chi}=\int dt \int dV \langle e^{-\tau_{\rm abs}(E_\chi,r)} \rangle e^{-\Gamma_\chi r_\nu/(\gamma\beta)}\frac{E_\chi p_\chi}{2\pi^2}\frac{\partial f_\chi}{\partial t},
\label{Spectrumchi2}
\end{align}
where $\langle e^{-\tau_{\rm abs}(E_\chi,r)} \rangle$ is a directional average of the absorption factor \cite{Caputo:2021rux},
\begin{align}
\langle e^{-\tau_{\rm abs}(E_\chi,r)} \rangle=\frac{1}{2}\int^1_{-1} d\mu\ e^{-\int^{\infty}_0 ds\ \tau_{\rm abs}(E,\sqrt{r^2+s^2-2rs\mu})}.
\end{align}
Typically, the trapping case is already excluded by other cosmological observations and experiments (In fact, the upper limits we derived in section~\ref{sec5} highly overlap with their constraints). In addition, due to the exponential suppression, we would not improve significantly the constraints on the trapping regime from the SN energy loss argument.

Light particles will also be trapped by the gravitational potential of the core when the kinetic energy of $\chi$ satisfies \cite{Dreiner:2003wh,Mastrototaro:2019vug}
\begin{align}
E_{\rm kin} \leq K_{\rm tr}\equiv \frac{G_N M_r m_\chi}{r},
\end{align}
where $G_N$ is the Newton constant, $M_r$ is the enclosed mass of the core inside a radius $r$.
One can take into account this effect by modifying the energy spectrum for light particles,
\begin{align}
\frac{d^2N_\chi}{dE_\chi dr}\rightarrow \frac{d^2N_\chi}{dE_\chi dr}\Theta(E_\chi-K_{\rm tr}-m_\chi),
\label{GT}
\end{align}
where $\Theta(x)$ is the Heaviside step function.

All emitted $\chi$ with masses and couplings of our interest decay en route to Earth. 
The emitted number of $\nu_\alpha$ (per volume in the SN core) per $t_D$ and $\omega_\nu$, where $t_D$ and $\omega_\nu$ are the time at the $\chi$ decay in the laboratory frame and the neutrino energy in the rest frame of $\chi$, is given by \cite{Oberauer:1993yr,Mastrototaro:2019vug,Akita:2022etk}
\begin{align}
\frac{d^2N_{\nu_\alpha}}{dt_Dd\omega_\nu}={\rm Br}_{\nu_\alpha}\bar{N}_{\nu_\alpha}\int \cos\theta\int dE_\chi \frac{1}{\tau \gamma}\exp\left(-\frac{t_D}{\tau\gamma} \right)\frac{dN_\chi}{dE_\chi}(t_D=0)f(\omega_\nu,\cos\theta),
\end{align}
where $\theta$ is the emitted angle of $\nu_\alpha$ in the rest frame of $\chi$, $\tau  = \Gamma_\chi^{-1}$ the lifetime of $\chi$ in the rest frame, ${\rm Br}_{\nu_\alpha}$ denotes the branching ratio for a decay process, and $\bar{N}_{\nu_\alpha}$ is the total number of the emitted $\nu_\alpha$ by a $\chi$ decay. ${\rm Br}_{\nu_\alpha}\bar{N}_\nu/(\tau\gamma)\exp(-t_D/(\tau\gamma))$ is the production rate for $\nu_\alpha$ per time, i.e.,  $(dN_{\chi}/dt_D)/N_{\chi}(t_D=0)$. $dN_\chi/dE_\chi(t_D=0)$ is given by eq.~(\ref{Spectrumchi}). $f(\omega_\nu,\cos\theta)$ is a distribution function for $\omega_\nu$ and $\theta$ by a $\chi$ decay, normalized to be 1.

If $\chi$'s travel in a different direction of Earth and neutrinos are emitted by the $\chi$ decays with an angle relative to the $\chi$ momentum, the produced neutrinos can travel a triangle distance to Earth, which is longer than the travel distance of the standard SN neutrinos.
The time delay between neutrinos by the $\chi$ decays and the standard ones arriving at Earth is \cite{Mastrototaro:2019vug}
\begin{align}
t_{\rm delay}=\frac{t_D}{\gamma^2(1+\beta\cos\theta)},
\label{TD}
\end{align}
where the time delay is suppressed by the Lorentz boost factor, $\gamma$.
If $t_{\rm delay}$ is longer than the data-taking time in neutrino telescopes, the neutrinos by the light particle decays cannot be observed.
We take into account this by modifying
\begin{align}
\frac{d^3N_{\nu_\alpha}}{dt_Dd\omega_\nu d\cos\theta}\rightarrow \frac{d^3N_{\nu_\alpha}}{dt_Dd\omega_\nu d\cos\theta}\Theta(t_{\rm data}-t_{\rm delay}),
\end{align}
where $t_{\rm data}$ is the data-taking time after the first SN neutrino arrives at Earth.
We find the secondary neutrino fluence (time-integrated flux) with energy in the laboratory frame, $E_\nu=\gamma(1+\beta\cos\theta)\omega_\nu$,
\begin{align}
\frac{dN_{\nu_\alpha}}{dE_\nu}&={\rm Br}_{\nu_\alpha}\bar{N}_{\nu_\alpha}\int^1_{\cos\theta_{\rm min}} \cos\theta\int^\infty_{E_{\rm min}} dE_\chi \frac{1}{\gamma(1+\beta\cos\theta)}\left[1-\exp\left(-\frac{\gamma(1+\beta\cos\theta)t_{\rm data}}{\tau} \right) \right] \frac{dN_\chi}{dE_\chi} \nonumber \\
&\ \ \ \ \ \ \ \  \times f\left(\frac{E_\nu}{\gamma(1+\beta\cos\theta)},\cos\theta\right),
\label{SSpectrum}
\end{align}
where $E_{\rm min}$ is the minimal energy of $\chi$ to produce neutrinos with $E_\nu$ and $\cos\theta_{\rm min}$ is determined by the kinematically allowed region for $\nu$ in the rest frame of $\chi$. Several examples of neutrino spectra on Earth are shown in appendix~\ref{appa}.

\subsection{Heavy neutral lepton}
\label{sec4.2}
Heavy neutral leptons with mass above the keV range can have an impact on the SN dynamics \cite{Shi:1993ee,Nunokawa:1997ct,Abazajian:2001nj,Hidaka:2006sg,Hidaka:2007se,Fuller:2008erj,Raffelt:2011nc,Arguelles:2016uwb,Suliga:2019bsq,Warren:2014qza,Warren:2016slz,Mastrototaro:2019vug,Syvolap:2019dat,Rembiasz:2018lok,Ray:2023gtu,Carenza:2023old,Mori:2024vrf,Ray:2024jeu}.
We consider a single mass eigenstate of HNLs, $N$, mixing with a single flavor neutrino, $\nu_{\alpha}$, with mass of $m_N$ and mixing angle $U_{\alpha N}\ (\alpha=e,\mu,\tau)$.
We assume that HNLs are Majorana particles and will comment on the Dirac and Majorana nature of HNLs in the last paragraph of this section.
For HNLs mixing with $\nu_{\mu,\tau}$,
the range of mixing angle, $10^{-10}\lesssim |U_{\mu N, \tau N}|^2 \lesssim 10^{-4}$, for masses of $\mathcal{O}(100)\ {\rm MeV}$ is currently unconstrained. On the other hand, this parameter region is severely constrained for HNLs mixing with $\nu_{e}$ by the terrestrial experiments \cite{deGouvea:2015euy,Beacham:2019nyx,T2K:2019jwa,NA62:2020mcv,Alekhin:2015byh} 
and Cosmic Microwave Background (CMB) \& Big Bang Nucleosynthesis (BBN) \cite{Boyarsky:2009ix,Ruchayskiy:2012si,Sabti:2020yrt,Mastrototaro:2021wzl,Boyarsky:2020dzc,Carenza:2023old}, but HNLs with $|U_{eN}|^2\lesssim 10^{-6}$ and $m_N\gtrsim 500\ {\rm MeV}$ is less constrained.

The Boltzmann equation for the production of HNLs is given by \cite{Hannestad:1995rs}
\begin{align}
\frac{\partial f_N(p_N)}{\partial t}=\mathcal{C}[f]
\label{BE}
\end{align}
with
\begin{align}
\mathcal{C}[f]=\frac{1}{2E_N}\int \prod_{i=1,2,3}\frac{d^3p_i}{(2\pi)^32E_i}S|\mathcal{M}|^2(2\pi)^4 \delta^{(4)}(p_3+p_N-p_1-p_2)f_1f_2(1-f_3)(1-f_N),
\label{CT}
\end{align}
where $f_i\ (i=1,2,3)$ is the thermalized distribution function, which is the Fermi-Dirac distribution $f_i=[e^{(E_i-\mu_i)/T}+1]^{-1}$ in the core, $f_N$ is the distribution function of HNLs, $S$ is the symmetry factor, $|\mathcal{M}|^2$ is the squared matrix element summed over all spins except for HNLs. 
In the core, $e^-$ are highly degenerate while $n,p,\mu^-,\nu_e,\bar{\nu}_e$ are partially degenerate \cite{Bollig:2020xdr,Caputo:2021rux,Fiorillo:2022cdq,Carenza:2023old}. Pauli-blocking suppresses the production of HNLs via the weak interactions with $e^-$ while the Pauli-blocking effects for $n,p,\mu^-,\nu_e,\bar{\nu}_e$ are sub-dominant. The dominant production processes of HNLs in the core are neutrino scattering with nucleons via neutral current interactions,
$\overset{(-)}{\nu} n \rightarrow n \overset{(-)}{N}$ and $\overset{(-)}{\nu} p \rightarrow p \overset{(-)}{N}$ \cite{Carenza:2023old}.
For HNLs mixing with $\nu_e$ ($\nu_\mu$), the other dominant production processes are electron (muon) scattering with nucleons via charged current interactions, $e^- p \rightarrow N n$ ($\mu^- p \rightarrow N n$).
In particular, the most dominant process is $e^- p \rightarrow N n$ because $e^-$ is highly degenerated in the core but this process is less Pauli-blocked.
Their squared matrix elements summed over initial and final spin states are listed in Table~\ref{tb:Process}. 
The results are a factor of 2 different from ref.~\cite{Carenza:2023old} due to the Dirac and Majorana nature of HNLs.
Since the produced HNLs would not be largely populated (and not be degenerated) in the core, we neglect the Pauli-blocking effect by HNLs, $(1-f_N)\simeq 1$, and also neglect the inverse processes in eq.~(\ref{CT}).
The spectrum of HNLs per volume and time are given by eq.~(\ref{Spectrumchi2}) after numerically solving the Boltzmann equation (\ref{BE}).
The 9-dimensional integrals in the collision term of eq.~(\ref{CT}) can be analytically reduced to 2-dimensional integrals, following ref.~\cite{Dolgov:1997mb}.

The decay rates of HNLs to SM particles are given by refs.~\cite{Gorbunov:2007ak,Atre:2009rg,Helo:2010cw,Bondarenko:2018ptm,Coloma:2020lgy}.
For HNLs with masses of $\mathcal{O}(100)\ {\rm MeV}$ and $U_e\neq 0$ ($U_\mu\neq 0$), the dominant decay processes are $N\rightarrow \nu_{e(\mu)}\pi_0$, $N\rightarrow \nu_{e(\mu)} \nu_\alpha \bar{\nu}_\alpha\ (\alpha=e,\mu,\tau)$, $N\rightarrow \nu_{e(\mu)} \ell^+ \ell^-\ (\ell=e,\mu)$, $N\rightarrow \nu_{\mu(e)} e^{+(-)} \mu^{-(+)}$ and $N\rightarrow \ell^-\pi^+$ and their charge-conjugated processes.
For HNLs with masses of $\mathcal{O}(100)\ {\rm MeV}$ and $U_\tau\neq 0$, the dominant decay processes are $N\rightarrow \nu_\tau\pi_0$, $N\rightarrow \nu_\tau \nu_\alpha \bar{\nu}_\alpha\ (\alpha=e,\mu,\tau)$ and $N\rightarrow \nu_\tau \ell^+ \ell^-\ (\ell=e,\mu)$ and their charge-conjugated processes \footnote{We neglect HNLs decays to other mesons heavier than pions. Their branching ratios are sub-dominant up to $m_N\lesssim 800\ {\rm MeV}$ (see e.g., figure 2 in ref.~\cite{Sabti:2020yrt}). We also conservatively neglect secondary decays of muons and pions to neutrinos.}.
For Dirac HNLs, the charge-conjugated processes are not allowed.

For the 2-body decays of $N\rightarrow \nu_\alpha\pi_0$, the distribution function in eq.~(\ref{SSpectrum}) is given by $f_\nu(E)=(1/2)\delta(E-\bar{E})$ with $\bar{E}=(m_N^2-m_\pi^2)/2m_N$. We obtain the secondary neutrino fluence for $N\rightarrow \nu_\alpha\pi_0$,
\begin{align}
\frac{dN_{\nu_\alpha}}{dE_\nu}\biggl|_{\rm 2-body}=\frac{m_N^2}{m_N^2-m_\pi^2}{\rm Br}_{\nu_\alpha}\left[1-\exp\left(-\frac{E_\nu t_{\rm data}}{\bar{E}\tau}\right) \right]\int^{\infty}_{E_{\rm min}}dE_N\frac{1}{p_N}\frac{dN_N}{dE_N},
\label{2bodyS}
\end{align}
where
\begin{align}
\bar{E}=\frac{m_N^2-m_{\pi}^2}{2m_N},\ \ \ \ E_{\rm min}=m_N\frac{E_\nu^2+\bar{E}^2}{2E_\nu \bar{E}}.
\end{align}
For 3-body decays, 
we have 2 types of the distribution function $f(E,\cos\theta)$ in eq.~(\ref{SSpectrum}) in the rest frame of HNLs \footnote{These formulae are slightly different from those presented in refs.~\cite{Mastrototaro:2019vug,Syvolap:2023trc}.},
\begin{align}
f_I(E)&=\frac{1}{2}16\frac{E^2}{m_N^3}\left(3-4\frac{E}{m_N} \right), \label{Dist1} \\
f_{II}(E)&=\frac{1}{2}96\frac{E^2}{m_N^3}\left(1-2\frac{E}{m_N} \right).
\label{Dist2}
\end{align}
Eqs.~(\ref{Dist1}) and (\ref{Dist2}) are normalized to be 1/2 in the integral from 0 to $m_N/2$, which is the kinematically allowed range for $\nu$.
$f_I$ corresponds to the spectrum for both neutrinos in the decays $N\rightarrow \nu_\alpha\nu_\beta\bar{\nu}_\beta$ while $f_{II}$ is the one for antineutrinos in this process. For the charge-conjugated process $N\rightarrow \bar{\nu}_\alpha\bar{\nu}_\beta\nu_\beta$, $f_I$ is the spectrum for antineutrinos while $f_{II}$ is the one for neutrinos. 
Eqs.~\eqref{Dist1} and \eqref{Dist2} are only valid for massless decay products.
Then we use eqs.~\eqref{Dist1} and \eqref{Dist2} for $N\rightarrow \nu_\alpha e^+ e^-$ and conservatively neglect the signals for $N\rightarrow \nu_e \ell^+ \ell'^-$ including $\mu^+$ or $\mu^-$.
The branching ratio for $N\rightarrow \nu_\alpha \ell^+ \ell'^-$ is also smaller than those for $N\rightarrow \nu_\alpha\pi_0$ and/or $N\rightarrow \nu_\alpha\nu_\beta\bar{\nu}_\beta$ (see figure~2 in ref.\cite{Sabti:2020yrt}).


Finally we comment on the Dirac and Majorana nature of HNLs. We consider Majorana HNLs with one degree of freedom (and anti HNLs).
In a realistic model to successfully explain small neutrino masses and a baryon asymmetry, the masses and mixing angles of the some Majorana HNLs need to be degenerate \cite{Kersten:2007vk,Antusch:2017pkq,Drewes:2021nqr}. In this case, our limits on $U_{\alpha N}$, presented in section~\ref{sec5.2}, is projected to $|U_{\alpha N}|^2\rightarrow |U_{\alpha N}|^2=\sum_i |U_{\alpha i}|^2$, where $i$ denotes the number of degenerate species of HNLs. Our limits do not simply apply to Dirac HNLs. One reason is HNLs and anti HNLs are produced differently in the core as in Table~\ref{tb:Process}. Another reason is that the charge-conjugated decay processes are allowed for Majorana HNLs while they are not allowed for Dirac HNLs. We leave a detailed study for Dirac HNLs as future work.

\begin{table}[htbp]
\begin{center}
  \begin{tabular}{c|c} \hline \hline
    Process  & $S|\mathcal{M}|^2/(32G_F^2V_{ud}^2U_{eN,\mu N,\tau N}^2)$  \\ \hline 
    $\overset{(-)}{\nu} n\rightarrow n \overset{(-)}{N}$   & $(G_V^n+G_A^n)^2(p_1\cdot p_2)(p_3\cdot p_4)+(G_V^n-G_A^n)^2(p_1\cdot p_3)(p_2\cdot p_4)-(G_V^{n2}+G_A^{n2})m_n^2(p_1\cdot p_4)$          \\ 
    $\overset{(-)}{\nu} p\rightarrow p \overset{(-)}{N}$ & $(G_V^p+G_A^p)^2(p_1\cdot p_2)(p_3\cdot p_4)+(G_V^p-G_A^p)^2(p_1\cdot p_3)(p_2\cdot p_4)-(G_V^{p2}+G_A^{p2})m_p^2(p_1\cdot p_4)$        \\
     $\ell^- p\rightarrow n N$ & $(g_V+g_A)^2(p_1\cdot p_2)(p_3\cdot p_4)+(g_V-g_A)^2(p_1\cdot p_3)(p_2\cdot p_4)-(g_V^2+g_A^2)m_nm_p(p_1\cdot p_4)$        \\
     \hline \hline 
  \end{tabular}
  \caption{\small{The squared matrix elements $S|\mathcal{M}|^2$ summed over initial and final spin states for the dominant production processes of HNLs mixing in the SN core. Subscript momentum number 
 correspond to reactions labeled as $1+2\rightarrow 3+4$. The last process with $\ell=e$ ($\ell=\mu$) is valid only for HNLs mixing with $\nu_e$ ($\nu_{\mu}$). $g_V=1$, $g_A=1.27$, $G_V^n=1/2$, $G_V^p=1/2-2\sin^2\theta_W$ and $G_A^n=G_A^p=g_A/2$ \cite{Bruenn:1985en}. $V_{ud}$ and $\sin\theta_W$ are the up-down element of the Cabibbo-Kobayashi-Maskawa matrix and the Weinberg angle, respectively.}}
  \label{tb:Process}
  \end{center}
\end{table}

\subsection{${\rm U(1)}_{L_\mu-L_\tau}$ gauge boson}
\label{sec4.3}

We consider the extension of the SM to include a spontaneously broken ${\rm U(1)}_{L_\mu-L_\tau}$ gauge symmetry.
Then a massive ${\rm U(1)}_{L_\mu-L_\tau}$ gauge boson, denoted by $Z'$, is introduced and the effective Lagrangian for ${\rm U(1)}_{L_\mu-L_\tau}$ gauge bosons is described by
\begin{align}
\mathcal{L}_{L_\mu-L_\tau}&=\mathcal{L}_{SM} -\frac{1}{4}Z'_{\mu\nu}Z'^{\mu\nu}-\frac{\varepsilon}{2}Z'_{\mu\nu}F^{\mu\nu}+\frac{m_{Z'}^2}{2}Z'_\mu Z'^{\mu}\nonumber \\
&\ \ \ \ + g_{\mu-\tau}Z'_\alpha(\bar{\mu}\gamma^\alpha\mu+\bar{\nu}_\mu\gamma^\alpha P_L \nu_\mu-\bar{\tau}\gamma^\alpha\tau-\bar{\nu}_\tau\gamma^\alpha P_L\nu_\tau),
\end{align}
where $g_{\mu-\tau}$ is the gauge coupling, $P_L=\frac{1}{2}(1-\gamma_5)$ is the left chirality projector, $\varepsilon$ is the kinetic mixing parameter, and $F_{\alpha\beta}$ and $Z'_{\alpha\beta}$ are the field strength for the photon and $Z'$, respectively.
The decay rates for $Z'\rightarrow \bar{\nu}_\alpha\nu_\alpha\ (\alpha=\mu,\tau)$ and $Z'\rightarrow \ell^+\ell^-\ (\ell=\mu,\tau)$ in the rest frame are
\begin{align}
\Gamma_{Z'\rightarrow \bar{\nu}_\alpha\nu_\alpha}=\frac{g_{\mu-\tau}^2}{24\pi}m_{Z'},\ \ \ \ \Gamma_{Z'\rightarrow \ell^+\ell^-}=\frac{g_{\mu-\tau}^2}{12\pi}m_{Z'}\left(1+\frac{2m_l^2}{m_{Z'}^2} \right)\sqrt{1-\frac{4m_\ell^2}{m_{Z'}^2}} \,.
\label{DRZ}
\end{align}

In the canonically diagonalized basis, we obtain the coupling of $Z'$ to the electromagnetic current
\begin{align}
\mathcal{L}\supset \varepsilon Z'_\alpha J_{\rm EM}^\alpha,\ \ \ \ J_{\rm EM}^\alpha=e\sum_f Q_f \bar{f}\gamma^\alpha f,
\end{align}
where $e$ is the electric charge and $f$ is a SM fermion with charge $Q_f$.
We assume at the tree level this kinetic mixing vanishes.
Nevertheless, the muon and tau loops lead to this kinetic mixing.
Such irreducible contributions to $\varepsilon$ below energy scales of the muon mass are given by \cite{Kamada:2015era}
\begin{align}
\varepsilon=-\frac{eg_{\mu-\tau}}{2\pi^2}\int^1_0dx\ x(1-x)\log\left[\frac{m_\tau^2-x(1-x)q^2}{m_\mu^2-x(1-x)q^2} \right] \xrightarrow[m_\mu \gg q]{}
-\frac{eg_{\mu-\tau}}{12\pi^2}\log\frac{m_\tau^2}{m_\mu^2}\simeq -\frac{g_{\mu-\tau}}{70}.
\label{epsilonnatural}
\end{align}
Note that model-dependent contributions could potentially arise if there exist exotic particles charged under both ${\rm U}(1)_{\rm EM}$ and ${\rm U(1)}_{L_\mu-L_\tau}$~\cite{Holdom:1985ag}.

The kinetic mixing can induce a decay channel of $Z'\rightarrow e^+e^-$ whose decay rate is given by
\begin{align}
\Gamma_{Z'\rightarrow e^+e^-}=\frac{(\varepsilon e)^2}{12\pi^2}m_{Z'}\left(1+\frac{2m_e^2}{m_{Z'}^2}\sqrt{1-\frac{4m_e^2}{m_{Z'}^2}}\right) \, ,
\end{align}
and the branching ratio of $Z'\rightarrow e^+e^-$ for $m_e\ll m_{Z'} \ll m_\mu$ is
\begin{align}
{\rm Br}_{Z'\rightarrow e^+e^-}=\frac{\Gamma_{Z'\rightarrow e^+e^-}}{\Gamma_{Z'\rightarrow \bar{\nu}_\mu\nu_\mu}+\Gamma_{Z'\rightarrow \bar{\nu}_\tau\nu_\tau}}\simeq \left(\frac{\varepsilon e}{g_{\mu-\tau}}\right)^2\simeq 2\times 10^{-5},
\end{align}
where we use the value of $\varepsilon$ in Eq.~\eqref{epsilonnatural}. Effects of the relatively small kinetic mixing of $|\varepsilon|\lesssim g_{\mu-\tau}/70$ on supernovae would be sub-dominant \cite{Croon:2020lrf}. 
In the following we consider the kinetic mixing of $|\varepsilon| \lesssim g_{\mu-\tau}/70$.

The main production processes for $Z'$ in the SN core are the neutrino-pair coalescence ($\bar{\nu}\nu\rightarrow Z'$) and the semi-Compton scattering ($\mu \gamma\rightarrow \mu Z'$) if $|\varepsilon| \lesssim g_{\mu-\tau}/70$  \cite{Croon:2020lrf}. Subsequently, the decay processes of $Z'\rightarrow \bar{\nu}\nu$ (and $Z'\rightarrow \mu^+\mu^-$ if $m_Z'>2m_\mu$) occur.
The emitted $Z'$ spectrum per volume and time from SNe is given by
\begin{align}
\frac{dn_{Z'}}{dE_{Z'}dt}&=\sum_{\nu=\nu_\mu,\nu_\tau}\frac{g_{\mu-\tau}^2m_{Z'}^2}{48\pi^3}\int_{E_\nu^{\rm min}}^{E_\nu^{\rm max}} dE_\nu\ f_{\nu}(E_\nu)f_{\bar{\nu}}(E_{Z'}-E_\nu) + \frac{E_{Z'}p_{Z'}}{2\pi^2}\mathcal{C}_{\mu \gamma\rightarrow \mu Z'}[f],
\label{BEZ}
\end{align}
where $f_\nu(E)$ is the Fermi-Dirac distribution for $\nu$, $E_\nu^{\rm max}=\frac{1}{2}(E_{Z'}+p_{Z'})$ and $E_\nu^{\rm min}=\frac{1}{2}(E_{Z'}-p_{Z'})$. $\mathcal{C}_{\mu \gamma\rightarrow \mu Z'}[f]\propto g_{\mu-\tau}^2$ is the collision term for $\mu \gamma\rightarrow \mu Z'$, which is given in the same manner of Eq.~\eqref{CT} except for the spin of particles.
The first term in Eq.~\eqref{BEZ} stems from neutrino-pair coalescence, while the second term stems from the semi-Compton scattering.
To calculate the second term, we use an analogy of the Compton scattering ($\gamma\mu \rightarrow \gamma\mu$), assuming $m_{Z'} \simeq 0$. This approximation is valid since as production of $Z'$, the semi-Compton scattering only dominates at small $m_{Z'}$. 
In the matrix elements for $\mu \gamma\rightarrow \mu Z'$, the $Z'$ mass only enters the polarization vector for $Z'$ and the muon propagator due to the momentum conservation. From this reason, for heavy $Z'$, the order of the squared matrix elements for the semi-Compton scattering is independent of $m_{Z'}$ while that for neutrino-pair coalescence is proportional to $m_{Z'}^2$ as in the first term of eq.~\eqref{BEZ}.
The squared matrix element for $\mu \gamma\rightarrow \mu Z'$ with $m_{Z'}=0$ is listed in Table~\ref{tb:Process2}. The 9-dimensional integrals in the collision term can be analytically reduced to 3-dimensional integrals, following ref.~\cite{Hannestad:1995rs}.

The secondary neutrino fluence for $\nu_\alpha$ ($\alpha=\mu,\tau$) produced by their decays, $Z'\rightarrow \nu\bar{\nu}$, is given by, as in eq.~(\ref{2bodyS}),
\begin{align}
\frac{dN_{\nu_\alpha}}{dE_\nu}={\rm Br}_{\nu_\alpha}\left[1-\exp\left(-\frac{2E_\nu t_{\rm data}}{m_{Z'}\tau} \right) \right]\int^{\infty}_{E_{\rm min}} dE_{Z'} \frac{1}{p_{Z'}} \frac{dN_{Z'}}{dE_{Z'}}
\simeq 
{\rm Br}_{\nu_\alpha}\int^{\infty}_{E_{\rm min}} dE_{Z'} \frac{1}{p_{Z'}} \frac{dN_{Z'}}{dE_{Z'}},
\label{SZ}
\end{align}
where $E_{\rm min}=E_\nu+m_{Z'}^2/(4E_\nu)$.
The time delay in eq.~(\ref{TD}) is well suppressed by the Lorentz boost factor for $Z'\rightarrow \nu\bar{\nu}$ and the suppression factor by the time delay in eq.~(\ref{SZ}) is negligible in the parameter region of our interest with $t_{\rm data}\gtrsim 10\ {\rm s}$ for SN 1987A neutrino observations and $t_{\rm data}\gtrsim 10^3\ {\rm s}$ for HK. 

\begin{table}[htbp]
\begin{center}
  \begin{tabular}{c|c} \hline \hline
    Process  & $S|\mathcal{M}|^2/(8e^2g_{\mu-\tau}^2)$  \\ \hline 
    $\mu \gamma\rightarrow \mu Z'$  &{\large $\frac{p_1\cdot p_4}{p_3\cdot p_4} + \frac{p_3\cdot p_4}{p_1 \cdot p_4} + 2m_\mu^2\left(\frac{1}{p_3\cdot p_4}-\frac{1}{p_1\cdot p_4}\right) + m_\mu^4\left(\frac{1}{p_3\cdot p_4}-\frac{1}{p_1\cdot p_4}\right)^2$ }         \\ 
     \hline \hline
  \end{tabular}
  \caption{\small{The squared matrix elements $S|\mathcal{M}|^2$ for $\mu \gamma\rightarrow \mu Z'$ assuming $m_{Z'}\simeq 0$. Notation as in Table~\ref{tb:Process}. }}
  \label{tb:Process2}
  \end{center}
\end{table}

\subsection{${\rm U(1)}_{B-L}$ gauge boson}
\label{sec4.4}

The other $Z'$ model of our interest is the ${\rm U(1)}_{B-L}$ gauge extension of the SM. The Lagrangian for the ${\rm U(1)}_{B-L}$ model is
\begin{align}
\mathcal{L}_{B-L}&=\mathcal{L}_{SM}-\frac{1}{4}Z'_{\mu\nu}Z'^{\mu\nu}-\frac{\varepsilon}{2}Z'_{\mu\nu}F^{\mu\nu}+\frac{m_{Z'}^2}{2}Z_\mu'Z'^{\mu}
+ g_{B-L}Z'_\mu\left(\frac{1}{3}\bar{q}\gamma^\mu q-\bar{\ell}\gamma^\mu \ell \right),
\end{align}
where $g_{B-L}$ is the ${\rm U(1)}_{B-L}$ gauge coupling, and $q$ and $\ell$ denote all the quark and lepton fields, respectively; see refs.~\cite{Shin:2021bvz,Shin:2022ulh} for the details of the low-energy effective interactions in a medium.
Right-handed neutrinos coupled to $Z'$ with $g_{B-L}$ have to be introduced to satisfy the anomaly cancellation condition.
In this analysis, we assume that they are heavy enough to be integrated out, thus it gives no phenomenological consequence in the low-energy regime.
If right-handed neutrinos are light enough that $Z'$ decays to them, the branching ratio to left-handed neutrinos is smaller by a factor of $\sim 2$.
The decay rates for $Z'\rightarrow \bar{\nu}_\alpha\nu_\alpha$ ($\alpha=e,\mu,\tau$) and $Z'\rightarrow \ell^+\ell^-$ ($\ell=e,\mu,\tau$) are the same with eq.~(\ref{DRZ}). Their decay to neutrinos occurs in a flavor-universal way, so that the effects of neutrino oscillations are averaged out. 

The dominant $Z^\prime$ production processes within the core involves nucleon-nucleon bremsstrahlung mediated by strong interactions ($pp\rightarrow pp Z'$, $nn\rightarrow nnZ'$ and $np\rightarrow npZ'$)~\cite{Shin:2021bvz} and pion-induced reaction  ($\pi^-p\rightarrow Z'n$)~\cite{Shin:2022ulh}, supported by the non-negligible abundance of negatively-charged pions.
In this study, we exclusively consider pion-induced reactions as the source of $Z'$ production, following Ref.~\cite{Shin:2022ulh}.
Although including the $Z'$ productions via nucleon-nucleon bremsstrahlung~\cite{Shin:2021bvz,Shin:2022ulh} would enhance our newly derived limits, we anticipate that in the energy range higher than the temperature as our interest, the pion-induced reaction becomes the more efficient channel to generate such an energetic $Z^\prime$ due to the sizable pion mass.
Moreover, the formulation of $Z'$ production via nucleon bremsstrahlung relies on the one-pion exchange approximation, in which nucleons interact themselves via a one-pion mediator, leading to theoretical uncertainties that could potentially reduce its contribution to the $Z^\prime$ luminosity by an order of magnitude (see ref.~\cite{Carenza:2019pxu} for the detailed discussions in the axion case). 
A comprehensive analysis of the effects of bremsstrahlung processes on our limits is left for future work.
The secondary neutrino fluence for $\nu_{\alpha}$ ($\alpha=e,\mu,\tau$) by their decays is described in the same way as in Eq.~(\ref{SZ}).

\subsection{Majoron}
\label{sec4.5}

We consider a (pseudo-)scalar boson $\phi$ coupled to neutrinos.
Its effective Lagrangian can be described by
\begin{align}
\mathcal{L}=\frac{1}{2}m_\phi^2\phi^2 +\frac{1}{2}g_{\alpha\beta}\bar{\nu}_\beta\nu_\alpha\phi,
\end{align}
where $\nu$ is a 4-component field, and $\alpha,\beta=e,\mu,\tau$. In the following, we assume Majorana neutrinos.
For Majorana neutrinos, $\nu$ consists of left-handed active neutrinos and right-handed active neutrinos.

The dominant production processes in the SN core are $\nu_\alpha\nu_\beta\rightarrow \phi$ and $\bar{\nu}_\alpha\bar{\nu}_\beta\rightarrow \phi$ ~\cite{Farzan:2002wx} \footnote{As mentioned in ref.~\cite{Farzan:2002wx}, $\nu\nu \rightarrow \phi$ might stall because neutrinos might be depleted in the core due to this process. However, the SM weak interaction processes would supply neutrinos. In our analysis, we neglect any feedback to the core by hypothesis particle productions as the typical analysis of SN energy loss argument.}. Subsequently, the decay processes $\phi\rightarrow \nu_\alpha\nu_\beta$ and $\phi\rightarrow \bar{\nu}_\alpha\bar{\nu}_\beta$ occur for massive $\phi$.
The decay rate for $\phi\rightarrow \nu_\alpha \nu_\beta$ in the rest frame of $\phi$ in vacuum are
\begin{align}
    \Gamma_{\phi\rightarrow \nu_\alpha\nu_\beta}
    &=S\frac{g_{\alpha\beta}^2}{16\pi}m_\phi,
    \label{DecayMajoron}
\end{align}
where $S=1/2$ ($S=1$) for $\alpha=\beta$ ($\alpha\neq \beta$) and $m_\phi$ is the mass of $\phi$.
The same expressions hold for the processes $\phi\rightarrow \bar{\nu}_\alpha \bar{\nu}_\beta$.

The emitted spectrum per volume and time for $\phi$ from SNe is, following refs.~\cite{Farzan:2002wx,Akita:2022etk},
\begin{align}
    \frac{dn_\phi}{dE_\phi dt}= \sum\frac{Sg_{\alpha\beta}^2 }{32\pi^3}\int dE_\nu\ [m_\phi^2-2E_\nu(V_{\nu_\alpha}+V_{\nu_\beta})] f_{\nu_\alpha}(E_\nu)f_{\nu_\beta}(E_\phi-E_\nu),
    \label{phispectrum}
\end{align}
where the summation is over all flavor neutrinos and antineutrinos, $V_{\nu_\alpha}=-V_{\bar{\nu}_\alpha}$ is the effective potential for neutrinos, which is $V_{\nu_\alpha}=V_{\rm CC}\delta_{\alpha e}+ V_{\rm NC}$ with $V_{\rm CC}=\sqrt{2}G_F n_e$ and $V_{\rm NC}=-\frac{1}{2}\sqrt{2}G_F n_n$. The contribution from protons is canceled out by the contribution from electrons in an electrically neutral medium. Typical values of $V_{\nu_\alpha}$ in the SN core are $V_{\nu_e}\simeq-10\ {\rm eV}$ and $V_{\nu_{\mu,\tau}}\simeq -20\ {\rm eV}$. The integration range must satisfy the kinematically allowed relation with $V_{\nu_\alpha}\ll E_\nu$ as follows
\begin{align}
0\leq1-\cos\theta=\frac{m_\phi^2-2E_\phi(V_{\nu_\alpha}+V_{\nu_\beta})}{2E_\nu (E_\phi-E_\nu)}\leq2,
\end{align}
where $\theta$ is the relative angle between the momentums of a majoron and a neutrino.

The secondary neutrino fluence for $\nu_\alpha$ produced by their decays is given by, as in eq.~(\ref{2bodyS}),
\begin{align}
\frac{dN_{\nu_\alpha}}{dE_\nu}
= 
{\rm Br}_{\nu_\alpha}\bar{N}_{\nu_\alpha}\int^{\infty}_{E_{\rm min}} dE_\phi \frac{1}{p_\phi}  \frac{dN_\phi}{dE_\phi},
\end{align}
where $E_{\rm min}=E_\nu+m_\phi^2/4E_\nu$.
The suppression factor by the time delay as in eq.~(\ref{SZ}) is also negligible 
for $\phi\rightarrow \nu\nu$ in the parameter region of our interest.

\section{Limits from SN 1987A and a future galactic supernova}
\label{sec5}

\subsection{Event rates and statistical analysis}
\label{sec5.1}

As observations of SN 1987A neutrinos, we use the data from 
Kamiokande-II (with a fiducial mass of 0.78 kton) \cite{Fiorillo:2022cdq} and IMB (6.8 kton) \cite{IMB:1988suc}. For future galactic SN neutrino observations, we consider the detectors of Hyper-Kamiokande (HK) (220 kton) \cite{Hyper-Kamiokande:2018ofw}.
SN neutrinos are dominantly captured by the inverse beta decay (IBD), $\bar{\nu}_e + p \rightarrow e^+ + n$, up to $E_\nu \lesssim 70\ {\rm MeV}$ while for $E_\nu \gtrsim 70\ {\rm MeV}$, they are captured by the charged current (CC) processes on oxygen, $\bar{\nu}_e + \mathrm{O}\rightarrow e^+ + \mathrm{X}$ and $\nu_e + \mathrm{O}\rightarrow e^- + \mathrm{Y}$, where X and Y are nuclei at the final state.
We also account for the CC processes of $\nu_\mu$ and $\bar{\nu}_\mu$ above the muon production threshold, $E_\nu \gtrsim m_\mu=106\ {\rm MeV}$.
The produced muons rapidly go to rest by ionization and decay into $e^\pm$, whose spectrum is the known Michel $e^{\pm}$ spectrum ending at $m_\mu/2=53\ {\rm MeV}$. 
Below the muon Cherenkov threshold of $E_{\mu^\pm}=160\ {\rm MeV}$, these events are called invisible muons. 
This signal might be sub-dominant because its energy range overlaps with that for the standard SN neutrinos.
Above this threshold, visible $\mu^\pm$ can contribute to the SN events. 
However, the Cherenkov threshold behavior and the detection efficiency for visible $\mu^\pm$ and Michel $e^\pm$ above $E_\mu= 160\ {\rm MeV}$ in the IMB and Kamiokande-II are not well known. We neglect the visible $\mu^\pm$ signal as in ref.~\cite{Fiorillo:2022cdq}, but the result would not change significantly because the secondary neutrino fluence is suppressed in the related energy region.
We conservatively assume the data-taking time for high-energy SN neutrino events in Kamiokande-II and IMB is $t_{\rm data}= 10\ {\rm s}$.
For HK, we assume $t_{\rm data}=10^3\ {\rm s}$ and neglect background except for the standard SN neutrino events for simplicity. We expect the other backgrounds such as atmospheric neutrino events in the energy region of $1\textbf{--}1000\ {\rm MeV}$ are negligible for $t_{\rm data}=10^3\ {\rm s}$ by rough rescaling a rate of 2/day in the region of $20\textbf{--}2000\ {\rm MeV}$ for atmospheric neutrino events \cite{Bionta:1987qt}.
As a result, the value of $t_{\rm data}$ affects only the limit on HNLs in this work.

The number of the events in an earth-based detector is approximately given by 
\begin{align}
\frac{dN_e}{dE_e}(E_e)&=\frac{1}{4\pi d_{SN}^2}\sum_{x=p,\mathrm{O}}N_x \nonumber \\
&\ \ \ \ \times\left[\sum_{\nu=\nu_e,\bar{\nu}_e}\frac{dN_\nu}{dE_\nu}(E_\nu)\sigma_{\nu-x}(E_e,E_\nu) +\frac{dn_e}{dE_e}\sum_{\nu'=\nu_\mu,\bar{\nu}_\mu}\int dE_\nu' \frac{dN_{\nu'}}{dE_{\nu'}}(E_{\nu'})\sigma_{\nu'-x}(E_e,E_{\nu'}) \right],
\end{align}
where $N_{p}=6.69\times 10^{31}$ and $N_{\mathrm{O}}
=3.35\times 10^{31}$ are respectively the number of free proton and oxygen targets in 1~kton of the water Cerenkov detector. $1/(4\pi d_{SN}^2)$ is the geometrical dilution factor with $d_{SN}=50\ {\rm kpc}$ for SN 1987A. We assume $d_{SN}=10\ {\rm kpc}$ for a future galactic SN, which is the average distance~\cite{Mirizzi:2006xx}. $dN_\nu/dE_\nu$ is the neutrino spectrum from a supernova. $dn_e/dE_e$ is the Michel $e^\pm$ spectrum given by
\begin{align}
\frac{dn_{e}}{dE_{e}} = \frac{4}
{m_{\mu}}\left(\frac{2E_{e}}{m_{\mu}} \right)^{2}\left(3-\frac{4E_{e}}{m_{\mu}} \right)
\end{align}
with a cutoff at $E_e=m_\mu/2=53\ {\rm MeV}$.
$\sigma_{\nu-x}$ is the cross section between a neutrino and a nucleus.
We take the IBD cross sections for $\bar{\nu}_e$ from ref.~\cite{Strumia:2003zx} and for $\bar{\nu}_\mu$ from ref.~\cite{Formaggio:2012cpf}, and the CC cross sections for $\overset{(-)}{\nu}_e$ with $\mathrm{^{16}O}$ from ref.~\cite{Kolbe:2002gk} and for $\overset{(-)}{\nu}_\mu$ from ref.~\cite{Marteau:1999zp} (see also FIG. S1 in ref.~\cite{Fiorillo:2022cdq}).
The number of events with detected (reconstructed) energy with an energy resolution $\delta(E)$ and a detection efficiency $\epsilon(E)$ is given by
\begin{align}
\frac{dN_{e}}{dE_{e}^{\text{det}}} (E_{e}^{\text{det}}) = 
\int dE_{e}\ \epsilon(E_e) \frac{1}{\sqrt{2\pi}\delta(E_{e})} 
\text{exp}\bigg[-\frac{\big(E_{e}-E_{e}^{\text{det}}\big)^{2}}{2\delta^{2}(E_{e})} \bigg] 
\frac{dN_{e}}{dE_{e}} (E_{e}). 
\label{Espectrum}
\end{align} 
The energy resolution $\delta(E)$ is taken as $\sqrt{(1.35\,{\rm MeV})E}$ for IMB and $\sqrt{(0.75\,{\rm MeV})E}$ for Kamiokande-II \cite{Jegerlehner:1996kx}. For HK, we assume that $\delta(E)$ is the same with the SK-IV observation, $\delta(E)/E=0.0397(E/{\rm MeV})+0.349\sqrt{(E/{\rm MeV})}-0.0839$~\cite{Super-Kamiokande:2016yck}.
The detection efficiencies $\epsilon(E)$ are taken from ref.~\cite{Jegerlehner:1996kx,Fiorillo:2022cdq} for IMB and Kamiokande-II. For HK, we take $\epsilon(E)=1$. Some examples of $e^\pm$ spectra are shown in appendix~\ref{appa}.

To impose limits on light particle decays from SN 1987A and future galactic SNe, we perform a maximum likelihood analysis, following refs.~\cite{Jegerlehner:1996kx,Mirizzi:2005tg,Fiorillo:2022cdq}.
For water Cherenkov detectors, we define a likelihood,
\begin{align}
\mathcal{L}&\propto \exp\left[-\int^{E_{\rm high}}_{E_{\rm low}}\frac{dN_e}{dE_e^{\rm det}}dE_e^{\rm det} \right]\prod_i^{N_{\rm bin}} N_e^{N_i}  \nonumber \\
&\propto \exp\left[-\int^{E_{\rm high}}_{E_{\rm low}}\frac{dN_e}{dE_e^{\rm det}}dE_e^{\rm det} \right]\prod_i^{N_{\rm obs}}\frac{dN_e}{dE_e^{\rm det}}(E_i)
\end{align}
where $N_e$ is the number of events with an energy bin, $E_i$ and $N_i$ are the actually observed energies and number of events with an energy bin. $N_{\rm bin}$ and $N_{\rm obs}$ are the number of energy bins and the total number of events. The normalization constant is not important because we will consider only the likelihood ratio. 
For the SN 1987A, we consider the product of the individual likelihoods of IMB and Kamiokande-II.
According to the SN 1987A data, the relevant regions are $(E_{\rm low}, E_{\rm high})=(7.5, 50)\ {\rm MeV}$ for Kamiokande-II and $(19, 75)\ {\rm MeV}$ for IMB \cite{Jegerlehner:1996kx,Fiorillo:2022cdq}.
However, no other triggers except for muon-trigger events were observed in the IMB \cite{Fiorillo:2022cdq}. We do not impose a upper cutoff of energy for the IMB.
For HK, we consider $(E_{\rm low}, E_{\rm high})=(10, 1000)\ {\rm MeV}$ and an energy bin of 10 MeV.
As the actual observational data for SN 1987A, we use FIG.~S2 in ref.~\cite{Fiorillo:2022cdq}. As the expected data for a galactic SN, we use eq.~(\ref{STSN}) with $(E_{\rm tot},E_0,\alpha)=(1.98\times 10^{53}\ {\rm erg},\ 12.7\ {\rm MeV},\ 2.39)$ for all $\nu$ and $\bar{\nu}$ for simplicity.

SN simulations might have a tension with the SN 1987A data \cite{Li:2023ulf,Fiorillo:2023frv}. In addition, the data is not informative about the pinching parameter $\alpha$ \cite{Mirizzi:2005tg}. So we fix $\alpha=2.39$ based on the SN model SFHo-18.8 and conservatively marginalize the likelihood over the other SN model parameters, $E_0$ and $E_{\rm tot}$,
\begin{align}
\tilde{\mathcal{L}}(g,m)=\max_{E_0, E_{\rm tot}}\mathcal{L}(g,m,E_0,E_{\rm tot}),
\end{align}
where $g$ and $m$ are the coupling and mass for hypothesis particles.
We confirm the marginalized likelihood $\tilde{L}$ is maximum at $g=0$, i.e., the new signal is not preferred as in ref.~\cite{Fiorillo:2022cdq}.
Then we define a test statistics,
\begin{align}
\chi^2=2\left[\log \tilde{L}(0,m)-\log\tilde{L}(g,m_\phi) \right].
\end{align}
We compute threshold values of $(g, m)$ for 95\% C.L. exclusion limits at $\chi^2=2.7$. 

\subsection{Heavy neutral lepton}
\label{sec5.2}

\begin{figure}
\vspace{-28mm}
 \begin{minipage}{1\hsize}
   \begin{center}
     \includegraphics[clip,width=16cm]{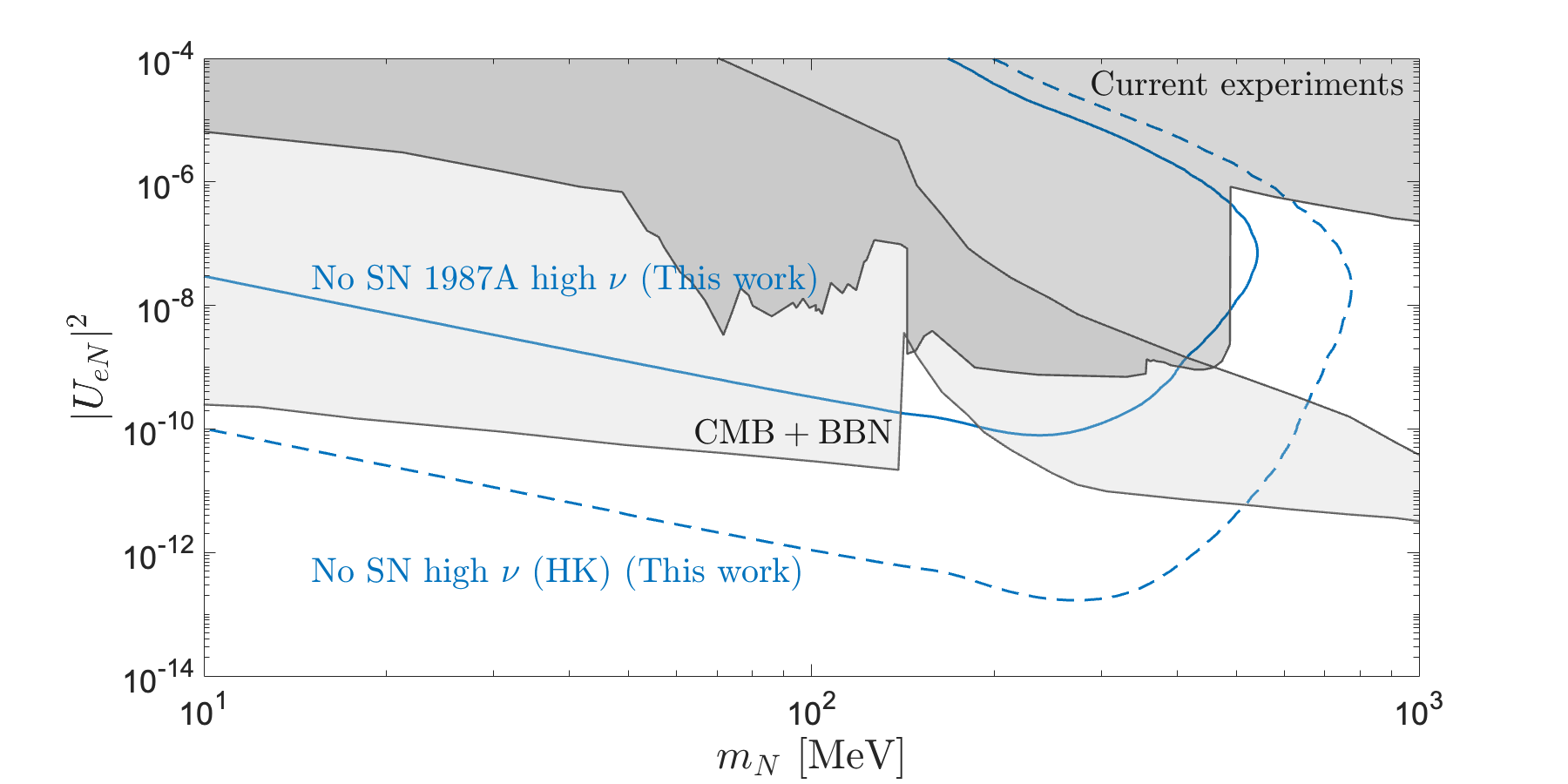}
    \end{center}
     \end{minipage} \vspace{-1mm} \\
    \begin{minipage}{1\hsize}
   \begin{center}
     \includegraphics[clip,width=16cm]{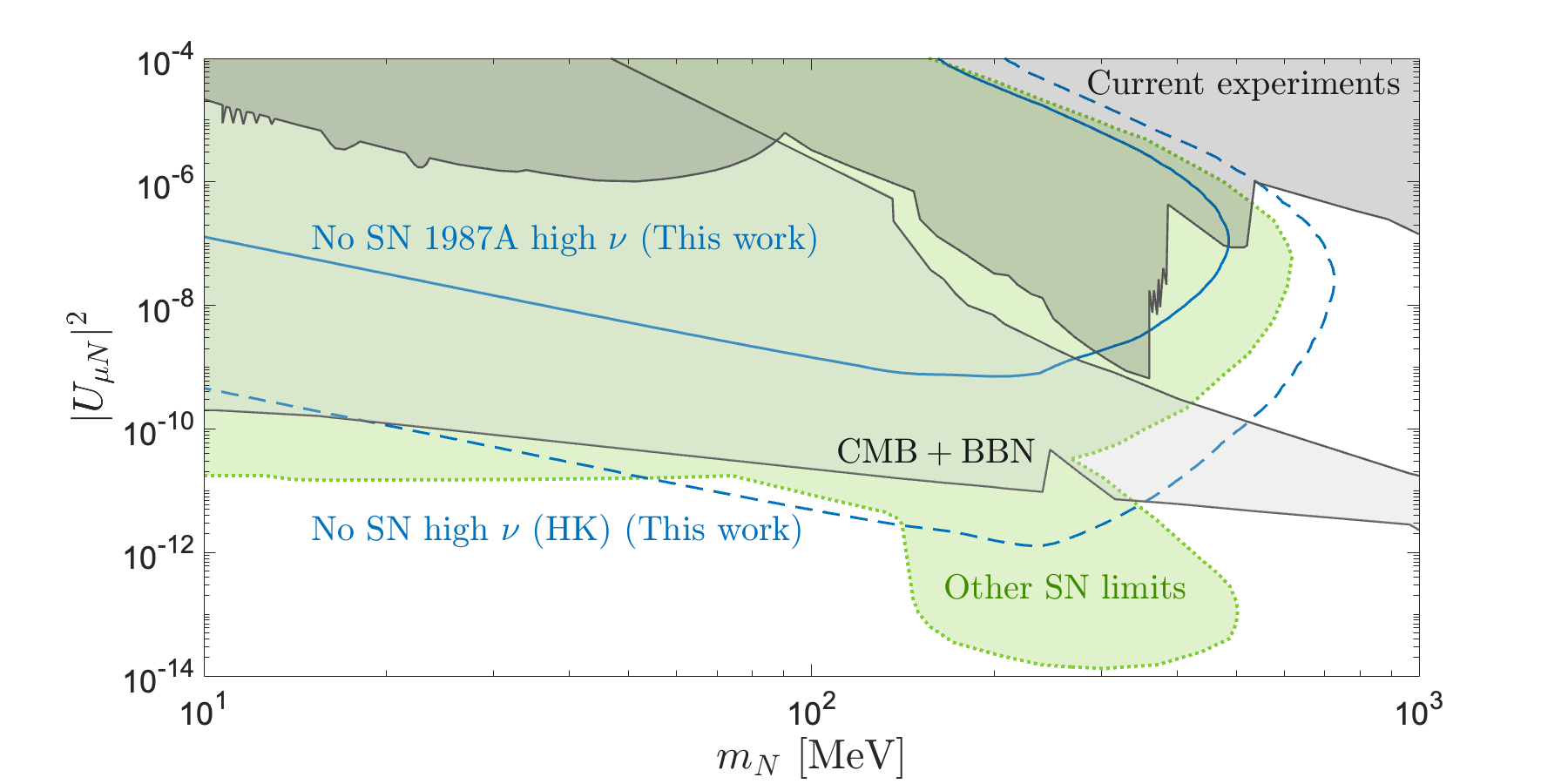}
    \end{center}
     \end{minipage} \vspace{-1mm} \\
      \begin{minipage}{1\hsize}
      \begin{center}
     \includegraphics[clip,width=16cm]{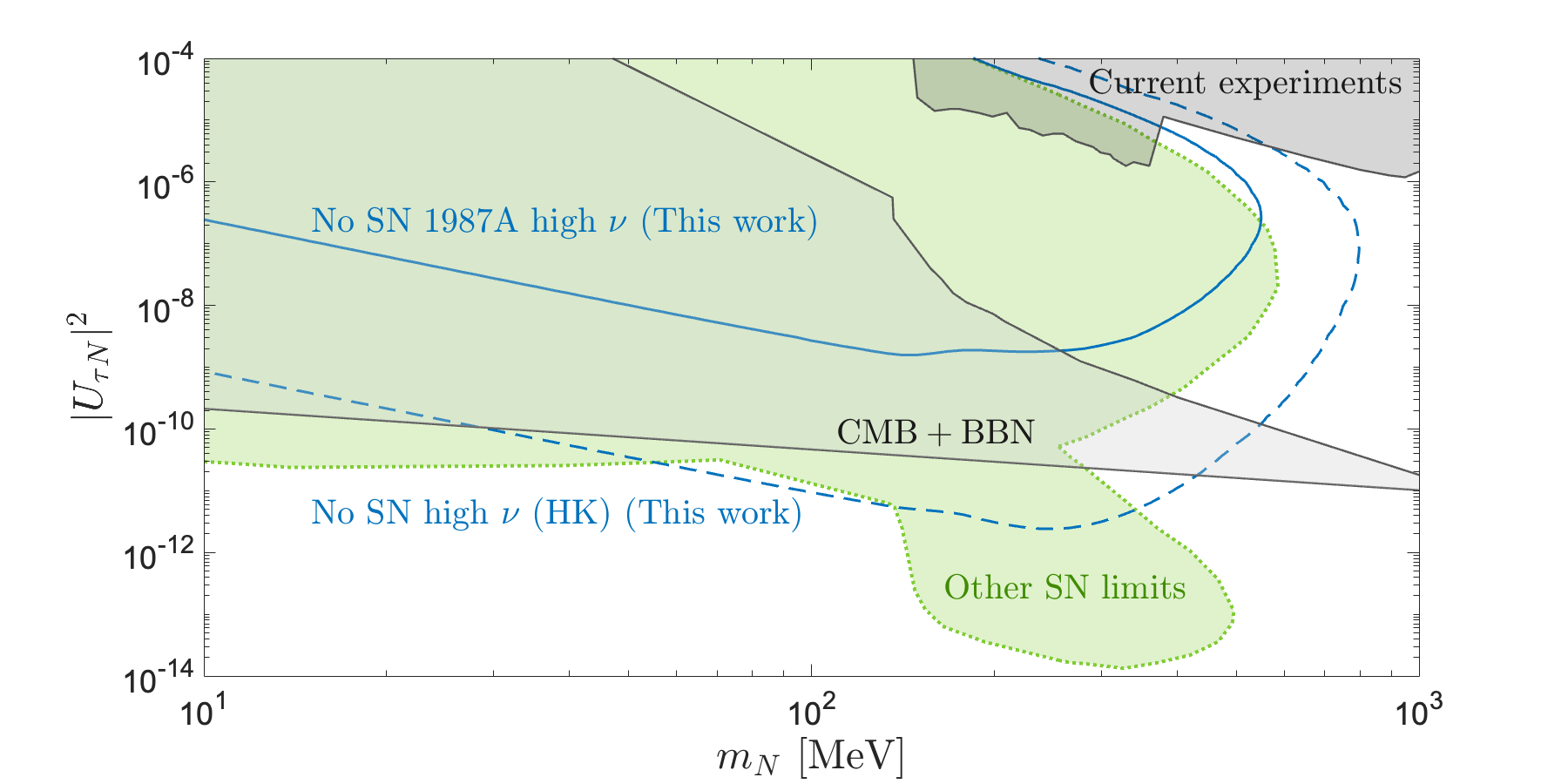}
    \end{center}
     \end{minipage}
    \vspace{-4mm}
	\caption{\small{SN1987A constraint on heavy neutral leptons mixing with $\nu_e$ (upper panel), $\nu_\mu$ (middle panel) and $\nu_\tau$ (bottom panel) from no observations of high energy neutrino (high $\nu$) events (blue solid line) and future sensitivity from observations of galactic high energy SN neutrinos in HK with $d_{\rm SN}=10\ {\rm kpc}$ (blue dashed line). The results in the NO and IO cases are the same within $20\%$ level. We also show the other SN limits (green shaded region enclosed by dotted line) \cite{Carenza:2023old} and other relevant constraints (see text for details).}}
 \label{fig:Limit_HNL_SN1987A}
\end{figure}

Figure~\ref{fig:Limit_HNL_SN1987A} presents our limits (blue solid line) on HNLs mixing with $\nu_e$ (upper panel), $\nu_\mu$ (middle panel) and $\nu_\tau$ (bottom panel) from no observations for the high-energy events of SN 1987A neutrinos in Kamiokande-II and IMB. We also show the future sensitivity (blue dashed line) on HNLs from the observations of galactic supernova neutrinos in HK with $d_{\rm SN}=10\ {\rm kpc}$.
We confirm the results in the NO and IO cases are the same within 20 $\%$ level and show the NO case in figure~\ref{fig:Limit_HNL_SN1987A}.
In this figure we include other relevant constraints from the other supernova arguments \footnote{When this work appeared on the arXiv, no work imposes constraints on HNLs mixing with $\nu_e$, accounting for the dominant production process of HNLs in the core, $e^-p\rightarrow nN$. Using other arguments from SN observations as ref.~\cite{Carenza:2023old} and considering the production process, $e^-p\rightarrow nN$, we may further constrain HNLs mixing with $\nu_e$.} (green shaded region enclosed by dotted line) \cite{Carenza:2023old}, BBN \& CMB assuming the standard cosmology (light-gray shaded region) \cite{Boyarsky:2009ix,Ruchayskiy:2012si,Sabti:2020yrt,Mastrototaro:2021wzl,Boyarsky:2020dzc,Carenza:2023old} and experiments (dark-gray shaded region) \cite{Bolton:2019pcu,Ema:2023buz} (see also the sensitivities for upcoming and proposed experiments, DUNE \cite{Krasnov:2019kdc,Ballett:2019bgd}, MATHUSLA \cite{Curtin:2018mvb}, SHiP \cite{SHiP:2018xqw,Gorbunov:2020rjx}, PIP2-BD \cite{Ema:2023buz} and PIONEER \cite{PIONEER:2022alm}).
We should note that the BBN constraints can be significantly weakened in presence of a large neutrino asymmetry \cite{Gelmini:2020ekg} or with low reheating temperatures \cite{Gelmini:2004ah,Gelmini:2008fq}.

Our constraints are one or two orders of magnitude weaker than the other SN limits (in particular, the observations of $\gamma$-ray by HNL decays and the excessive energy deposition by HNL decays in the SN envelope) \cite{Carenza:2023old}. 
Observations of future galactic SN neutrinos in HK with $d_{\rm SN}=10\ {\rm kpc}$ would have an order stronger sensitivities on $|U_{\tau N,\tau N}|^2$ than the current SN limits in the mass region of $300\ {\rm MeV}\lesssim m_N \lesssim 800\ {\rm MeV}$. These future observations would also be complementary to the current limits and future sensitivities of experiments. These sensitivities would improve our limits on $|U_{eN,\mu N,\tau N} |^2$ by about 3 orders of magnitude in this mass region, roughly corresponding to the ratios of the dilution factor due to the SN distance and the detector mass, $(10\ {\rm kpc}/50\ {\rm kpc})^{-2} \times (220\ {\rm kton}/6.8\ {\rm kton}) \simeq   8\times 10^2$.
For $m_{N}\lesssim 100\ {\rm MeV}$, the limit is suppressed because the time delay of the secondary flux events from the standard ones begins to exceed the data-taking time. For $m_{N}\gtrsim 100\ {\rm MeV}$, the limit is also suppressed because the production of heavy $N$ is suppressed by the Boltzmann factor.

We should note that the SN constraints on HNLs highly depend on the SN core model (temperature, chemical potentials, etc.) because the production rate of HNLs in the core includes 9 phase-space integrals as in eq.~(\ref{BE}). The difference between the SN models would induce a large uncertainty and our reference SN core model is different from ref.~\cite{Carenza:2023old}. We find that the constraints on HNLs mixing with $\nu_\mu$ or $\nu_\tau$ from the SN energy loss with our reference core model are a several factor weaker than those with the SN model in ref.~\cite{Carenza:2023old} (and we show the result of ref.~\cite{Carenza:2023old} in figure~\ref{fig:Limit_HNL_SN1987A}) \footnote{We adopt that $L_{\rm BSM}$ must not exceed  $L_\nu^{\rm st}= 4\times 10^{52}\ {\rm erg\ s^{-1}}$ at 1 s of the post bounce \cite{Fiorillo:2022cdq} as the SN energy loss argument, $L_{\rm BSM}< L_\nu^{\rm st}$, where $L_{\rm BSM}$ is the luminosity carried away by new particles to the outside of the neutrinosphere and $L_{\nu}^{\rm st}$ is the luminosity that would carried away by neutrinos.}.
Though our argument is different from the SN energy loss argument, this will also contain a similar uncertainty from the SN core models. Our SN core model is more conservative than in ref.~\cite{Carenza:2023old} (at least for the SN energy loss argument).

\subsection{${\rm U(1)}_{L_\mu-L_\tau}$ gauge boson}
\label{sec5.3}

\begin{figure}[t]
	\begin{center}
	\includegraphics[width=17cm]{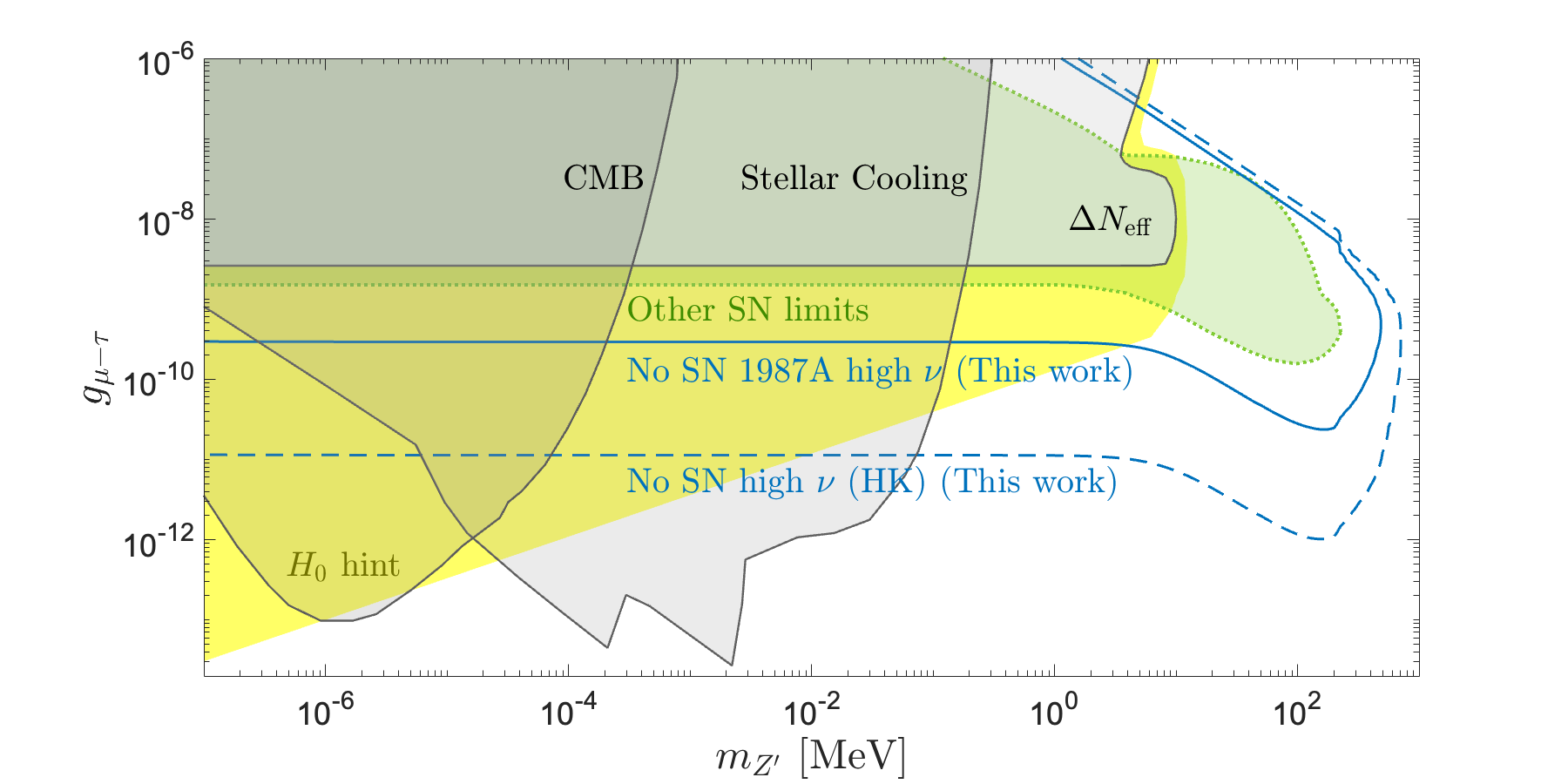}
	\end{center}
	 \vspace{-8mm}
	\caption{\small{SN1987A constraint on ${\rm U(1)}_{L_\mu-L_\tau}$ gauge bosons from no observations of high energy neutrino (high $\nu$) events (blue solid line) and future sensitivity from observations of galactic high energy SN neutrinos in HK with $d_{\rm SN}=10\ {\rm kpc}$ (blue dashed line). The results are the same in the NO and IO cases within 20\% level. Our results are applicable for the kinetic mixing up to a natural one $|\varepsilon|\lesssim g_{\mu-\tau}/70$. We show the constraint from SN 1987A energy loss argument (green shaded region enclosed by dotted line) \cite{Croon:2020lrf,Cerdeno:2023kqo} and other constraints in the case of a natural kinetic mixing, $\varepsilon=-g_{\mu-\tau}/70$ (see text for details). The preferred regions to explain the $H_0$ tension (yellow band) \cite{Escudero:2019gvw} are also shown.}}
 \label{fig:Limit_LmuLtau_SN1987A}
\end{figure}

Figure~\ref{fig:Limit_LmuLtau_SN1987A} presents our limit (blue solid lines) and future sensitivity in HK with $d_{\rm SN}=10\ {\rm kpc}$ (blue dashed lines) on ${\rm U(1)}_{L_\mu-L_\tau}$ gauge bosons, $Z'$.
Our constraint and future sensitivity are applicable for the small kinetic mixing up to a natural one, $|\varepsilon| \lesssim g_{\mu-\tau}/70$ in eq.~(\ref{epsilonnatural}).
We confirm the results in the NO and IO cases are the same within 20\% level.
We include other relevant constraints from the other SN 1987A arguments (green shaded region enclosed by dotted line) \cite{Croon:2020lrf,Cerdeno:2023kqo}, contribution of $Z'$ bosons to the effective number of neutrino species $\Delta N_{\rm eff}$ (light-gray shaded region) \cite{Escudero:2019gvw}, stellar cooling (mid-gray shaded region), which is rescaled from \cite{An:2013yfc,Hardy:2016kme}, and CMB (dark-gray shaded region) \cite{Sandner:2023ptm}.
We also show the preferred regions to explain the Hubble constant tension (yellow band) \cite{Escudero:2019gvw}.
Note that constraints by the stellar cooling stem from kinetic mixing of $Z'$ with photons and electrons and depend on the value of the kinetic mixing, $\varepsilon$. In figure~~\ref{fig:Limit_LmuLtau_SN1987A}, the constraints by the stellar cooling with a natural value of $\varepsilon=-g_{\mu-\tau}/70$ are shown. For $\varepsilon=0$, these constraints are invalid.

Our lower constraint is the strongest in the mass region of $0.1\ {\rm MeV}\lesssim m_{Z'}\lesssim 400\ {\rm MeV}$. Compared with the constraint from SN 1987A energy loss \cite{Croon:2020lrf}, our constraint on the coupling $g_{\mu-\tau}$ is improved by a factor of 3--10. Observations of future galactic SN neutrinos in HK with $d_{SN}=10\ {\rm kpc}$ would improve our limit on the coupling by a factor of 26, corresponding to the improvements of the SN distance squared times the detector mass. 
The results do not depend on the data-taking time $t_{\rm data}$ because of the short time delay of the secondary fluxes.
For $m_{Z'}\gtrsim 10\ {\rm MeV}$, the production processes of neutrino-pair coalescence ($\nu\bar{\nu}\rightarrow Z'$) are dominant while the production process of semi-Compton scattering ($\mu \gamma\rightarrow \mu Z'$) is dominant below  $Z'$ mass of 10 MeV. This is because the production rate for neutrino-pair coalescence in eq.~(\ref{BEZ}) is proportional to $m_{Z'}^2$ while that for constant semi-Compton scattering with light $Z'$ does not depend on $m_{Z'}$. Our upper limit is relaxed by a factor of a few compared with the SN limit in ref.~\cite{Croon:2020lrf}. This is because our treatment of $Z'\rightarrow \nu\bar{\nu}$ in the trapping case is different from that in ref.~\cite{Croon:2020lrf}.
 
We find the lower limit on ${\rm U(1)}_{L_\mu-L_\tau}$ gauge bosons from the SN energy loss with our SN model are weaker than the conservative limit in ref.~\cite{Croon:2020lrf} by a factor of 1.6 (and we show the result of ref.~\cite{Croon:2020lrf} in figure~\ref{fig:Limit_LmuLtau_SN1987A}). 
In ref.~\cite{Croon:2020lrf}, this limit is derived in the two conservative and optimistic SN models and their difference is a factor of $\sim 4$. The limit from our new argument will also contain a similar uncertainty but our SN reference model is conservative.

\begin{figure}[t]
	\begin{center}
	\includegraphics[clip,width=17cm]{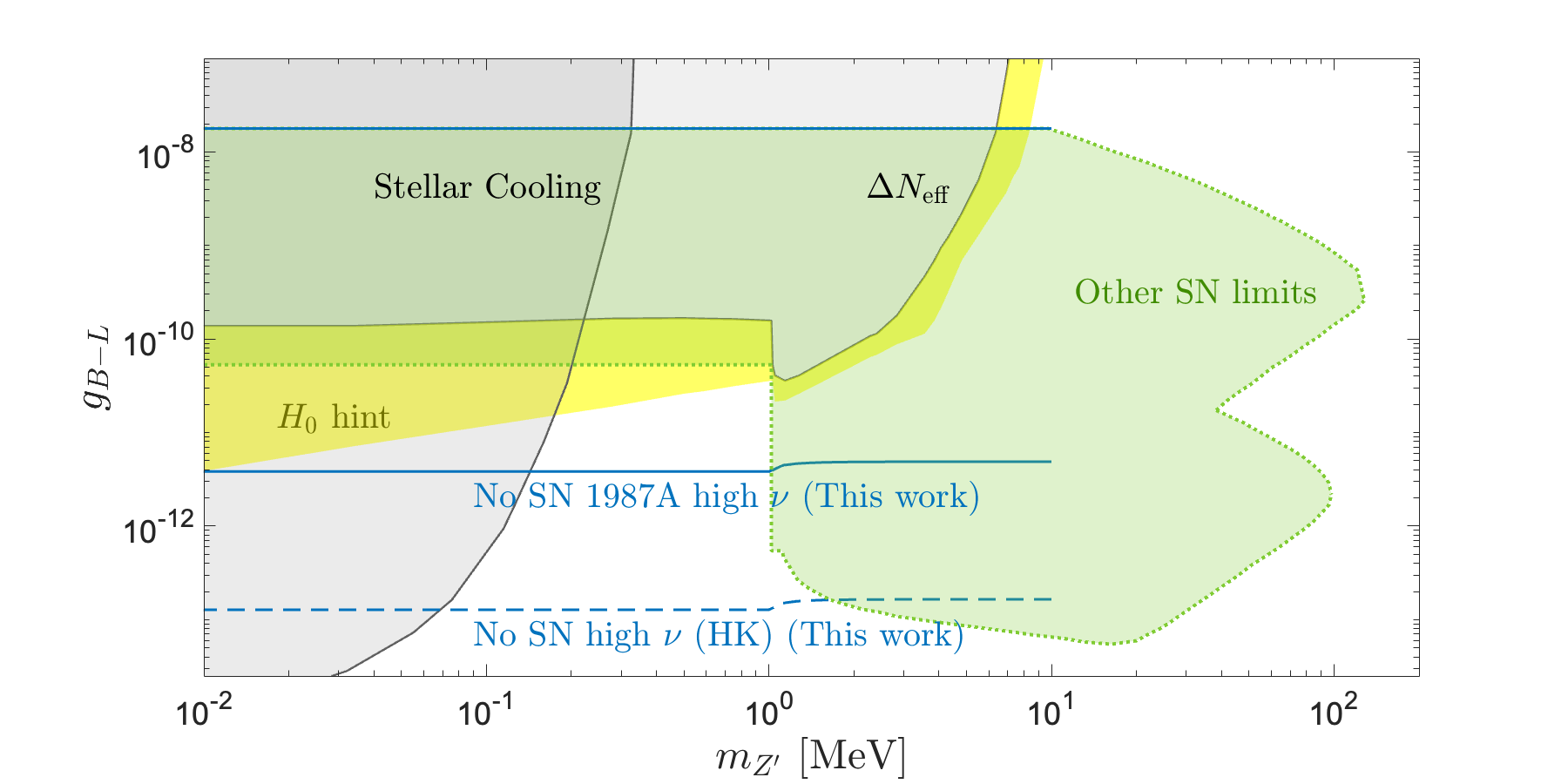}
	\end{center}
	 \vspace{-8mm}
	\caption{\small{SN1987A constraints on ${\rm U(1)}_{B-L}$ gauge bosons from no observations of high energy neutrino (high $\nu$) events (solid lines) and future sensitivity from observations of galactic high energy SN neutrinos in HK with $d_{\rm SN}=10\ {\rm kpc}$ (dashed lines). We also show the other SN limits (green shaded region enclosed by dotted lines) \cite{Shin:2022ulh}, and other constraints from $\Delta N_{\rm eff}$ assuming Majorana neutrinos \cite{Esseili:2023ldf}(see also figure 6 in ref.~\cite{Esseili:2023ldf} for Dirac neutrinos) and stellar cooling \cite{An:2014twa,Hardy:2016kme,Hong:2020bxo}. The preferred regions to explain the $H_0$ tension (yellow band) \cite{Esseili:2023ldf} are also shown.}}
 \label{fig:Limit_B_L_SN1987A}
\end{figure}

\subsection{${\rm U(1)}_{B-L}$ gauge boson}
\label{sec555}

Figure~\ref{fig:Limit_B_L_SN1987A} presents our limits (blue solid lines) and future sensitivity in HK with $d_{\rm SN}=10\ {\rm kpc}$ (blue dashed lines) on ${\rm U(1)}_{B-L}$ gauge bosons. In this case, $Z'$ bosons decay to neutrinos flavor-universally so that there is no effect of neutrino oscillations and our results are independent of the neutrino mass ordering. In this figure we include other relevant constraints from the other SN arguments (green shaded region enclosed by dotted line) \cite{Shin:2022ulh}, contribution of $Z'$ to the effective number of neutrino species $\Delta N_{\rm eff}$ assuming Majorana neutrinos (light-gray shaded region) \cite{Heeck:2014zfa,Knapen:2017xzo,Esseili:2023ldf} (see ref.~\cite{Esseili:2023ldf} for the case of Dirac neutrinos) and stellar cooling (dark-gray shaded region) \cite{An:2014twa,Hardy:2016kme,Hong:2020bxo}.
We also show the preferred regions to explain the Hubble constant tension (yellow band) \cite{Esseili:2023ldf}.
We do not show the SN limit by the $Z'$ production from nucleon-nucleon bremsstrahlung because of the large uncertainty of the production rate. See ref.~\cite{Shin:2021bvz} for interested readers.
For $m_{Z'}\geq 2 m_e$, the observations of $\gamma$-ray by $Z'$ decays produced in the SN core and the excessive energy deposition by $Z'$ decays in the SN envelope have already imposed stringent constraints on $Z'$ \cite{Shin:2022ulh}. For $m_{Z'}\geq T$, our constraint becomes weaker by the Boltzmann suppression $e^{-m_{Z'}/T}$. We only compute and show our constraints and future sensitivity for $m_{Z'}\lesssim 2 m_e$ \footnote{For $m_{Z'}\lesssim 2 m_e$, the kinetic mixing coupling $\varepsilon$ is constrained much more stringently by cosmology than SN 1987A observations \cite{Kazanas:2014mca,Shin:2022ulh}. We assume $\varepsilon=0$ in the ${\rm U(1)}_{B-L}$ model and impose a supernova limit on $g_{B-L}$ in this work.}.
However, the future sensitivity for $m_{Z'}\geq T$ in HK may still reach the unconstrained region.
For the strong coupling in a trapping regime, our constraint is only connected to the upper bound for the trapping regime in ref.~\cite{Shin:2022ulh}.

Our lower limits are the most stringent in the region of $0.1\ {\rm MeV}\lesssim m_{Z'}\lesssim 1\ {\rm MeV}$.
Compared with the SN energy loss argument \cite{Shin:2022ulh}, our constraint on the coupling $g_{B-L}$ is improved by a factor of 14.
Observations of future galactic SN neutrinos in HK with $d_{SN}=10\ {\rm kpc}$ would improve our limit on the coupling by a factor of 30, corresponding to the improvements of the SN distance squared and the detector mass. The results are approximately independent of $t_{\rm data}$.
The previous SN limits are weaker for $m_{Z'}\lesssim 2m_e \sim 1\ {\rm MeV}$. This is because $Z'$ bosons with $m_{Z'}\lesssim 2m_e$ cannot decay to charged particles and $\gamma$-ray telescopes cannot observe a secondary flux from the produced charged particles.

Our improvement on ${\rm U(1)}_{B-L}$ gauge bosons is stronger than that on ${\rm U(1)}_{L_\mu-L_\tau}$ gauge bosons. This might be because the produced ${\rm U(1)}_{B-L}$ gauge bosons would have a typical energy of $\sim m_{\pi^-}+3T+\alpha_{n,p}\mu_{n,p}$ due to the production process of $\pi^- p \rightarrow Z'n$ while ${\rm U(1)}_{L_\mu-L_\tau}$ gauge bosons would have a typical energy of $\sim 3T+\alpha_\mu \mu_\mu$. $\alpha_{n,p,\mu}$ is a suppression factor of the chemical potential due to the Pauli-blocking effects, but we could not estimate $\alpha_{n,p,\mu}$ quantitatively. Since the detection rate in the Water Cherenkov detectors is roughly $\sigma_\nu\propto E_\nu^2$, neutrinos produced by the decays of ${\rm U(1)}_{B-L}$ gauge bosons might be detected more efficiently.

We find the limit on ${\rm U}(1)_{B-L}$ gauge bosons from the SN energy loss with our SN reference model are 30 \% stronger than that with the SN model in ref.~\cite{Shin:2022ulh} (We show the result of ref.~\cite{Shin:2022ulh} in figure~\ref{fig:Limit_B_L_SN1987A}). Comparison with other SN models and a detailed discussion of uncertainty in our arguments are left for future work because there is no other previous work on pion-induced production of ${\rm U}(1)_{\rm B-L}$ gauge bosons in SNe.

\subsection{Majoron}
\label{sec5.4}

\begin{figure}
	\begin{center}
	\includegraphics[clip,width=17cm]{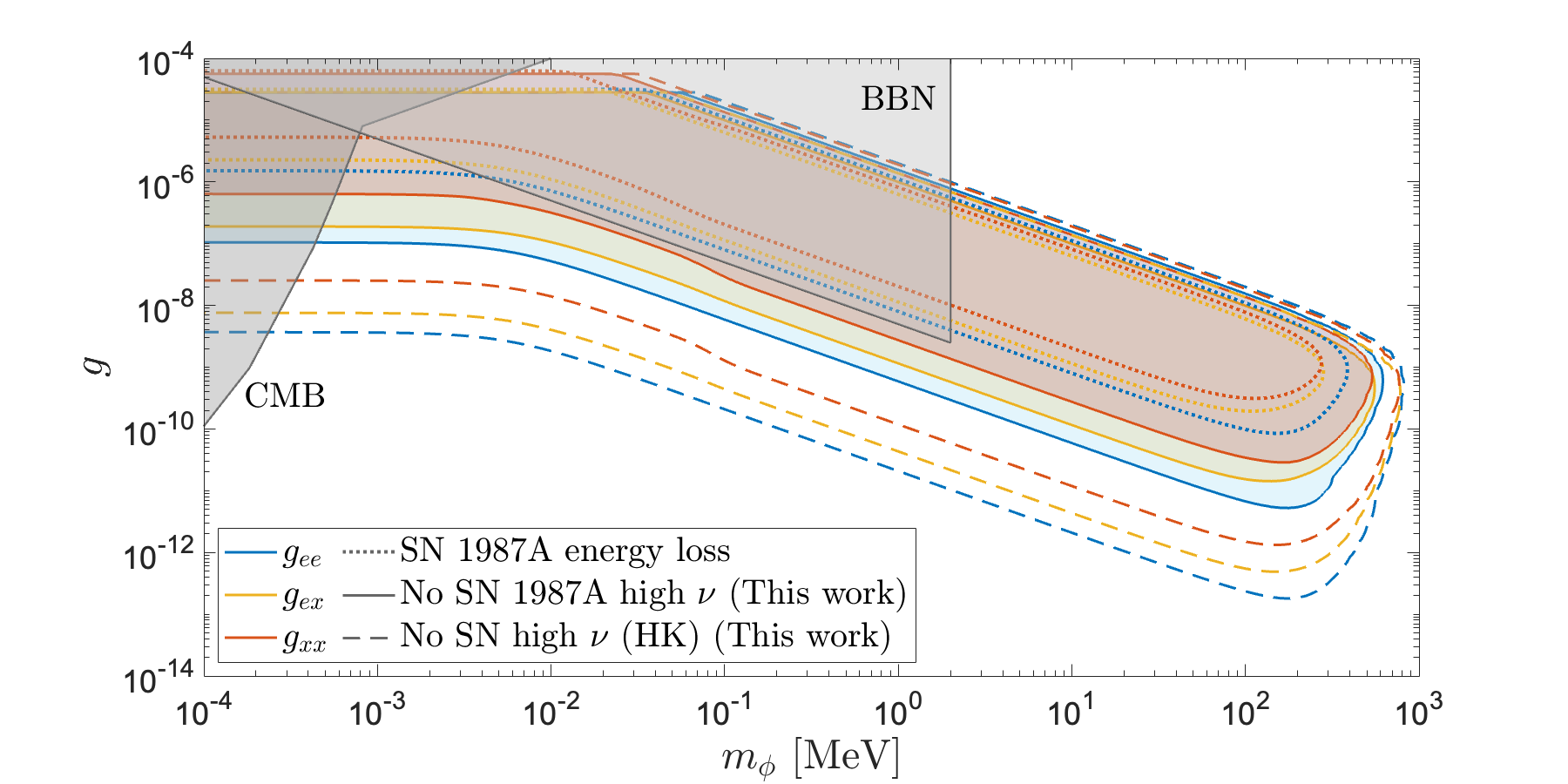}
	\end{center}
	 \vspace{-8mm}
	\caption{\small{SN1987A constraints on flavored majorons with couplings of $g_{ee}$ (blue), $g_{ex}$ (yellow) or $g_{xx}$ $(x=\mu,\tau)$ (red) from no observations of high energy neutrino (high $\nu$) events (solid lines) and future sensitivity from observations of galactic high energy SN neutrinos in HK with $d_{\rm SN}=10\ {\rm kpc}$ (dashed lines). The results are the same in the NO and IO cases and the cases of $x=\mu$ and $x=\tau$ within 20\% level. We also show constraints from the SN 1987A energy loss argument (dotted lines) (We recalculate them based on the SN model SFHo 18.8) \cite{Heurtier:2016otg,Farzan:2002wx}, BBN \cite{Escudero:2019gvw,Huang:2017egl} and CMB \cite{Archidiacono:2013dua,Sandner:2023ptm}.}}
 \label{fig:Limit_majoron_SN1987A}
\end{figure}

Figure~\ref{fig:Limit_majoron_SN1987A} presents our limits (solid lines) and future sensitivities in HK with $d_{\rm SN}=10\ {\rm kpc}$ (dashed lines) on flavored majorons coupled to neutrinos only with $g_{ee}$ (blue), $g_{ex}$ (yellow) or $g_{xx}$ ($x=\mu,\tau$) (red). We confirm the constraints in the NO and IO cases are the same within 20\% level and the constraints for $x=\mu$ and $\tau$ are also the same within 20\% level.
We include other relevant constraints from the SN 1987A energy loss argument (dotted lines) \cite{Heurtier:2016otg,Farzan:2002wx} (we recalculate this argument based on the SN model SFHo 18.8.), CMB (dark-gray shaded region) \cite{Archidiacono:2013dua,Sandner:2023ptm} and BBN (light-gray shaded region) \cite{Escudero:2019gvw,Huang:2017egl} \footnote{If majorons are thermalized in the early universe, CMB and BBN constrain majoron parameter space more severely and complementary to our limits \cite{Li:2023kuz,Chang:2024mvg}.}.
At $g\sim 10^{-4}$, majorons would enter a trapping regime by rescattering $\nu\phi \rightarrow \nu\phi$ \cite{Heurtier:2016otg}. Our constraints are connected to the upper limits for the trapping case in ref.~\cite{Heurtier:2016otg}.
For $m_\phi \lesssim 10\ {\rm keV}$, the effective potential for neutrinos induced by the electron and 
nucleus background in the core dominates the production rates of majorons, changing the dispersion relation for neutrinos. 
For $m_\phi \lesssim 10\ {\rm keV}$, the production rates and lower bounds do not depend on the majoron mass $m_\phi$.

Our limits are the strongest in the region of $1\ {\rm keV}\lesssim m_{\phi}\lesssim 500\ {\rm MeV}$. Compared with the constraint from SN 1987A energy loss, our constraints on the couplings are improved by a factor of 13 for $g_{ee}$, 10 for $g_{ex}$, and 7 for $g_{xx}$. Observations of future galactic SN neutrinos in HK with $d_{SN}=10\ {\rm kpc}$ would improve our limits on the couplings by a factor of 30, corresponding to the improvements of the SN distance squared and the detector mass.
The results do not depend on the data-taking time $t_{\rm data}$ because of the short time delay of the secondary fluxes.
In each flavor, the lower limit on $g_{ee}$ is stronger than those on $g_{ex}$ and $g_{xx}$ because a huge number of $\nu_e$ is emitted in the SN core due to the large chemical potential for $\mu_e$ in the duration of 2 s. 
The improvement for $g_{ee}$ is also stronger than those for $g_{ex}$ and $g_{xx}$. This would be because the detection cross sections are roughly proportional to $\sigma_\nu \propto E_\nu^2$ and the process $\nu_e\nu_e\rightarrow \phi \rightarrow \nu_e \nu_e$ produces the most energetic secondary neutrinos due to the large chemical potential of $\nu_e$.
The upper limit on $g_{ex}$ is weaker than that on $g_{xx}$ because the decay rate of $\Gamma_{\phi\rightarrow \nu_e\nu_x}$ is larger than $\Gamma_{\phi\rightarrow \nu_x\nu_x}$ due to the symmetric factor in eq.~(\ref{DecayMajoron}).

Our constraint on $g_{ee}$ is very similar to the constraint on majoron coupled to neutrinos flavor-universally \cite{Fiorillo:2022cdq} because in both cases, the dominant production process of majoron in the core is $\nu_e\nu_e\rightarrow \phi$. An uncertainty of our limit on $g_{ee}$ from the SN models would be very small as discussed in ref.~\cite{Fiorillo:2022cdq}. For $g_{ex}$ and $g_{xx}$, the uncertainties of our argument might be large because the production process and rate for $g_{ex}$ and $g_{xx}$ in the core are different from those for $g_{ee}$. We leave an detailed discussion of uncertainties of our argument on $g_{ex}$ and $g_{xx}$, but our SN model is basically conservative as commented in section~\ref{sec2}.
We also recalculate the SN energy loss argument based on the SN model SFHo 18.8 and show these results in figure~\ref{fig:Limit_majoron_SN1987A}. Compared with the results with refs.~\cite{Heurtier:2016otg,Farzan:2002wx}, the limits on majorons from the SN energy loss with our SN model is a few factor weaker or almost the same because of a smaller chemical potential of $\nu_e$ and/or different estimation methods for majoron production rates in the SN core.

\section{Summary}
\label{sec6}

The secondary neutrino fluxes by decays of light hypothetical particles produced in the SN core can modify the high energy tail of the standard SN neutrino flux.
The lack of high energy SN 1987A neutrino events imposes the strong limit on flavor-universal neutrino non-standard interactions with light bosons from SN 1987A neutrino observations \cite{Fiorillo:2022cdq} and can significantly improve the sensitivities on the interactions from future galactic SN neutrino observations \cite{Akita:2022etk}.

In this work, we extend refs. \cite{Akita:2022etk,Fiorillo:2022cdq} to more diverse particle physics models, studying neutrino oscillation effects in the SN envelope on the secondary neutrino fluxes for the first time.
In particular, we obtain the strong constraints and future sensitivity for HNLs, ${\rm U(1)}_{L_\mu-L_\tau}$ and ${\rm U(1)}_{B-L}$ gauge bosons, and flavored majorons.
In refs.~\cite{Mastrototaro:2019vug,Syvolap:2023trc}, the authors considered the secondary fluxes by decays of HNLs and the future sensitivity from observations of galactic supernovae. We improve many calculations such as the production rate of HNLs in the SN core, neutrino oscillations and the event rates at the detector. We also obtain the current limit on HNLs from SN 1987A observations using the above argument for the first time.

For HNLs, we find limits as in figure~\ref{fig:Limit_HNL_SN1987A} from the absence of high energy SN 1987A neutrino events but these limits are one or two orders of magnitude weaker than the current other SN limits \cite{Carenza:2023old}. Future observations of galactic SN neutrino events in HK with $d_{\rm SN}=10\ {\rm kpc}$ will have an order stronger sensitivities on mixing with $\nu_{\tau}$ than the current SN constraints in the mass region of $300\ {\rm MeV}\lesssim m_N \lesssim 800\ {\rm MeV}$.
Note that the SN constraints on HNLs highly depend on the SN model because of the high dependence of the production of HNLs on the energy of incoming and outgoing particles. 

For ${\rm U(1)}_{L_\mu-L_\tau}$ gauge bosons, the absence of high energy SN 1987A neutrinos imposes the strongest limit as in figure~\ref{fig:Limit_LmuLtau_SN1987A}.
We exclude the parameter region with masses of $\lesssim400\ {\rm MeV}$ and coupling down to $g_{\mu-\tau}\sim10^{-10}$. For $m_{Z'}\lesssim 1\ {\rm MeV}$, our limit is an order of magnitude stronger than the limit from observations of $\Delta N_{\rm eff}$ \cite{Escudero:2019gzq}.
Future observations of galactic SN neutrinos in HK will improve the limit on coupling down to $g_{\mu-\tau}\sim 10^{-11}$ in this mass range.

For ${\rm U(1)}_{B-L}$ gauge bosons, the absence of high energy SN 1987A neutrinos imposes the strongest limit as in figure~\ref{fig:Limit_B_L_SN1987A}.
We newly exclude the parameter region with masses of $\lesssim 1\ {\rm MeV}$ and coupling down to $g_{B-L}= 4\times 10^{-12}$. Future observations of galactic SN neutrinos in HK will improve the limit on coupling down to $g_{B-L}=10^{-13}$ in this mass range.

For flavored majorons, the absence of high energy SN 1987A neutrinos imposes the strongest limit as in figure~\ref{fig:Limit_majoron_SN1987A}.
We exclude the parameter region with $g\sim 10^{-7}$ for $m_\phi \lesssim 10\ {\rm keV}$ and $m_{\phi}g\sim 10^{-9}\ {\rm MeV}$ for $10\ {\rm keV} \lesssim m_\phi \lesssim 500\ {\rm MeV}$.
Future observations of galactic SN neutrinos in HK will improve the limit down to $g\sim 10^{-8}$ for $m_\phi \lesssim 10\ {\rm keV}$ and $m_{\phi}g\sim 10^{-11}\ {\rm MeV}$ for $10\ {\rm keV} \lesssim m_\phi \lesssim 800\ {\rm MeV}$.

\section*{Acknowledgments}
This work was supported by IBS under the project code, IBS-R018-D1. 
MM also acknowledges support from the grant NRF-2022R1A2C1009686.


\appendix

\section{Examples of neutrino and electron spectra on Earth}
\label{appa}

We show several examples of $\nu$ and $\bar{\nu}$ spectra (fluences) on Earth and their $e^\pm$ signal spectrum from the time-integrated emission of our SN reference model, accounting for the effect of neutrino oscillations. For the $e^\pm$ spectrum, we assume 1 kton volume of a water Cherenkov detector and the detector efficiency is unity. We do not take into account the energy resolution of the detector. We consider the normal ordering of neutrino masses and the case that a supernova occur at a distance to Earth of 10 kpc.

\begin{figure}
 \begin{minipage}{0.5\hsize}
  \begin{center}
   \includegraphics[width=80mm]{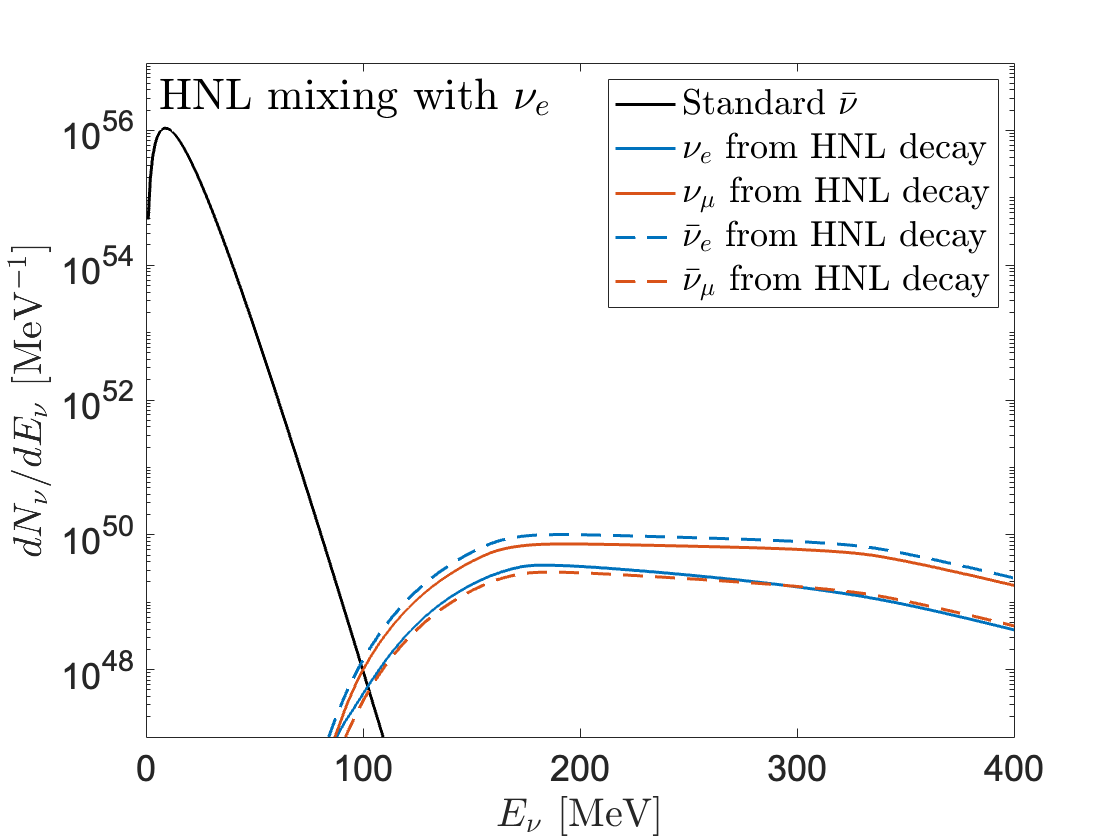}
  \end{center}
 \end{minipage}
 \begin{minipage}{0.5\hsize}
 \begin{center}
  \includegraphics[width=80mm]{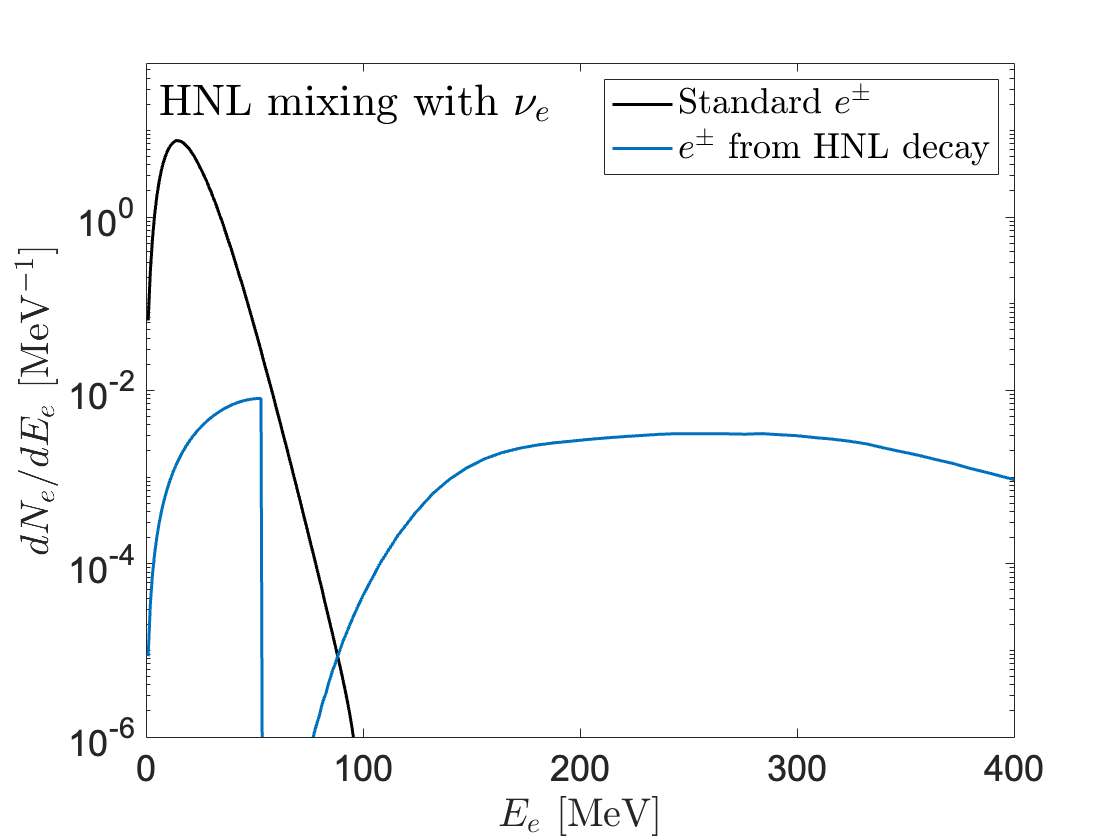}
 \end{center}
 \end{minipage} \\
 \begin{minipage}{0.5\hsize}
 \begin{center}
  \includegraphics[width=80mm]{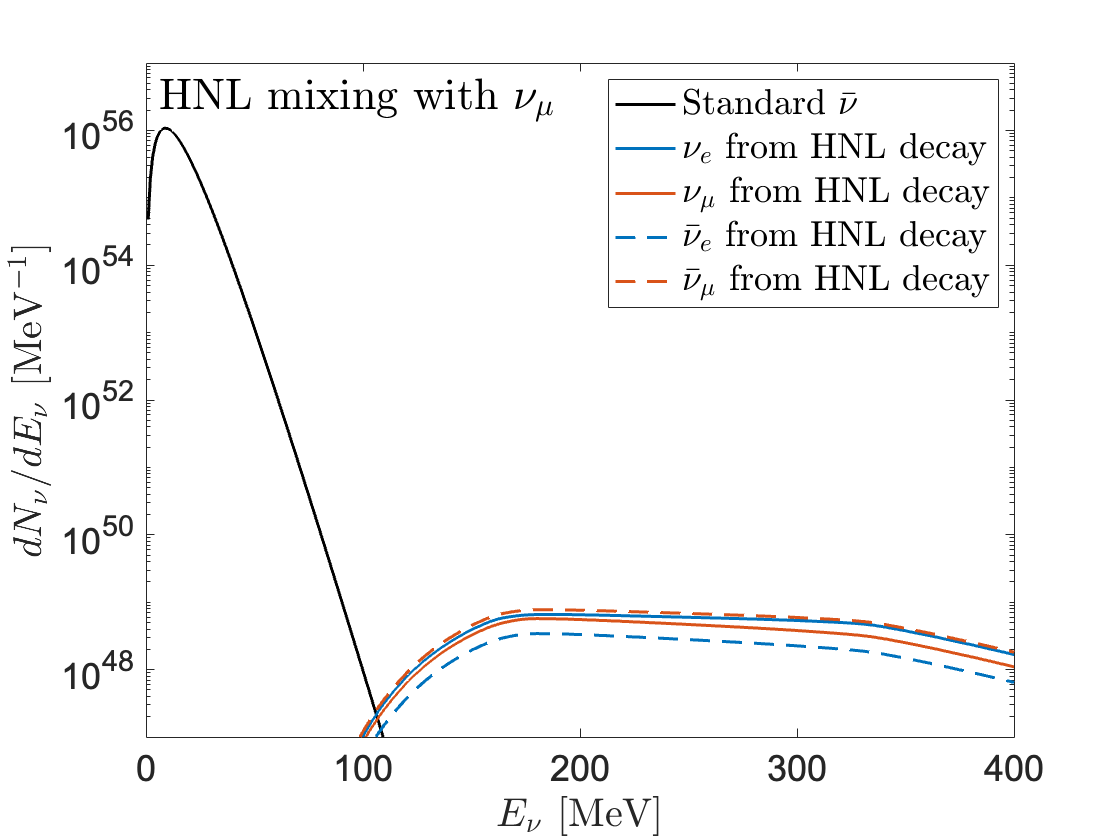}
 \end{center}
 \end{minipage}
 \begin{minipage}{0.5\hsize}
 \begin{center}
  \includegraphics[width=80mm]{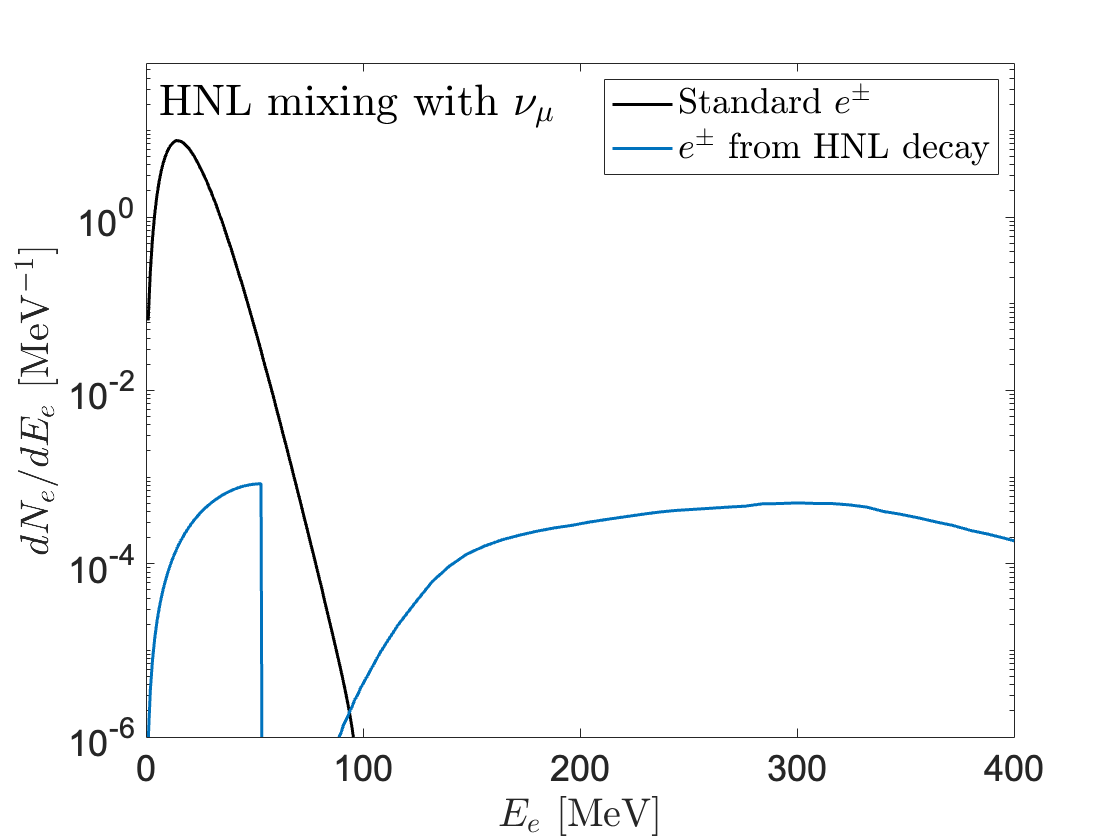}
 \end{center}
 \end{minipage} \\
 \begin{minipage}{0.5\hsize}
 \begin{center}
  \includegraphics[width=80mm]{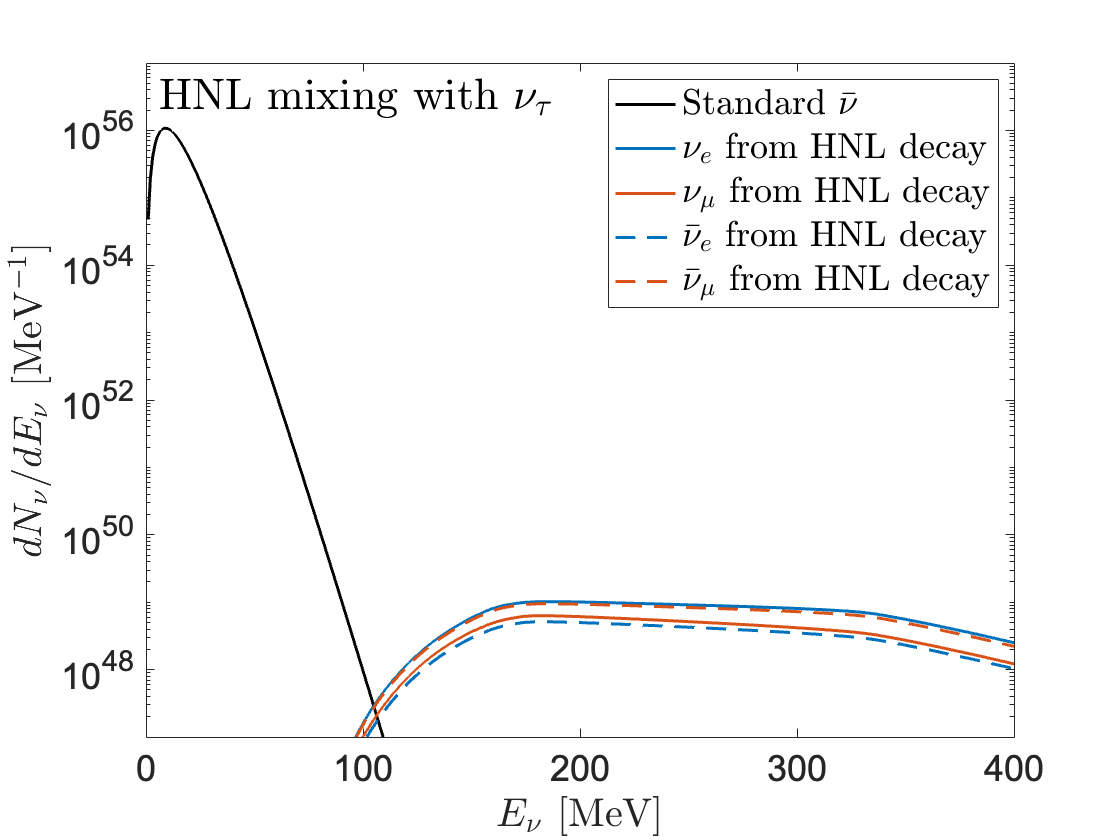}
 \end{center}
 \end{minipage}
 \begin{minipage}{0.5\hsize}
 \begin{center}
  \includegraphics[width=80mm]{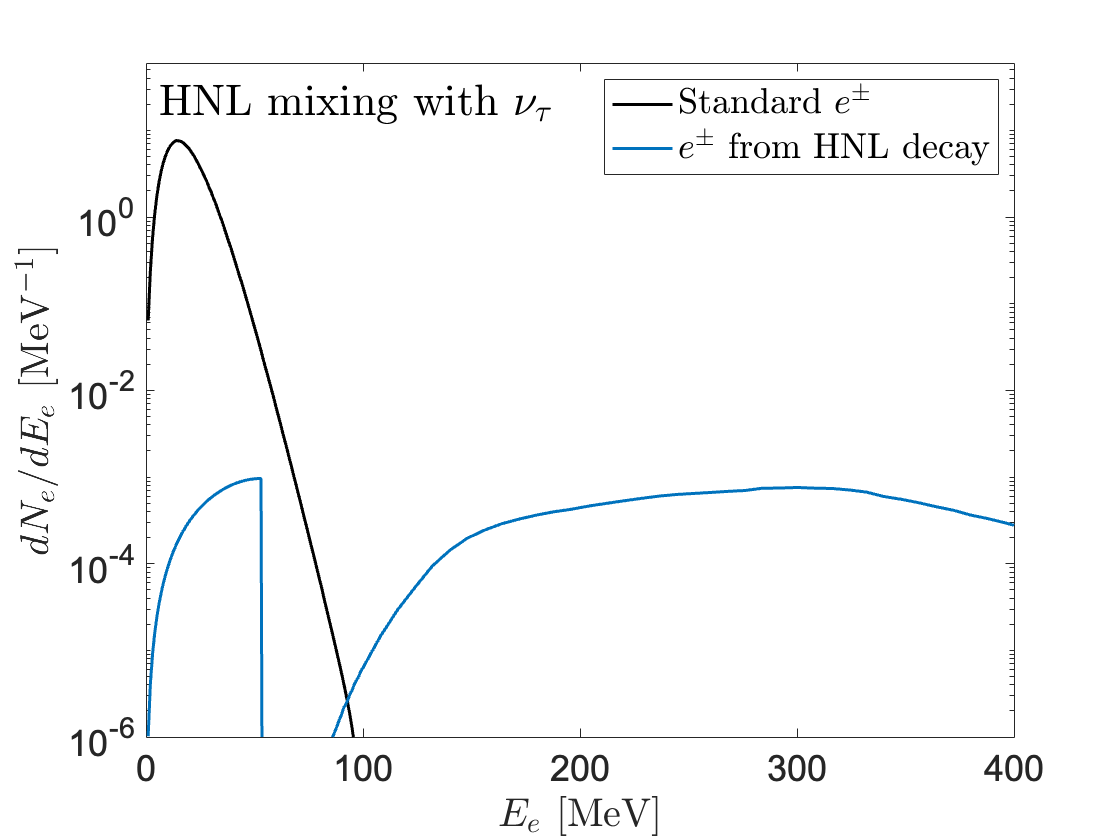}
 \end{center}
 \end{minipage}
 \caption{\small{Energy spectra on Earth from the time-integrated emission of a supernova in the HNL case only mixing with $\nu_e$ (top), $\nu_\mu$ (middle) and $\nu_\tau$ (bottom) with $m_N=500\ {\rm MeV}$ and $|U_\alpha|^2=10^{-9}\ (\alpha=e,\mu,\tau)$. The left panels show $\nu$ and $\bar{\nu}$ spectra while the right panels show $e^\pm$ spectra. For the standard anti-neutrinos and their $e^\pm$ signal (black lines), the spectra are averaged over all flavors. The $e^\pm$ spectra include Michel $e^\pm$ from $\mu^\pm$ decays at rest, whose endpoint is $m_\mu/2=53\ {\rm MeV}$, produced by the CC interactions of $\nu_\mu$ and $\bar{\nu}_\mu$. See text for the detailed normalization and setup.}}
    \label{fig:Spectrum_HNL}
\end{figure}

Figure~\ref{fig:Spectrum_HNL} presents the HNL case. This figure show the energy spectra for neutrinos and anti-neutrinos (left panels) and electron and positron (right panels) on Earth. We consider HNLs with only mixing $\nu_e$ (top panels), $\nu_\mu$ (middle panels) and $\nu_\tau$ (bottom panels). In all of these cases, we take $m_N=500\ {\rm MeV}$ and $|U_\alpha|^2=10^{-9}$, which are still unconstrained by experiments and observations. The two black lines show the standard $\bar{\nu}$ spectrum, averaged over all flavors, and its $e^\pm$ signal spectrum, respectively.

\begin{figure}
 \begin{minipage}{0.5\hsize}
  \begin{center}
   \includegraphics[width=80mm]{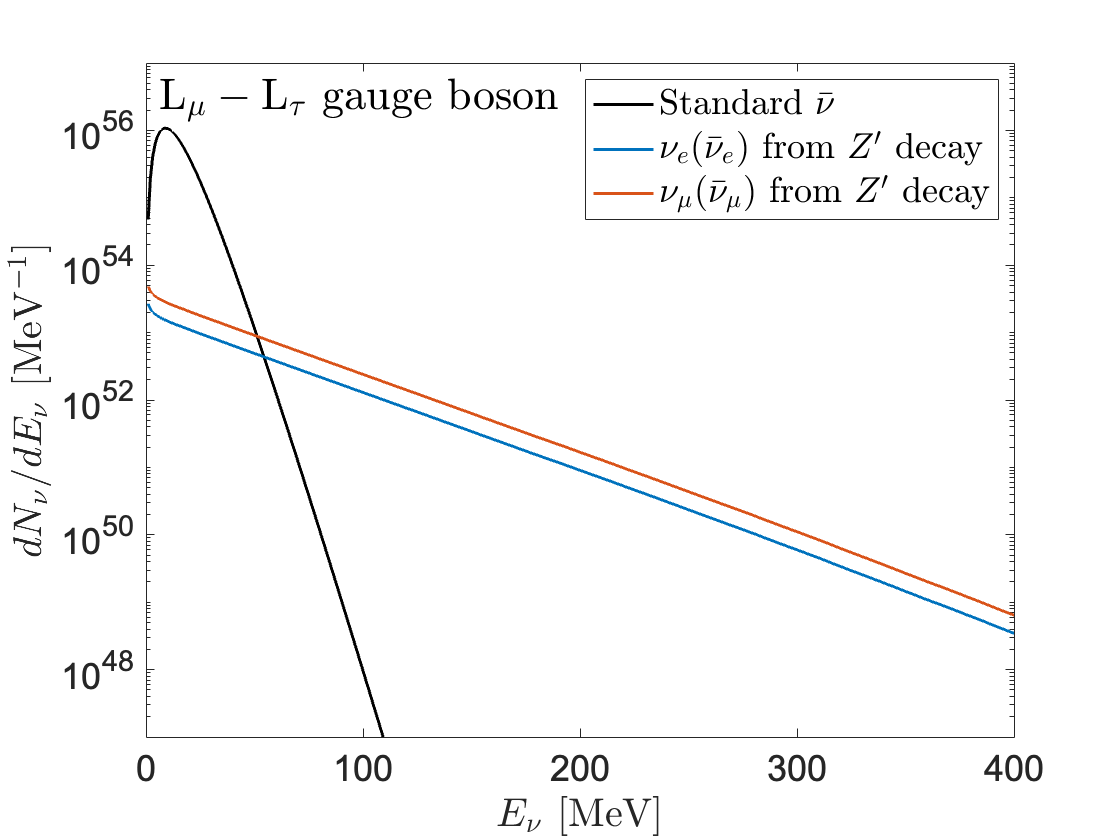}
  \end{center}
 \end{minipage}
 \begin{minipage}{0.5\hsize}
 \begin{center}
  \includegraphics[width=80mm]{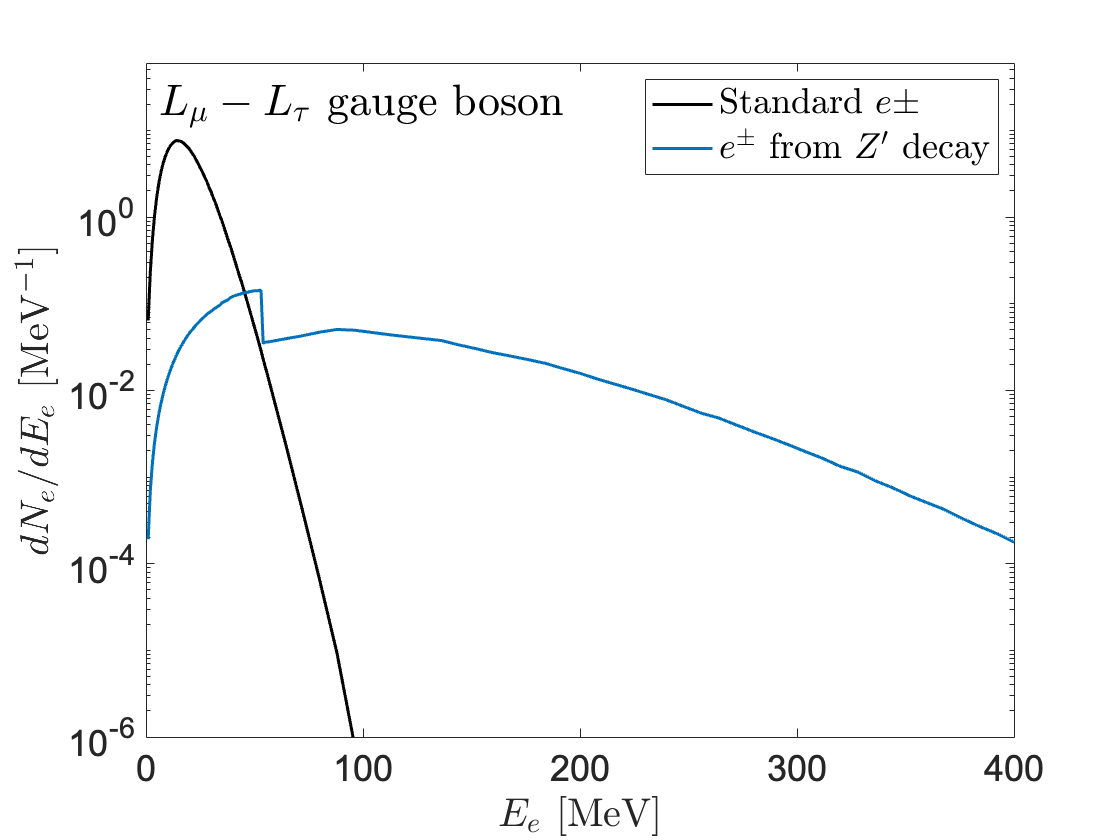}
 \end{center}
 \end{minipage}
 \caption{\small{Energy spectra on Earth from the time-integrated emission of a supernova in the case of ${\rm U}(1)_{L_\mu-L_\tau}$ gauge boson with $m_{Z'}=1\ {\rm MeV}$ and $g_{L_\mu-L_\tau}=3\times 10^{-10}$. The others are the same with figure~\ref{fig:Spectrum_HNL}.}}
    \label{fig:Spectrum_LmuLtau}
\end{figure}

\begin{figure}
 \begin{minipage}{0.5\hsize}
  \begin{center}
   \includegraphics[width=80mm]{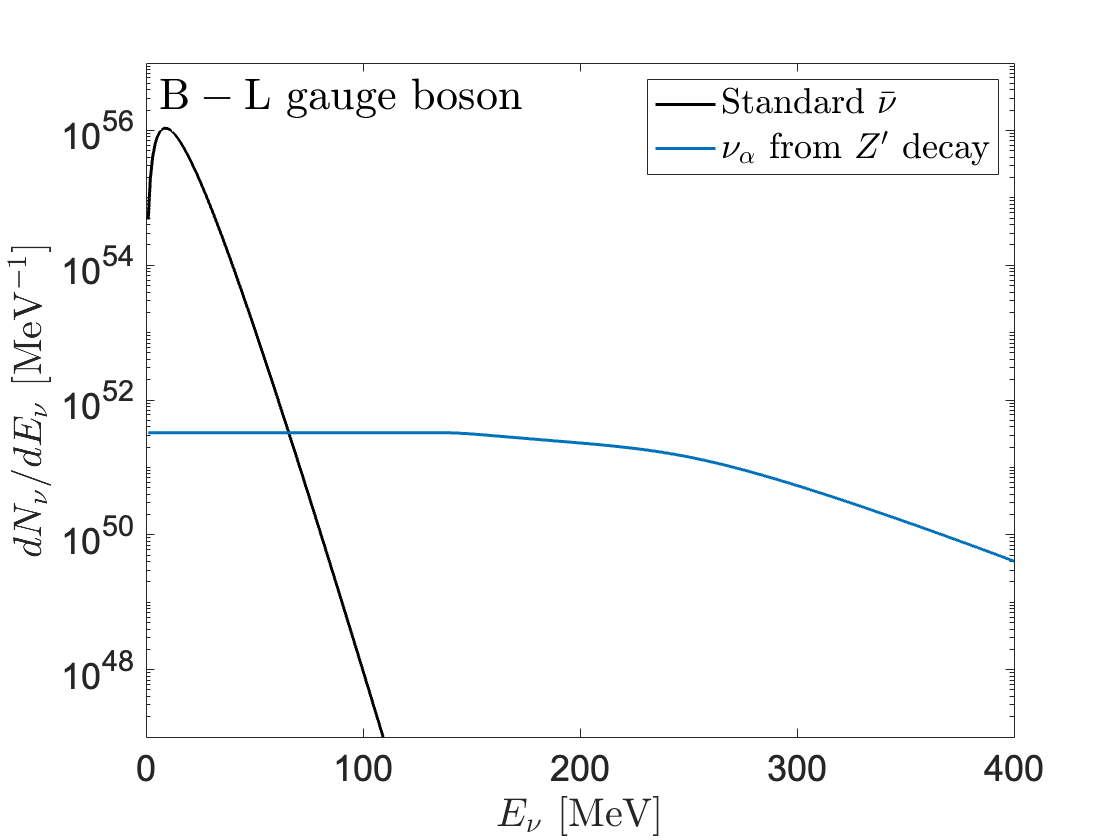}
  \end{center}
 \end{minipage}
 \begin{minipage}{0.5\hsize}
 \begin{center}
  \includegraphics[width=80mm]{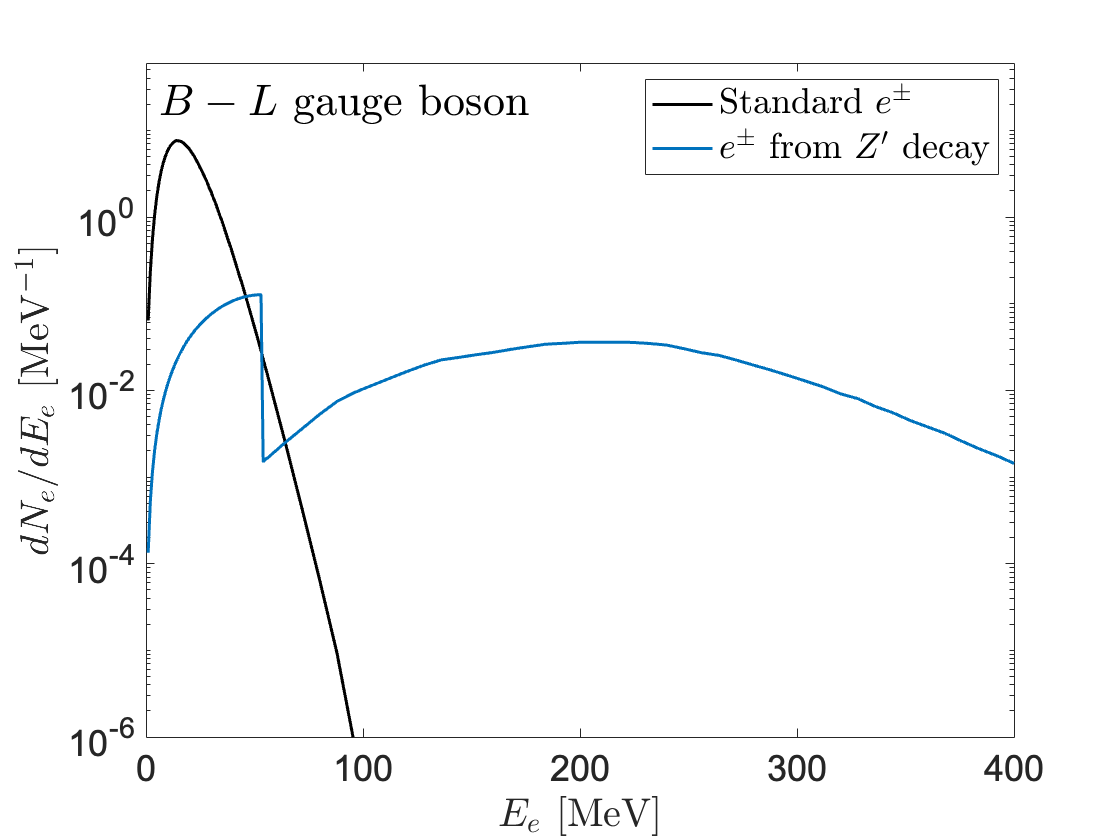}
 \end{center}
 \end{minipage}
 \caption{\small{Energy spectra on Earth from the time-integrated emission of a supernova in the case of ${\rm U}(1)_{B-L}$ gauge boson with $m_{Z'}=1\ {\rm MeV}$ and $g_{\rm B-L}=4\times 10^{-12}$. The others are the same with figure~\ref{fig:Spectrum_HNL}.}}
    \label{fig:Spectrum_B_L}
\end{figure}

Figure~\ref{fig:Spectrum_LmuLtau} gives the case of ${\rm U}(1)_{L_\mu-L_\tau}$ gauge boson. We take $m_{Z'}=1\ {\rm MeV}$ and $g_{\mu-\tau}=3\times 10^{-10}$ for the parameters of ${\rm U}(1)_{L_\mu-L_\tau}$ gauge boson. Figure~\ref{fig:Spectrum_B_L} presents the case of ${\rm U}(1)_{B-L}$ gauge boson and the parameters are fixed as $m_{Z'}=1\ {\rm MeV}$ and $g_{B-L}=4\times 10^{-12}$.

\begin{figure}
 \begin{minipage}{0.5\hsize}
  \begin{center}
   \includegraphics[width=80mm]{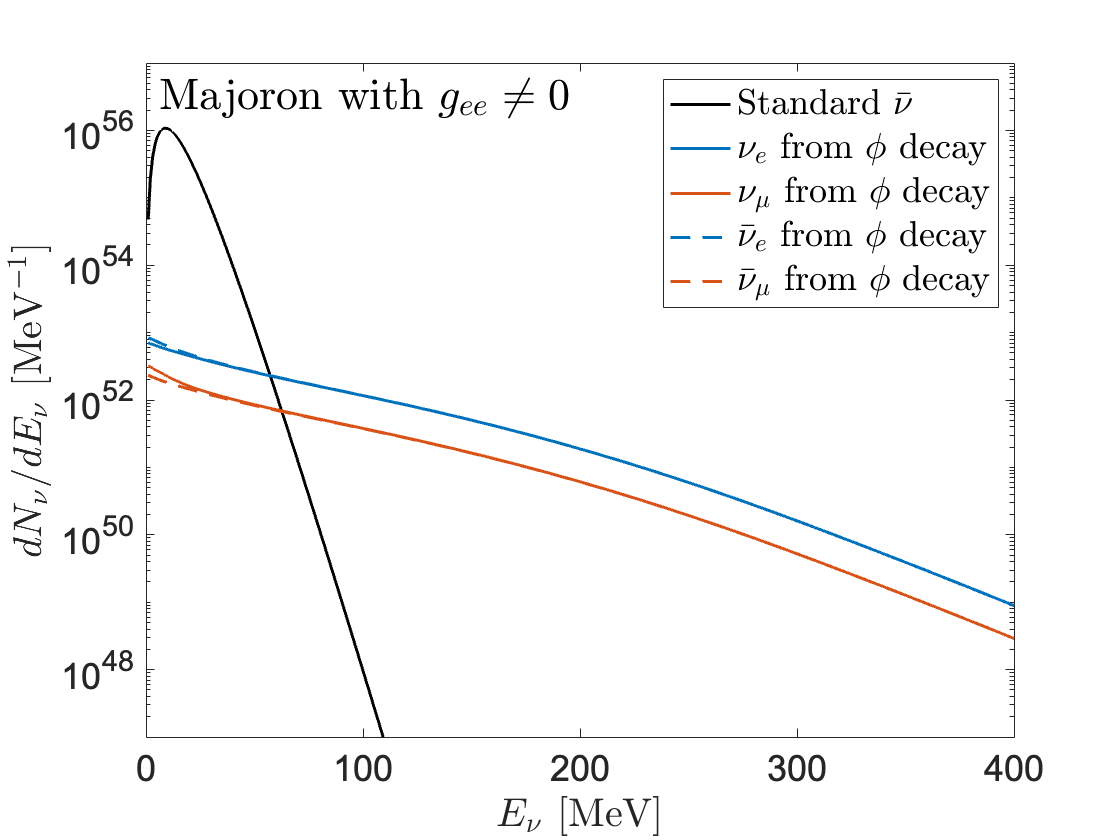}
  \end{center}
 \end{minipage}
 \begin{minipage}{0.5\hsize}
 \begin{center}
  \includegraphics[width=80mm]{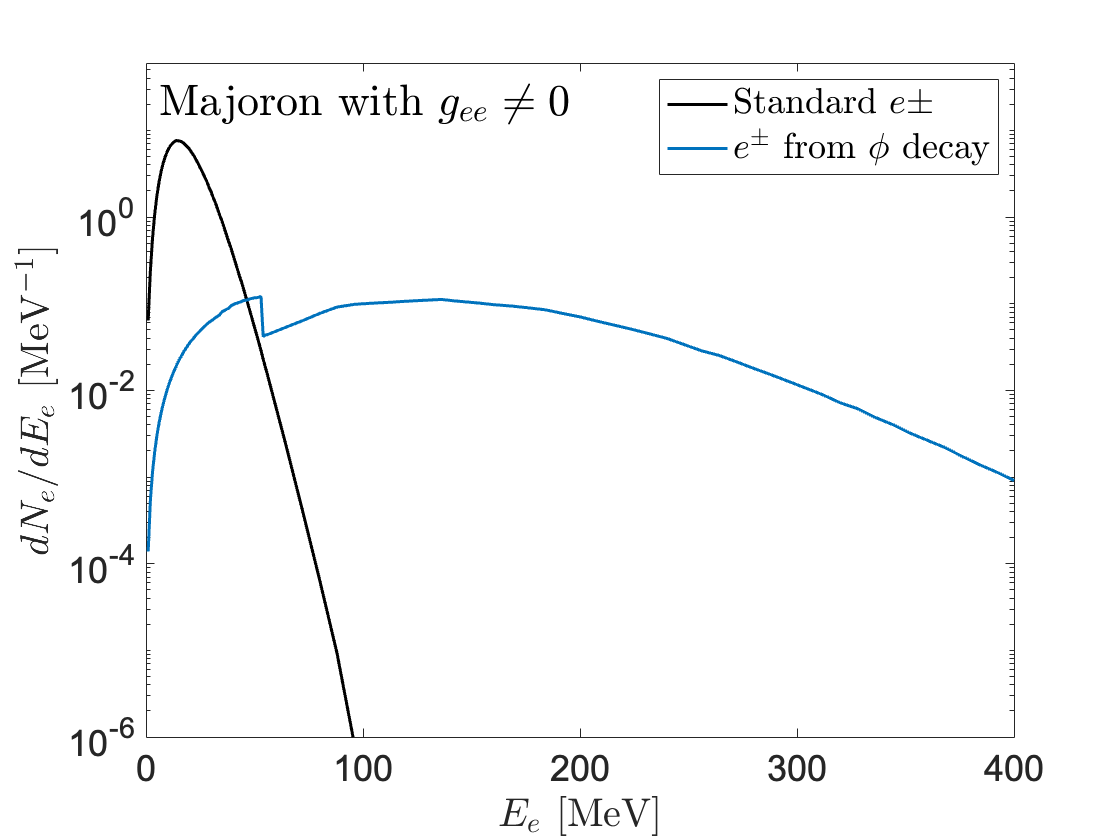}
 \end{center}
 \end{minipage} \\
 \begin{minipage}{0.5\hsize}
 \begin{center}
  \includegraphics[width=80mm]{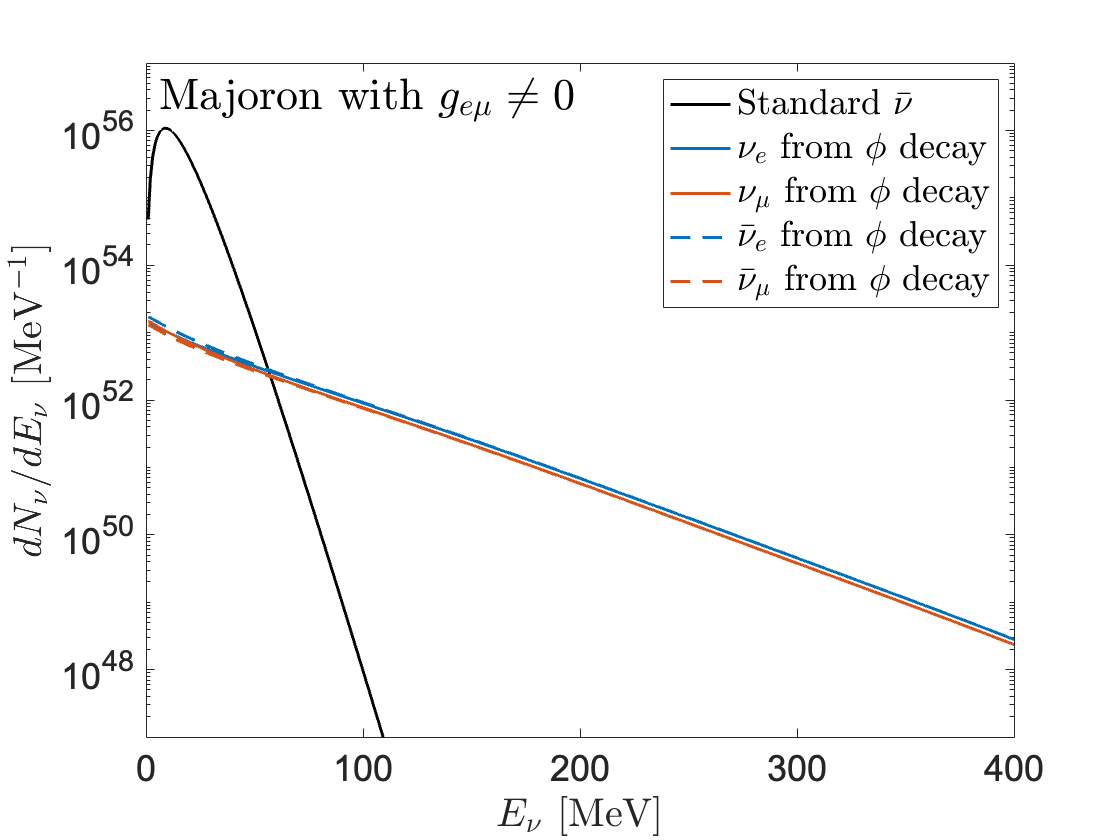}
 \end{center}
 \end{minipage}
 \begin{minipage}{0.5\hsize}
 \begin{center}
  \includegraphics[width=80mm]{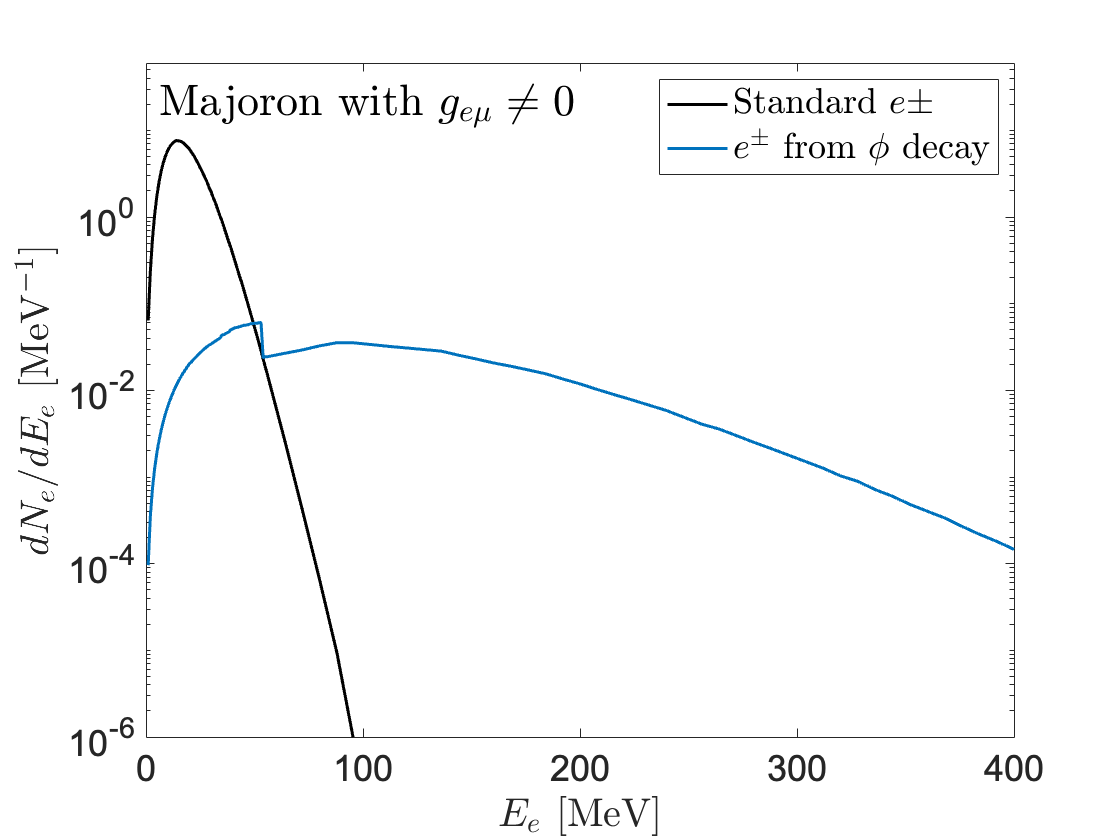}
 \end{center}
 \end{minipage} \\
 \begin{minipage}{0.5\hsize}
 \begin{center}
  \includegraphics[width=80mm]{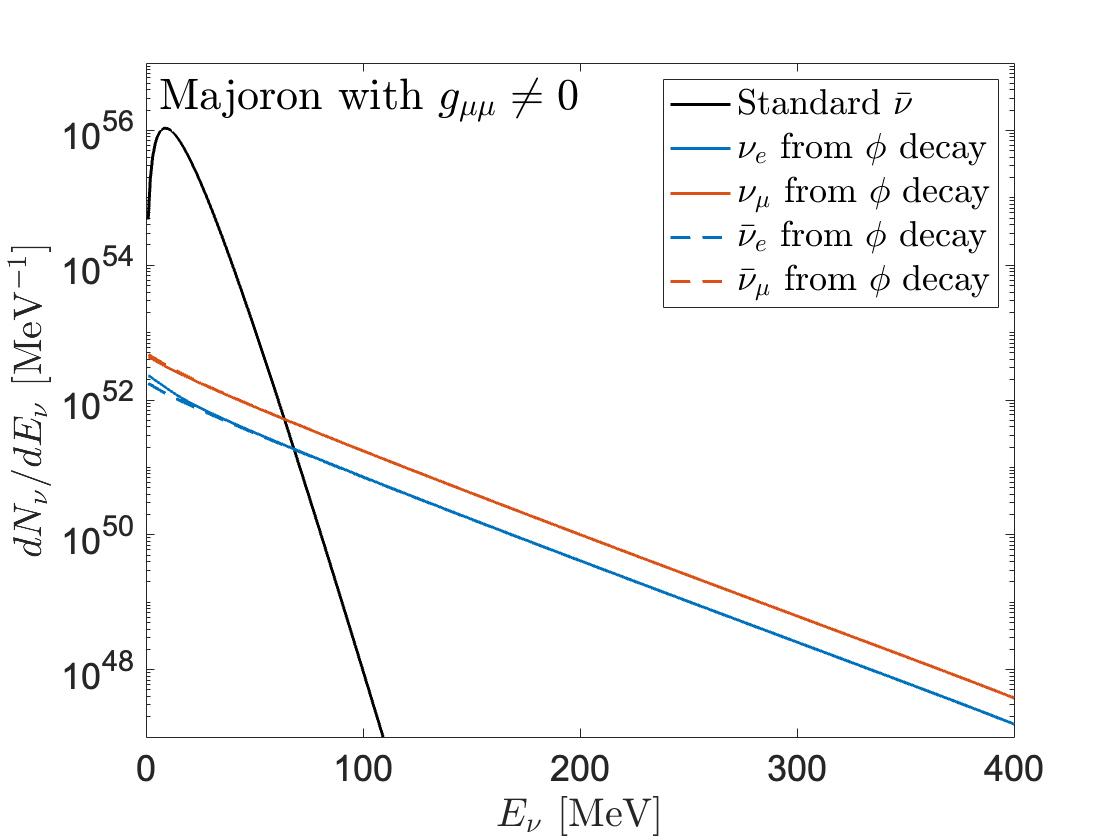}
 \end{center}
 \end{minipage}
 \begin{minipage}{0.5\hsize}
 \begin{center}
  \includegraphics[width=80mm]{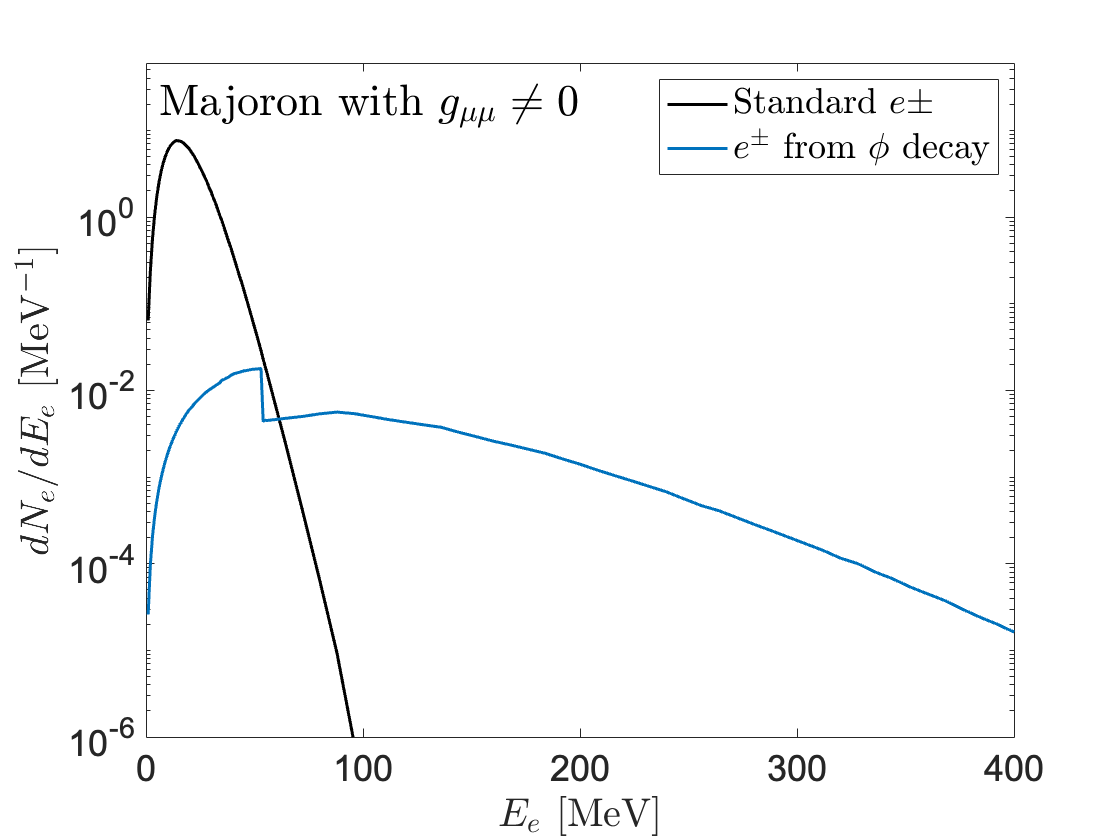}
 \end{center}
 \end{minipage}
 \caption{\small{Energy spectra on Earth from the time-integrated emission of a supernova in the case of majoron with $g_{ee}\neq 0$ (top), $g_{e\mu}\neq 0$ (middle) and $g_{\mu\mu}\neq 0$ (bottom) with $m_{\phi}=1\ {\rm MeV}$ and $g=10^{-9}$. The others are the same with figure~\ref{fig:Spectrum_HNL}.}}
    \label{fig:Spectrum_majoron}
\end{figure}

Figure~\ref{fig:Spectrum_majoron} presents the case of majoron. We consider majorons with $g_{ee}\neq 0$ (top panels), $g_{e\mu}\neq0$ (middle panels) and $g_{\mu\mu}\neq0$ (bottom panels). In all cases, we take $m_\phi=1\ {\rm MeV}$ and $g=10^{-9}$. 
The shape of the spectra for $g_{ee}\neq 0$ is flatter than for $g_{e\mu}\neq 0$ and $g_{\mu\mu}\neq 0$. This is because high energy spectra are produced by $\nu_e$ with large chemical potential through $\nu_e\nu_e\rightarrow \phi$.  

\bibliographystyle{JHEP}
\bibliography{reference}

\providecommand{\href}[2]{#2}\begingroup\raggedright\begin{thebibliography}{100}

\bibitem{Raffelt:1996wa}
G.G.~Raffelt, \emph{{Stars as laboratories for fundamental physics}: {The astrophysics of neutrinos, axions, and other weakly interacting particles}}, University of Chicago Press (1996).

\bibitem{Raffelt:2012kt}
G.G.~Raffelt, \emph{{Neutrinos and the stars}}, \href{https://doi.org/10.3254/978-1-61499-173-1-61}{\emph{Proc. Int. Sch. Phys. Fermi} {\bfseries 182} (2012) 61} [\href{https://arxiv.org/abs/1201.1637}{{\ttfamily 1201.1637}}].

\bibitem{PhysRevLett.60.1793}
G.~Raffelt and D.~Seckel, \emph{Bounds on exotic-particle interactions from sn1987a}, \href{https://doi.org/10.1103/PhysRevLett.60.1793}{\emph{Phys. Rev. Lett.} {\bfseries 60} (1988) 1793}.

\bibitem{Keil:1996ju}
W.~Keil, H.-T.~Janka, D.N.~Schramm, G.~Sigl, M.S.~Turner and J.R.~Ellis, \emph{{A Fresh look at axions and SN-1987A}}, \href{https://doi.org/10.1103/PhysRevD.56.2419}{\emph{Phys. Rev. D} {\bfseries 56} (1997) 2419} [\href{https://arxiv.org/abs/astro-ph/9612222}{{\ttfamily astro-ph/9612222}}].

\bibitem{Chang:2018rso}
J.H.~Chang, R.~Essig and S.D.~McDermott, \emph{{Supernova 1987A Constraints on Sub-GeV Dark Sectors, Millicharged Particles, the QCD Axion, and an Axion-like Particle}}, \href{https://doi.org/10.1007/JHEP09(2018)051}{\emph{JHEP} {\bfseries 09} (2018) 051} [\href{https://arxiv.org/abs/1803.00993}{{\ttfamily 1803.00993}}].

\bibitem{Carenza:2019pxu}
P.~Carenza, T.~Fischer, M.~Giannotti, G.~Guo, G.~Mart\'\i{}nez-Pinedo and A.~Mirizzi, \emph{{Improved axion emissivity from a supernova via nucleon-nucleon bremsstrahlung}}, \href{https://doi.org/10.1088/1475-7516/2019/10/016}{\emph{JCAP} {\bfseries 10} (2019) 016} [\href{https://arxiv.org/abs/1906.11844}{{\ttfamily 1906.11844}}].

\bibitem{Carenza:2020cis}
P.~Carenza, B.~Fore, M.~Giannotti, A.~Mirizzi and S.~Reddy, \emph{{Enhanced Supernova Axion Emission and its Implications}}, \href{https://doi.org/10.1103/PhysRevLett.126.071102}{\emph{Phys. Rev. Lett.} {\bfseries 126} (2021) 071102} [\href{https://arxiv.org/abs/2010.02943}{{\ttfamily 2010.02943}}].

\bibitem{Bollig:2020xdr}
R.~Bollig, W.~DeRocco, P.W.~Graham and H.-T.~Janka, \emph{{Muons in Supernovae: Implications for the Axion-Muon Coupling}}, \href{https://doi.org/10.1103/PhysRevLett.125.051104}{\emph{Phys. Rev. Lett.} {\bfseries 125} (2020) 051104} [\href{https://arxiv.org/abs/2005.07141}{{\ttfamily 2005.07141}}].

\bibitem{Croon:2020lrf}
D.~Croon, G.~Elor, R.K.~Leane and S.D.~McDermott, \emph{{Supernova Muons: New Constraints on $Z$' Bosons, Axions and ALPs}}, \href{https://doi.org/10.1007/JHEP01(2021)107}{\emph{JHEP} {\bfseries 01} (2021) 107} [\href{https://arxiv.org/abs/2006.13942}{{\ttfamily 2006.13942}}].

\bibitem{Caputo:2021rux}
A.~Caputo, G.~Raffelt and E.~Vitagliano, \emph{{Muonic boson limits: Supernova redux}}, \href{https://doi.org/10.1103/PhysRevD.105.035022}{\emph{Phys. Rev. D} {\bfseries 105} (2022) 035022} [\href{https://arxiv.org/abs/2109.03244}{{\ttfamily 2109.03244}}].

\bibitem{Choi:2021ign}
K.~Choi, H.J.~Kim, H.~Seong and C.S.~Shin, \emph{{Axion emission from supernova with axion-pion-nucleon contact interaction}}, \href{https://doi.org/10.1007/JHEP02(2022)143}{\emph{JHEP} {\bfseries 02} (2022) 143} [\href{https://arxiv.org/abs/2110.01972}{{\ttfamily 2110.01972}}].

\bibitem{Mastrototaro:2019vug}
L.~Mastrototaro, A.~Mirizzi, P.D.~Serpico and A.~Esmaili, \emph{{Heavy sterile neutrino emission in core-collapse supernovae: Constraints and signatures}}, \href{https://doi.org/10.1088/1475-7516/2020/01/010}{\emph{JCAP} {\bfseries 01} (2020) 010} [\href{https://arxiv.org/abs/1910.10249}{{\ttfamily 1910.10249}}].

\bibitem{Carenza:2023old}
P.~Carenza, G.~Lucente, L.~Mastrototaro, A.~Mirizzi and P.D.~Serpico, \emph{{Comprehensive constraints on heavy sterile neutrinos from core-collapse supernovae}},  \href{https://arxiv.org/abs/2311.00033}{{\ttfamily 2311.00033}}.

\bibitem{Chang:2016ntp}
J.H.~Chang, R.~Essig and S.D.~McDermott, \emph{{Revisiting Supernova 1987A Constraints on Dark Photons}}, \href{https://doi.org/10.1007/JHEP01(2017)107}{\emph{JHEP} {\bfseries 01} (2017) 107} [\href{https://arxiv.org/abs/1611.03864}{{\ttfamily 1611.03864}}].

\bibitem{Knapen:2017xzo}
S.~Knapen, T.~Lin and K.M.~Zurek, \emph{{Light Dark Matter: Models and Constraints}}, \href{https://doi.org/10.1103/PhysRevD.96.115021}{\emph{Phys. Rev. D} {\bfseries 96} (2017) 115021} [\href{https://arxiv.org/abs/1709.07882}{{\ttfamily 1709.07882}}].

\bibitem{Shin:2021bvz}
C.S.~Shin and S.~Yun, \emph{{Dark gauge boson production from neutron stars via nucleon-nucleon bremsstrahlung}}, \href{https://doi.org/10.1007/JHEP02(2022)133}{\emph{JHEP} {\bfseries 02} (2022) 133} [\href{https://arxiv.org/abs/2110.03362}{{\ttfamily 2110.03362}}].

\bibitem{Shin:2022ulh}
C.S.~Shin and S.~Yun, \emph{{Dark gauge boson emission from supernova pions}}, \href{https://doi.org/10.1103/PhysRevD.108.055014}{\emph{Phys. Rev. D} {\bfseries 108} (2023) 055014} [\href{https://arxiv.org/abs/2211.15677}{{\ttfamily 2211.15677}}].

\bibitem{Cerdeno:2023kqo}
D.G.~Cerde\~no, M.~Cerme\~no and Y.~Farzan, \emph{{Constraints from the duration of supernova neutrino burst on on-shell light gauge boson production by neutrinos}}, \href{https://doi.org/10.1103/PhysRevD.107.123012}{\emph{Phys. Rev. D} {\bfseries 107} (2023) 123012} [\href{https://arxiv.org/abs/2301.00661}{{\ttfamily 2301.00661}}].

\bibitem{Choi:1987sd}
K.~Choi, C.W.~Kim, J.~Kim and W.P.~Lam, \emph{{Constraints on the Majoron Interactions From the Supernova {SN1987A}}}, \href{https://doi.org/10.1103/PhysRevD.37.3225}{\emph{Phys. Rev. D} {\bfseries 37} (1988) 3225}.

\bibitem{BEREZHIANI1989279}
Z.~Berezhiani and A.~Smirnov, \emph{Matter-induced neutrino decay and supernova 1987a}, \href{https://doi.org/https://doi.org/10.1016/0370-2693(89)90052-X}{\emph{Physics Letters B} {\bfseries 220} (1989) 279}.

\bibitem{Choi:1989hi}
K.~Choi and A.~Santamaria, \emph{{Majorons and Supernova Cooling}}, \href{https://doi.org/10.1103/PhysRevD.42.293}{\emph{Phys. Rev. D} {\bfseries 42} (1990) 293}.

\bibitem{Chang:1993yp}
S.~Chang and K.~Choi, \emph{{Constraints from nucleosynthesis and SN1987A on majoron emitting double beta decay}}, \href{https://doi.org/10.1103/PhysRevD.49.R12}{\emph{Phys. Rev. D} {\bfseries 49} (1994) 12} [\href{https://arxiv.org/abs/hep-ph/9303243}{{\ttfamily hep-ph/9303243}}].

\bibitem{Kachelriess:2000qc}
M.~Kachelriess, R.~Tomas and J.W.F.~Valle, \emph{{Supernova bounds on Majoron emitting decays of light neutrinos}}, \href{https://doi.org/10.1103/PhysRevD.62.023004}{\emph{Phys. Rev. D} {\bfseries 62} (2000) 023004} [\href{https://arxiv.org/abs/hep-ph/0001039}{{\ttfamily hep-ph/0001039}}].

\bibitem{Tomas:2001dh}
R.~Tomas, H.~Pas and J.W.F.~Valle, \emph{{Generalized bounds on Majoron - neutrino couplings}}, \href{https://doi.org/10.1103/PhysRevD.64.095005}{\emph{Phys. Rev. D} {\bfseries 64} (2001) 095005} [\href{https://arxiv.org/abs/hep-ph/0103017}{{\ttfamily hep-ph/0103017}}].

\bibitem{Hannestad:2002ff}
S.~Hannestad, P.~Keranen and F.~Sannino, \emph{{A Supernova constraint on bulk Majorons}}, \href{https://doi.org/10.1103/PhysRevD.66.045002}{\emph{Phys. Rev. D} {\bfseries 66} (2002) 045002} [\href{https://arxiv.org/abs/hep-ph/0204231}{{\ttfamily hep-ph/0204231}}].

\bibitem{Farzan:2002wx}
Y.~Farzan, \emph{{Bounds on the coupling of the Majoron to light neutrinos from supernova cooling}}, \href{https://doi.org/10.1103/PhysRevD.67.073015}{\emph{Phys. Rev. D} {\bfseries 67} (2003) 073015} [\href{https://arxiv.org/abs/hep-ph/0211375}{{\ttfamily hep-ph/0211375}}].

\bibitem{Heurtier:2016otg}
L.~Heurtier and Y.~Zhang, \emph{{Supernova Constraints on Massive (Pseudo)Scalar Coupling to Neutrinos}}, \href{https://doi.org/10.1088/1475-7516/2017/02/042}{\emph{JCAP} {\bfseries 02} (2017) 042} [\href{https://arxiv.org/abs/1609.05882}{{\ttfamily 1609.05882}}].

\bibitem{Brune:2018sab}
T.~Brune and H.~P\"as, \emph{{Massive Majorons and constraints on the Majoron-neutrino coupling}}, \href{https://doi.org/10.1103/PhysRevD.99.096005}{\emph{Phys. Rev. D} {\bfseries 99} (2019) 096005} [\href{https://arxiv.org/abs/1808.08158}{{\ttfamily 1808.08158}}].

\bibitem{Grifols:1996id}
J.A.~Grifols, E.~Masso and R.~Toldra, \emph{{Gamma-rays from SN1987A due to pseudoscalar conversion}}, \href{https://doi.org/10.1103/PhysRevLett.77.2372}{\emph{Phys. Rev. Lett.} {\bfseries 77} (1996) 2372} [\href{https://arxiv.org/abs/astro-ph/9606028}{{\ttfamily astro-ph/9606028}}].

\bibitem{Brockway:1996yr}
J.W.~Brockway, E.D.~Carlson and G.G.~Raffelt, \emph{{SN1987A gamma-ray limits on the conversion of pseudoscalars}}, \href{https://doi.org/10.1016/0370-2693(96)00778-2}{\emph{Phys. Lett. B} {\bfseries 383} (1996) 439} [\href{https://arxiv.org/abs/astro-ph/9605197}{{\ttfamily astro-ph/9605197}}].

\bibitem{Payez:2014xsa}
A.~Payez, C.~Evoli, T.~Fischer, M.~Giannotti, A.~Mirizzi and A.~Ringwald, \emph{{Revisiting the SN1987A gamma-ray limit on ultralight axion-like particles}}, \href{https://doi.org/10.1088/1475-7516/2015/02/006}{\emph{JCAP} {\bfseries 02} (2015) 006} [\href{https://arxiv.org/abs/1410.3747}{{\ttfamily 1410.3747}}].

\bibitem{Meyer:2016wrm}
M.~Meyer, M.~Giannotti, A.~Mirizzi, J.~Conrad and M.A.~S\'anchez-Conde, \emph{{Fermi Large Area Telescope as a Galactic Supernovae Axionscope}}, \href{https://doi.org/10.1103/PhysRevLett.118.011103}{\emph{Phys. Rev. Lett.} {\bfseries 118} (2017) 011103} [\href{https://arxiv.org/abs/1609.02350}{{\ttfamily 1609.02350}}].

\bibitem{Jaeckel:2017tud}
J.~Jaeckel, P.C.~Malta and J.~Redondo, \emph{{Decay photons from the axionlike particles burst of type II supernovae}}, \href{https://doi.org/10.1103/PhysRevD.98.055032}{\emph{Phys. Rev. D} {\bfseries 98} (2018) 055032} [\href{https://arxiv.org/abs/1702.02964}{{\ttfamily 1702.02964}}].

\bibitem{Calore:2020tjw}
F.~Calore, P.~Carenza, M.~Giannotti, J.~Jaeckel and A.~Mirizzi, \emph{{Bounds on axionlike particles from the diffuse supernova flux}}, \href{https://doi.org/10.1103/PhysRevD.102.123005}{\emph{Phys. Rev. D} {\bfseries 102} (2020) 123005} [\href{https://arxiv.org/abs/2008.11741}{{\ttfamily 2008.11741}}].

\bibitem{Ferreira:2022xlw}
R.Z.~Ferreira, M.C.D.~Marsh and E.~M\"uller, \emph{{Strong supernovae bounds on ALPs from quantum loops}}, \href{https://doi.org/10.1088/1475-7516/2022/11/057}{\emph{JCAP} {\bfseries 11} (2022) 057} [\href{https://arxiv.org/abs/2205.07896}{{\ttfamily 2205.07896}}].

\bibitem{Diamond:2023scc}
M.~Diamond, D.F.G.~Fiorillo, G.~Marques-Tavares and E.~Vitagliano, \emph{{Axion-sourced fireballs from supernovae}}, \href{https://doi.org/10.1103/PhysRevD.107.103029}{\emph{Phys. Rev. D} {\bfseries 107} (2023) 103029} [\href{https://arxiv.org/abs/2303.11395}{{\ttfamily 2303.11395}}].

\bibitem{DeRocco:2019njg}
W.~DeRocco, P.W.~Graham, D.~Kasen, G.~Marques-Tavares and S.~Rajendran, \emph{{Observable signatures of dark photons from supernovae}}, \href{https://doi.org/10.1007/JHEP02(2019)171}{\emph{JHEP} {\bfseries 02} (2019) 171} [\href{https://arxiv.org/abs/1901.08596}{{\ttfamily 1901.08596}}].

\bibitem{Calore:2021lih}
F.~Calore, P.~Carenza, M.~Giannotti, J.~Jaeckel, G.~Lucente, L.~Mastrototaro et~al., \emph{{511~keV line constraints on feebly interacting particles from supernovae}}, \href{https://doi.org/10.1103/PhysRevD.105.063026}{\emph{Phys. Rev. D} {\bfseries 105} (2022) 063026} [\href{https://arxiv.org/abs/2112.08382}{{\ttfamily 2112.08382}}].

\bibitem{Akita:2022etk}
K.~Akita, S.H.~Im and M.~Masud, \emph{{Probing non-standard neutrino interactions with a light boson from next galactic and diffuse supernova neutrinos}}, \href{https://doi.org/10.1007/JHEP12(2022)050}{\emph{JHEP} {\bfseries 12} (2022) 050} [\href{https://arxiv.org/abs/2206.06852}{{\ttfamily 2206.06852}}].

\bibitem{Fiorillo:2022cdq}
D.F.G.~Fiorillo, G.G.~Raffelt and E.~Vitagliano, \emph{{Strong Supernova 1987A Constraints on Bosons Decaying to Neutrinos}}, \href{https://doi.org/10.1103/PhysRevLett.131.021001}{\emph{Phys. Rev. Lett.} {\bfseries 131} (2023) 021001} [\href{https://arxiv.org/abs/2209.11773}{{\ttfamily 2209.11773}}].

\bibitem{Syvolap:2023trc}
V.~Syvolap, \emph{{Testing heavy neutral leptons produced in the supernovae explosions with future neutrino detectors}},  \href{https://arxiv.org/abs/2301.07052}{{\ttfamily 2301.07052}}.

\bibitem{Falk:1978kf}
S.W.~Falk and D.N.~Schramm, \emph{{Limits From Supernovae on Neutrino Radiative Lifetimes}}, \href{https://doi.org/10.1016/0370-2693(78)90417-3}{\emph{Phys. Lett. B} {\bfseries 79} (1978) 511}.

\bibitem{Sung:2019xie}
A.~Sung, H.~Tu and M.-R.~Wu, \emph{{New constraint from supernova explosions on light particles beyond the Standard Model}}, \href{https://doi.org/10.1103/PhysRevD.99.121305}{\emph{Phys. Rev. D} {\bfseries 99} (2019) 121305} [\href{https://arxiv.org/abs/1903.07923}{{\ttfamily 1903.07923}}].

\bibitem{Caputo:2022mah}
A.~Caputo, H.-T.~Janka, G.~Raffelt and E.~Vitagliano, \emph{{Low-Energy Supernovae Severely Constrain Radiative Particle Decays}}, \href{https://doi.org/10.1103/PhysRevLett.128.221103}{\emph{Phys. Rev. Lett.} {\bfseries 128} (2022) 221103} [\href{https://arxiv.org/abs/2201.09890}{{\ttfamily 2201.09890}}].

\bibitem{Chauhan:2023sci}
G.~Chauhan, S.~Horiuchi, P.~Huber and I.M.~Shoemaker, \emph{{Low-Energy Supernovae Bounds on Sterile Neutrinos}},  \href{https://arxiv.org/abs/2309.05860}{{\ttfamily 2309.05860}}.

\bibitem{Chang:2022aas}
P.-W.~Chang, I.~Esteban, J.F.~Beacom, T.A.~Thompson and C.M.~Hirata, \emph{{Toward Powerful Probes of Neutrino Self-Interactions in Supernovae}}, \href{https://doi.org/10.1103/PhysRevLett.131.071002}{\emph{Phys. Rev. Lett.} {\bfseries 131} (2023) 071002} [\href{https://arxiv.org/abs/2206.12426}{{\ttfamily 2206.12426}}].

\bibitem{Fiorillo:2023cas}
D.F.G.~Fiorillo, G.~Raffelt and E.~Vitagliano, \emph{{Supernova Emission of Secretly Interacting Neutrino Fluid: Theoretical Foundations}},  \href{https://arxiv.org/abs/2307.15122}{{\ttfamily 2307.15122}}.

\bibitem{Fiorillo:2023ytr}
D.F.G.~Fiorillo, G.~Raffelt and E.~Vitagliano, \emph{{Large Neutrino Secret Interactions, Small Impact on Supernovae}},  \href{https://arxiv.org/abs/2307.15115}{{\ttfamily 2307.15115}}.

\bibitem{Diamond:2023cto}
M.~Diamond, D.F.G.~Fiorillo, G.~Marques-Tavares, I.~Tamborra and E.~Vitagliano, \emph{{Multimessenger Constraints on Radiatively Decaying Axions from GW170817}},  \href{https://arxiv.org/abs/2305.10327}{{\ttfamily 2305.10327}}.

\bibitem{Abdullahi:2022jlv}
A.M.~Abdullahi et~al., \emph{{The present and future status of heavy neutral leptons}}, \href{https://doi.org/10.1088/1361-6471/ac98f9}{\emph{J. Phys. G} {\bfseries 50} (2023) 020501} [\href{https://arxiv.org/abs/2203.08039}{{\ttfamily 2203.08039}}].

\bibitem{Foot:1990mn}
R.~Foot, \emph{{New Physics From Electric Charge Quantization?}}, \href{https://doi.org/10.1142/S0217732391000543}{\emph{Mod. Phys. Lett. A} {\bfseries 6} (1991) 527}.

\bibitem{He:1990pn}
X.G.~He, G.C.~Joshi, H.~Lew and R.R.~Volkas, \emph{{NEW Z-prime PHENOMENOLOGY}}, \href{https://doi.org/10.1103/PhysRevD.43.R22}{\emph{Phys. Rev. D} {\bfseries 43} (1991) 22}.

\bibitem{He:1991qd}
X.-G.~He, G.C.~Joshi, H.~Lew and R.R.~Volkas, \emph{{Simplest Z-prime model}}, \href{https://doi.org/10.1103/PhysRevD.44.2118}{\emph{Phys. Rev. D} {\bfseries 44} (1991) 2118}.

\bibitem{Davidson:1978pm}
A.~Davidson, \emph{{$B-L$ as the fourth color within an $\mathrm{SU}(2)_L \times \mathrm{U}(1)_R \times \mathrm{U}(1)$ model}}, \href{https://doi.org/10.1103/PhysRevD.20.776}{\emph{Phys. Rev. D} {\bfseries 20} (1979) 776}.

\bibitem{Mohapatra:1980qe}
R.N.~Mohapatra and R.E.~Marshak, \emph{{Local B-L Symmetry of Electroweak Interactions, Majorana Neutrinos and Neutron Oscillations}}, \href{https://doi.org/10.1103/PhysRevLett.44.1316}{\emph{Phys. Rev. Lett.} {\bfseries 44} (1980) 1316}.

\bibitem{Wetterich:1981bx}
C.~Wetterich, \emph{{Neutrino Masses and the Scale of B-L Violation}}, \href{https://doi.org/10.1016/0550-3213(81)90279-0}{\emph{Nucl. Phys. B} {\bfseries 187} (1981) 343}.

\bibitem{Buchmuller:1991ce}
W.~Buchmuller, C.~Greub and P.~Minkowski, \emph{{Neutrino masses, neutral vector bosons and the scale of B-L breaking}}, \href{https://doi.org/10.1016/0370-2693(91)90952-M}{\emph{Phys. Lett. B} {\bfseries 267} (1991) 395}.

\bibitem{Chikashige:1980ui}
Y.~Chikashige, R.N.~Mohapatra and R.D.~Peccei, \emph{{Are There Real Goldstone Bosons Associated with Broken Lepton Number?}}, \href{https://doi.org/10.1016/0370-2693(81)90011-3}{\emph{Phys. Lett. B} {\bfseries 98} (1981) 265}.

\bibitem{Gelmini:1980re}
G.B.~Gelmini and M.~Roncadelli, \emph{{Left-Handed Neutrino Mass Scale and Spontaneously Broken Lepton Number}}, \href{https://doi.org/10.1016/0370-2693(81)90559-1}{\emph{Phys. Lett. B} {\bfseries 99} (1981) 411}.

\bibitem{Schechter:1981cv}
J.~Schechter and J.W.F.~Valle, \emph{{Neutrino Decay and Spontaneous Violation of Lepton Number}}, \href{https://doi.org/10.1103/PhysRevD.25.774}{\emph{Phys. Rev. D} {\bfseries 25} (1982) 774}.

\bibitem{Asaadi:2017bhx}
J.~Asaadi, E.~Church, R.~Guenette, B.J.P.~Jones and A.M.~Szelc, \emph{{New light Higgs boson and short-baseline neutrino anomalies}}, \href{https://doi.org/10.1103/PhysRevD.97.075021}{\emph{Phys. Rev. D} {\bfseries 97} (2018) 075021} [\href{https://arxiv.org/abs/1712.08019}{{\ttfamily 1712.08019}}].

\bibitem{Chauhan:2018dkd}
B.~Chauhan and S.~Mohanty, \emph{{Signature of light sterile neutrinos at IceCube}}, \href{https://doi.org/10.1103/PhysRevD.98.083021}{\emph{Phys. Rev. D} {\bfseries 98} (2018) 083021} [\href{https://arxiv.org/abs/1808.04774}{{\ttfamily 1808.04774}}].

\bibitem{Smirnov:2021zgn}
A.Y.~Smirnov and V.B.~Valera, \emph{{Resonance refraction and neutrino oscillations}}, \href{https://doi.org/10.1007/JHEP09(2021)177}{\emph{JHEP} {\bfseries 09} (2021) 177} [\href{https://arxiv.org/abs/2106.13829}{{\ttfamily 2106.13829}}].

\bibitem{Dentler:2019dhz}
M.~Dentler, I.~Esteban, J.~Kopp and P.~Machado, \emph{{Decaying Sterile Neutrinos and the Short Baseline Oscillation Anomalies}}, \href{https://doi.org/10.1103/PhysRevD.101.115013}{\emph{Phys. Rev. D} {\bfseries 101} (2020) 115013} [\href{https://arxiv.org/abs/1911.01427}{{\ttfamily 1911.01427}}].

\bibitem{deGouvea:2019qre}
A.~de~Gouv\^ea, O.L.G.~Peres, S.~Prakash and G.V.~Stenico, \emph{{On The Decaying-Sterile Neutrino Solution to the Electron (Anti)Neutrino Appearance Anomalies}}, \href{https://doi.org/10.1007/JHEP07(2020)141}{\emph{JHEP} {\bfseries 07} (2020) 141} [\href{https://arxiv.org/abs/1911.01447}{{\ttfamily 1911.01447}}].

\bibitem{Jeong:2018yts}
Y.S.~Jeong, S.~Palomares-Ruiz, M.H.~Reno and I.~Sarcevic, \emph{{Probing secret interactions of eV-scale sterile neutrinos with the diffuse supernova neutrino background}}, \href{https://doi.org/10.1088/1475-7516/2018/06/019}{\emph{JCAP} {\bfseries 06} (2018) 019} [\href{https://arxiv.org/abs/1803.04541}{{\ttfamily 1803.04541}}].

\bibitem{Abdallah:2022grs}
W.~Abdallah, R.~Gandhi and S.~Roy, \emph{{Requirements on common solutions to the LSND and MiniBooNE excesses: a post-MicroBooNE study}}, \href{https://doi.org/10.1007/JHEP06(2022)160}{\emph{JHEP} {\bfseries 06} (2022) 160} [\href{https://arxiv.org/abs/2202.09373}{{\ttfamily 2202.09373}}].

\bibitem{Muong-2:2006rrc}
{\scshape Muon g-2} collaboration, \emph{{Final Report of the Muon E821 Anomalous Magnetic Moment Measurement at BNL}}, \href{https://doi.org/10.1103/PhysRevD.73.072003}{\emph{Phys. Rev. D} {\bfseries 73} (2006) 072003} [\href{https://arxiv.org/abs/hep-ex/0602035}{{\ttfamily hep-ex/0602035}}].

\bibitem{Araki:2015mya}
T.~Araki, F.~Kaneko, T.~Ota, J.~Sato and T.~Shimomura, \emph{{MeV scale leptonic force for cosmic neutrino spectrum and muon anomalous magnetic moment}}, \href{https://doi.org/10.1103/PhysRevD.93.013014}{\emph{Phys. Rev. D} {\bfseries 93} (2016) 013014} [\href{https://arxiv.org/abs/1508.07471}{{\ttfamily 1508.07471}}].

\bibitem{Borsanyi:2020mff}
S.~Borsanyi et~al., \emph{{Leading hadronic contribution to the muon magnetic moment from lattice QCD}}, \href{https://doi.org/10.1038/s41586-021-03418-1}{\emph{Nature} {\bfseries 593} (2021) 51} [\href{https://arxiv.org/abs/2002.12347}{{\ttfamily 2002.12347}}].

\bibitem{Muong-2:2021ojo}
{\scshape Muon g-2} collaboration, \emph{{Measurement of the Positive Muon Anomalous Magnetic Moment to 0.46 ppm}}, \href{https://doi.org/10.1103/PhysRevLett.126.141801}{\emph{Phys. Rev. Lett.} {\bfseries 126} (2021) 141801} [\href{https://arxiv.org/abs/2104.03281}{{\ttfamily 2104.03281}}].

\bibitem{Carpio:2021jhu}
J.A.~Carpio, K.~Murase, I.M.~Shoemaker and Z.~Tabrizi, \emph{{High-energy cosmic neutrinos as a probe of the vector mediator scenario in light of the muon g-2 anomaly and Hubble tension}}, \href{https://doi.org/10.1103/PhysRevD.107.103057}{\emph{Phys. Rev. D} {\bfseries 107} (2023) 103057} [\href{https://arxiv.org/abs/2104.15136}{{\ttfamily 2104.15136}}].

\bibitem{vandenAarssen:2012vpm}
L.G.~van~den Aarssen, T.~Bringmann and C.~Pfrommer, \emph{{Is dark matter with long-range interactions a solution to all small-scale problems of \textbackslash{}Lambda CDM cosmology?}}, \href{https://doi.org/10.1103/PhysRevLett.109.231301}{\emph{Phys. Rev. Lett.} {\bfseries 109} (2012) 231301} [\href{https://arxiv.org/abs/1205.5809}{{\ttfamily 1205.5809}}].

\bibitem{Cyr-Racine:2013jua}
F.-Y.~Cyr-Racine and K.~Sigurdson, \emph{{Limits on Neutrino-Neutrino Scattering in the Early Universe}}, \href{https://doi.org/10.1103/PhysRevD.90.123533}{\emph{Phys. Rev. D} {\bfseries 90} (2014) 123533} [\href{https://arxiv.org/abs/1306.1536}{{\ttfamily 1306.1536}}].

\bibitem{Cherry:2014xra}
J.F.~Cherry, A.~Friedland and I.M.~Shoemaker, \emph{{Neutrino Portal Dark Matter: From Dwarf Galaxies to IceCube}},  \href{https://arxiv.org/abs/1411.1071}{{\ttfamily 1411.1071}}.

\bibitem{Chu:2015ipa}
X.~Chu, B.~Dasgupta and J.~Kopp, \emph{{Sterile neutrinos with secret interactions\textemdash{}lasting friendship with cosmology}}, \href{https://doi.org/10.1088/1475-7516/2015/10/011}{\emph{JCAP} {\bfseries 10} (2015) 011} [\href{https://arxiv.org/abs/1505.02795}{{\ttfamily 1505.02795}}].

\bibitem{Lancaster:2017ksf}
L.~Lancaster, F.-Y.~Cyr-Racine, L.~Knox and Z.~Pan, \emph{{A tale of two modes: Neutrino free-streaming in the early universe}}, \href{https://doi.org/10.1088/1475-7516/2017/07/033}{\emph{JCAP} {\bfseries 07} (2017) 033} [\href{https://arxiv.org/abs/1704.06657}{{\ttfamily 1704.06657}}].

\bibitem{Chu:2018gxk}
X.~Chu, B.~Dasgupta, M.~Dentler, J.~Kopp and N.~Saviano, \emph{{Sterile neutrinos with secret interactions\textemdash{}cosmological discord?}}, \href{https://doi.org/10.1088/1475-7516/2018/11/049}{\emph{JCAP} {\bfseries 11} (2018) 049} [\href{https://arxiv.org/abs/1806.10629}{{\ttfamily 1806.10629}}].

\bibitem{Kreisch:2019yzn}
C.D.~Kreisch, F.-Y.~Cyr-Racine and O.~Dor\'e, \emph{{Neutrino puzzle: Anomalies, interactions, and cosmological tensions}}, \href{https://doi.org/10.1103/PhysRevD.101.123505}{\emph{Phys. Rev. D} {\bfseries 101} (2020) 123505} [\href{https://arxiv.org/abs/1902.00534}{{\ttfamily 1902.00534}}].

\bibitem{Escudero:2019gzq}
M.~Escudero, D.~Hooper, G.~Krnjaic and M.~Pierre, \emph{{Cosmology with A Very Light L$_{\mu}$ \ensuremath{-} L$_{\tau}$ Gauge Boson}}, \href{https://doi.org/10.1007/JHEP03(2019)071}{\emph{JHEP} {\bfseries 03} (2019) 071} [\href{https://arxiv.org/abs/1901.02010}{{\ttfamily 1901.02010}}].

\bibitem{Grohs:2020xxd}
E.~Grohs, G.M.~Fuller and M.~Sen, \emph{{Consequences of neutrino self interactions for weak decoupling and big bang nucleosynthesis}}, \href{https://doi.org/10.1088/1475-7516/2020/07/001}{\emph{JCAP} {\bfseries 07} (2020) 001} [\href{https://arxiv.org/abs/2002.08557}{{\ttfamily 2002.08557}}].

\bibitem{RoyChoudhury:2020dmd}
S.~Roy~Choudhury, S.~Hannestad and T.~Tram, \emph{{Updated constraints on massive neutrino self-interactions from cosmology in light of the $H_0$ tension}}, \href{https://doi.org/10.1088/1475-7516/2021/03/084}{\emph{JCAP} {\bfseries 03} (2021) 084} [\href{https://arxiv.org/abs/2012.07519}{{\ttfamily 2012.07519}}].

\bibitem{Araki:2021xdk}
T.~Araki, K.~Asai, K.~Honda, R.~Kasuya, J.~Sato, T.~Shimomura et~al., \emph{{Resolving the Hubble tension in a U(1)$_{L_\mu-L_\tau}$ model with the Majoron}}, \href{https://doi.org/10.1093/ptep/ptab108}{\emph{PTEP} {\bfseries 2021} (2021) 103B05} [\href{https://arxiv.org/abs/2103.07167}{{\ttfamily 2103.07167}}].

\bibitem{Venzor:2023aka}
J.~Venzor, G.~Garcia-Arroyo, J.~De-Santiago and A.~P\'erez-Lorenzana, \emph{{Resonant neutrino self-interactions and the H0 tension}}, \href{https://doi.org/10.1103/PhysRevD.108.043536}{\emph{Phys. Rev. D} {\bfseries 108} (2023) 043536} [\href{https://arxiv.org/abs/2303.12792}{{\ttfamily 2303.12792}}].

\bibitem{Esseili:2023ldf}
H.~Esseili and G.D.~Kribs, \emph{{Cosmological Implications of Gauged $U(1)_{B-L}$ on $\Delta N_{\rm eff}$ in the CMB and BBN}},  \href{https://arxiv.org/abs/2308.07955}{{\ttfamily 2308.07955}}.

\bibitem{Asai:2023ajh}
K.~Asai, T.~Asano, J.~Sato and M.J.S.~Yang, \emph{{Contribution of Majoron to Hubble tension in gauged U(1)$_{L_\mu-L_\tau}$ Model}},  \href{https://arxiv.org/abs/2309.01162}{{\ttfamily 2309.01162}}.

\bibitem{Strumia:2003zx}
A.~Strumia and F.~Vissani, \emph{{Precise quasielastic neutrino/nucleon cross-section}}, \href{https://doi.org/10.1016/S0370-2693(03)00616-6}{\emph{Phys. Lett. B} {\bfseries 564} (2003) 42} [\href{https://arxiv.org/abs/astro-ph/0302055}{{\ttfamily astro-ph/0302055}}].

\bibitem{Formaggio:2012cpf}
J.A.~Formaggio and G.P.~Zeller, \emph{{From eV to EeV: Neutrino Cross Sections Across Energy Scales}}, \href{https://doi.org/10.1103/RevModPhys.84.1307}{\emph{Rev. Mod. Phys.} {\bfseries 84} (2012) 1307} [\href{https://arxiv.org/abs/1305.7513}{{\ttfamily 1305.7513}}].

\bibitem{Kolbe:2002gk}
E.~Kolbe, K.~Langanke and P.~Vogel, \emph{{Estimates of weak and electromagnetic nuclear decay signatures for neutrino reactions in Super-Kamiokande}}, \href{https://doi.org/10.1103/PhysRevD.66.013007}{\emph{Phys. Rev. D} {\bfseries 66} (2002) 013007}.

\bibitem{Marteau:1999zp}
J.~Marteau, J.~Delorme and M.~Ericson, \emph{{Neutrino oxygen interactions: Role of nuclear physics in the atmospheric neutrino anomaly}},  in \emph{{34th Rencontres de Moriond: Electroweak Interactions and Unified Theories}}, pp.~121--126, 1999 [\href{https://arxiv.org/abs/hep-ph/9906449}{{\ttfamily hep-ph/9906449}}].

\bibitem{Kamiokande-II:1987idp}
{\scshape Kamiokande-II} collaboration, \emph{{Observation of a Neutrino Burst from the Supernova SN 1987a}}, \href{https://doi.org/10.1103/PhysRevLett.58.1490}{\emph{Phys. Rev. Lett.} {\bfseries 58} (1987) 1490}.

\bibitem{Bionta:1987qt}
R.M.~Bionta et~al., \emph{{Observation of a Neutrino Burst in Coincidence with Supernova SN 1987a in the Large Magellanic Cloud}}, \href{https://doi.org/10.1103/PhysRevLett.58.1494}{\emph{Phys. Rev. Lett.} {\bfseries 58} (1987) 1494}.

\bibitem{Alekseev:1987ej}
E.N.~Alekseev, L.N.~Alekseeva, V.I.~Volchenko and I.V.~Krivosheina, \emph{{Possible Detection of a Neutrino Signal on 23 February 1987 at the Baksan Underground Scintillation Telescope of the Institute of Nuclear Research}}, {\emph{JETP Lett.} {\bfseries 45} (1987) 589}.

\bibitem{Rozwadowska:2020nab}
K.~Rozwadowska, F.~Vissani and E.~Cappellaro, \emph{{On the rate of core collapse supernovae in the milky way}}, \href{https://doi.org/10.1016/j.newast.2020.101498}{\emph{New Astron.} {\bfseries 83} (2021) 101498} [\href{https://arxiv.org/abs/2009.03438}{{\ttfamily 2009.03438}}].

\bibitem{Hyper-Kamiokande:2018ofw}
{\scshape Hyper-Kamiokande} collaboration, \emph{{Hyper-Kamiokande Design Report}},  \href{https://arxiv.org/abs/1805.04163}{{\ttfamily 1805.04163}}.

\bibitem{Garching}
\emph{{Garching core-collapse supernova research archive}}, {\emph{\href{https://wwwmpa.mpa-garching.mpg.de/ccsnarchive/}{https://wwwmpa.mpa-garching.mpg.de/ccsnarchive/}} }.

\bibitem{Fischer:2021jfm}
T.~Fischer, P.~Carenza, B.~Fore, M.~Giannotti, A.~Mirizzi and S.~Reddy, \emph{{Observable signatures of enhanced axion emission from protoneutron stars}}, \href{https://doi.org/10.1103/PhysRevD.104.103012}{\emph{Phys. Rev. D} {\bfseries 104} (2021) 103012} [\href{https://arxiv.org/abs/2108.13726}{{\ttfamily 2108.13726}}].

\bibitem{Fore:2019wib}
B.~Fore and S.~Reddy, \emph{{Pions in hot dense matter and their astrophysical implications}}, \href{https://doi.org/10.1103/PhysRevC.101.035809}{\emph{Phys. Rev. C} {\bfseries 101} (2020) 035809} [\href{https://arxiv.org/abs/1911.02632}{{\ttfamily 1911.02632}}].

\bibitem{Page:2020gsx}
D.~Page, M.V.~Beznogov, I.~Garibay, J.M.~Lattimer, M.~Prakash and H.-T.~Janka, \emph{{NS 1987A in SN 1987A}}, \href{https://doi.org/10.3847/1538-4357/ab93c2}{\emph{Astrophys. J.} {\bfseries 898} (2020) 125} [\href{https://arxiv.org/abs/2004.06078}{{\ttfamily 2004.06078}}].

\bibitem{Hempel:2014ssa}
M.~Hempel, \emph{{Nucleon self-energies for supernova equations of state}}, \href{https://doi.org/10.1103/PhysRevC.91.055807}{\emph{Phys. Rev. C} {\bfseries 91} (2015) 055807} [\href{https://arxiv.org/abs/1410.6337}{{\ttfamily 1410.6337}}].

\bibitem{Bollig:2017lki}
R.~Bollig, H.T.~Janka, A.~Lohs, G.~Martinez-Pinedo, C.J.~Horowitz and T.~Melson, \emph{{Muon Creation in Supernova Matter Facilitates Neutrino-driven Explosions}}, \href{https://doi.org/10.1103/PhysRevLett.119.242702}{\emph{Phys. Rev. Lett.} {\bfseries 119} (2017) 242702} [\href{https://arxiv.org/abs/1706.04630}{{\ttfamily 1706.04630}}].

\bibitem{Keil:2002in}
M.T.~Keil, G.G.~Raffelt and H.-T.~Janka, \emph{{Monte Carlo study of supernova neutrino spectra formation}}, \href{https://doi.org/10.1086/375130}{\emph{Astrophys. J.} {\bfseries 590} (2003) 971} [\href{https://arxiv.org/abs/astro-ph/0208035}{{\ttfamily astro-ph/0208035}}].

\bibitem{Tamborra:2012ac}
I.~Tamborra, B.~Muller, L.~Hudepohl, H.-T.~Janka and G.~Raffelt, \emph{{High-resolution supernova neutrino spectra represented by a simple fit}}, \href{https://doi.org/10.1103/PhysRevD.86.125031}{\emph{Phys. Rev. D} {\bfseries 86} (2012) 125031} [\href{https://arxiv.org/abs/1211.3920}{{\ttfamily 1211.3920}}].

\bibitem{Dighe:1999bi}
A.S.~Dighe and A.Y.~Smirnov, \emph{{Identifying the neutrino mass spectrum from the neutrino burst from a supernova}}, \href{https://doi.org/10.1103/PhysRevD.62.033007}{\emph{Phys. Rev. D} {\bfseries 62} (2000) 033007} [\href{https://arxiv.org/abs/hep-ph/9907423}{{\ttfamily hep-ph/9907423}}].

\bibitem{Duan:2010bg}
H.~Duan, G.M.~Fuller and Y.-Z.~Qian, \emph{{Collective Neutrino Oscillations}}, \href{https://doi.org/10.1146/annurev.nucl.012809.104524}{\emph{Ann. Rev. Nucl. Part. Sci.} {\bfseries 60} (2010) 569} [\href{https://arxiv.org/abs/1001.2799}{{\ttfamily 1001.2799}}].

\bibitem{Mirizzi:2015eza}
A.~Mirizzi, I.~Tamborra, H.-T.~Janka, N.~Saviano, K.~Scholberg, R.~Bollig et~al., \emph{{Supernova Neutrinos: Production, Oscillations and Detection}}, \href{https://doi.org/10.1393/ncr/i2016-10120-8}{\emph{Riv. Nuovo Cim.} {\bfseries 39} (2016) 1} [\href{https://arxiv.org/abs/1508.00785}{{\ttfamily 1508.00785}}].

\bibitem{Chakraborty:2016yeg}
S.~Chakraborty, R.~Hansen, I.~Izaguirre and G.~Raffelt, \emph{{Collective neutrino flavor conversion: Recent developments}}, \href{https://doi.org/10.1016/j.nuclphysb.2016.02.012}{\emph{Nucl. Phys. B} {\bfseries 908} (2016) 366} [\href{https://arxiv.org/abs/1602.02766}{{\ttfamily 1602.02766}}].

\bibitem{Tamborra:2020cul}
I.~Tamborra and S.~Shalgar, \emph{{New Developments in Flavor Evolution of a Dense Neutrino Gas}}, \href{https://doi.org/10.1146/annurev-nucl-102920-050505}{\emph{Ann. Rev. Nucl. Part. Sci.} {\bfseries 71} (2021) 165} [\href{https://arxiv.org/abs/2011.01948}{{\ttfamily 2011.01948}}].

\bibitem{Kuo:1989qe}
T.-K.~Kuo and J.T.~Pantaleone, \emph{{Neutrino Oscillations in Matter}}, \href{https://doi.org/10.1103/RevModPhys.61.937}{\emph{Rev. Mod. Phys.} {\bfseries 61} (1989) 937}.

\bibitem{Blennow:2013rca}
M.~Blennow and A.Y.~Smirnov, \emph{{Neutrino propagation in matter}}, \href{https://doi.org/10.1155/2013/972485}{\emph{Adv. High Energy Phys.} {\bfseries 2013} (2013) 972485} [\href{https://arxiv.org/abs/1306.2903}{{\ttfamily 1306.2903}}].

\bibitem{Mikheyev:1985zog}
S.P.~Mikheyev and A.Y.~Smirnov, \emph{{Resonance Amplification of Oscillations in Matter and Spectroscopy of Solar Neutrinos}}, {\emph{Sov. J. Nucl. Phys.} {\bfseries 42} (1985) 913}.

\bibitem{Woosley:1995ip}
S.E.~Woosley and T.A.~Weaver, \emph{{The Evolution and explosion of massive stars. 2. Explosive hydrodynamics and nucleosynthesis}}, \href{https://doi.org/10.1086/192237}{\emph{Astrophys. J. Suppl.} {\bfseries 101} (1995) 181}.

\bibitem{Nakazato:2012qf}
K.~Nakazato, K.~Sumiyoshi, H.~Suzuki, T.~Totani, H.~Umeda and S.~Yamada, \emph{{Supernova Neutrino Light Curves and Spectra for Various Progenitor Stars: From Core Collapse to Proto-neutron Star Cooling}}, \href{https://doi.org/10.1088/0067-0049/205/1/2}{\emph{Astrophys. J. Suppl.} {\bfseries 205} (2013) 2} [\href{https://arxiv.org/abs/1210.6841}{{\ttfamily 1210.6841}}].

\bibitem{Esteban:2020cvm}
I.~Esteban, M.C.~Gonzalez-Garcia, M.~Maltoni, T.~Schwetz and A.~Zhou, \emph{{The fate of hints: updated global analysis of three-flavor neutrino oscillations}}, \href{https://doi.org/10.1007/JHEP09(2020)178}{\emph{JHEP} {\bfseries 09} (2020) 178} [\href{https://arxiv.org/abs/2007.14792}{{\ttfamily 2007.14792}}].

\bibitem{deSalas:2020pgw}
P.F.~de~Salas, D.V.~Forero, S.~Gariazzo, P.~Mart\'\i{}nez-Mirav\'e, O.~Mena, C.A.~Ternes et~al., \emph{{2020 global reassessment of the neutrino oscillation picture}}, \href{https://doi.org/10.1007/JHEP02(2021)071}{\emph{JHEP} {\bfseries 02} (2021) 071} [\href{https://arxiv.org/abs/2006.11237}{{\ttfamily 2006.11237}}].

\bibitem{Sigl:1993ctk}
G.~Sigl and G.~Raffelt, \emph{{General kinetic description of relativistic mixed neutrinos}}, \href{https://doi.org/10.1016/0550-3213(93)90175-O}{\emph{Nucl. Phys. B} {\bfseries 406} (1993) 423}.

\bibitem{Akita:2022hlx}
K.~Akita and M.~Yamaguchi, \emph{{A Review of Neutrino Decoupling from the Early Universe to the Current Universe}}, \href{https://doi.org/10.3390/universe8110552}{\emph{Universe} {\bfseries 8} (2022) 552} [\href{https://arxiv.org/abs/2210.10307}{{\ttfamily 2210.10307}}].

\bibitem{Fischer:2016cyd}
T.~Fischer, S.~Chakraborty, M.~Giannotti, A.~Mirizzi, A.~Payez and A.~Ringwald, \emph{{Probing axions with the neutrino signal from the next galactic supernova}}, \href{https://doi.org/10.1103/PhysRevD.94.085012}{\emph{Phys. Rev. D} {\bfseries 94} (2016) 085012} [\href{https://arxiv.org/abs/1605.08780}{{\ttfamily 1605.08780}}].

\bibitem{Dreiner:2003wh}
H.K.~Dreiner, C.~Hanhart, U.~Langenfeld and D.R.~Phillips, \emph{{Supernovae and light neutralinos: SN1987A bounds on supersymmetry revisited}}, \href{https://doi.org/10.1103/PhysRevD.68.055004}{\emph{Phys. Rev. D} {\bfseries 68} (2003) 055004} [\href{https://arxiv.org/abs/hep-ph/0304289}{{\ttfamily hep-ph/0304289}}].

\bibitem{Oberauer:1993yr}
L.~Oberauer, C.~Hagner, G.~Raffelt and E.~Rieger, \emph{{Supernova bounds on neutrino radiative decays}}, \href{https://doi.org/10.1016/0927-6505(93)90004-W}{\emph{Astropart. Phys.} {\bfseries 1} (1993) 377}.

\bibitem{Shi:1993ee}
X.~Shi and G.~Sigl, \emph{{A Type II supernovae constraint on electron-neutrino - sterile-neutrino mixing}}, \href{https://doi.org/10.1016/0370-2693(94)91232-7}{\emph{Phys. Lett. B} {\bfseries 323} (1994) 360} [\href{https://arxiv.org/abs/hep-ph/9312247}{{\ttfamily hep-ph/9312247}}].

\bibitem{Nunokawa:1997ct}
H.~Nunokawa, J.T.~Peltoniemi, A.~Rossi and J.W.F.~Valle, \emph{{Supernova bounds on resonant active sterile neutrino conversions}}, \href{https://doi.org/10.1103/PhysRevD.56.1704}{\emph{Phys. Rev. D} {\bfseries 56} (1997) 1704} [\href{https://arxiv.org/abs/hep-ph/9702372}{{\ttfamily hep-ph/9702372}}].

\bibitem{Abazajian:2001nj}
K.~Abazajian, G.M.~Fuller and M.~Patel, \emph{{Sterile neutrino hot, warm, and cold dark matter}}, \href{https://doi.org/10.1103/PhysRevD.64.023501}{\emph{Phys. Rev. D} {\bfseries 64} (2001) 023501} [\href{https://arxiv.org/abs/astro-ph/0101524}{{\ttfamily astro-ph/0101524}}].

\bibitem{Hidaka:2006sg}
J.~Hidaka and G.M.~Fuller, \emph{{Dark matter sterile neutrinos in stellar collapse: Alteration of energy/lepton number transport and a mechanism for supernova explosion enhancement}}, \href{https://doi.org/10.1103/PhysRevD.74.125015}{\emph{Phys. Rev. D} {\bfseries 74} (2006) 125015} [\href{https://arxiv.org/abs/astro-ph/0609425}{{\ttfamily astro-ph/0609425}}].

\bibitem{Hidaka:2007se}
J.~Hidaka and G.M.~Fuller, \emph{{Sterile Neutrino-Enhanced Supernova Explosions}}, \href{https://doi.org/10.1103/PhysRevD.76.083516}{\emph{Phys. Rev. D} {\bfseries 76} (2007) 083516} [\href{https://arxiv.org/abs/0706.3886}{{\ttfamily 0706.3886}}].

\bibitem{Fuller:2008erj}
G.M.~Fuller, A.~Kusenko and K.~Petraki, \emph{{Heavy sterile neutrinos and supernova explosions}}, \href{https://doi.org/10.1016/j.physletb.2008.11.016}{\emph{Phys. Lett. B} {\bfseries 670} (2009) 281} [\href{https://arxiv.org/abs/0806.4273}{{\ttfamily 0806.4273}}].

\bibitem{Raffelt:2011nc}
G.G.~Raffelt and S.~Zhou, \emph{{Supernova bound on keV-mass sterile neutrinos reexamined}}, \href{https://doi.org/10.1103/PhysRevD.83.093014}{\emph{Phys. Rev. D} {\bfseries 83} (2011) 093014} [\href{https://arxiv.org/abs/1102.5124}{{\ttfamily 1102.5124}}].

\bibitem{Arguelles:2016uwb}
C.A.~Arg\"uelles, V.~Brdar and J.~Kopp, \emph{{Production of keV Sterile Neutrinos in Supernovae: New Constraints and Gamma Ray Observables}}, \href{https://doi.org/10.1103/PhysRevD.99.043012}{\emph{Phys. Rev. D} {\bfseries 99} (2019) 043012} [\href{https://arxiv.org/abs/1605.00654}{{\ttfamily 1605.00654}}].

\bibitem{Suliga:2019bsq}
A.M.~Suliga, I.~Tamborra and M.-R.~Wu, \emph{{Tau lepton asymmetry by sterile neutrino emission -- Moving beyond one-zone supernova models}}, \href{https://doi.org/10.1088/1475-7516/2019/12/019}{\emph{JCAP} {\bfseries 12} (2019) 019} [\href{https://arxiv.org/abs/1908.11382}{{\ttfamily 1908.11382}}].

\bibitem{Warren:2014qza}
M.L.~Warren, M.~Meixner, G.~Mathews, J.~Hidaka and T.~Kajino, \emph{{Sterile neutrino oscillations in core-collapse supernovae}}, \href{https://doi.org/10.1103/PhysRevD.90.103007}{\emph{Phys. Rev. D} {\bfseries 90} (2014) 103007} [\href{https://arxiv.org/abs/1405.6101}{{\ttfamily 1405.6101}}].

\bibitem{Warren:2016slz}
M.~Warren, G.J.~Mathews, M.~Meixner, J.~Hidaka and T.~Kajino, \emph{{Impact of sterile neutrino dark matter on core-collapse supernovae}}, \href{https://doi.org/10.1142/S0217751X16501372}{\emph{Int. J. Mod. Phys. A} {\bfseries 31} (2016) 1650137} [\href{https://arxiv.org/abs/1603.05503}{{\ttfamily 1603.05503}}].

\bibitem{Syvolap:2019dat}
V.~Syvolap, O.~Ruchayskiy and A.~Boyarsky, \emph{{Resonance production of keV sterile neutrinos in core-collapse supernovae and lepton number diffusion}}, \href{https://doi.org/10.1103/PhysRevD.106.015017}{\emph{Phys. Rev. D} {\bfseries 106} (2022) 015017} [\href{https://arxiv.org/abs/1909.06320}{{\ttfamily 1909.06320}}].

\bibitem{Rembiasz:2018lok}
T.~Rembiasz, M.~Obergaulinger, M.~Masip, M.A.~P\'erez-Garc\'\i{}a, M.-A.~Aloy and C.~Albertus, \emph{{Heavy sterile neutrinos in stellar core-collapse}}, \href{https://doi.org/10.1103/PhysRevD.98.103010}{\emph{Phys. Rev. D} {\bfseries 98} (2018) 103010} [\href{https://arxiv.org/abs/1806.03300}{{\ttfamily 1806.03300}}].

\bibitem{Ray:2023gtu}
A.~Ray and Y.-Z.~Qian, \emph{{Evolution of tau-neutrino lepton number in protoneutron stars due to active-sterile neutrino mixing}}, \href{https://doi.org/10.1103/PhysRevD.108.063025}{\emph{Phys. Rev. D} {\bfseries 108} (2023) 063025} [\href{https://arxiv.org/abs/2306.08209}{{\ttfamily 2306.08209}}].

\bibitem{Mori:2024vrf}
K.~Mori, T.~Takiwaki, K.~Kotake and S.~Horiuchi, \emph{{Two-dimensional models of core-collapse supernova explosions assisted by heavy sterile neutrinos}},  \href{https://arxiv.org/abs/2402.14333}{{\ttfamily 2402.14333}}.

\bibitem{Ray:2024jeu}
A.~Ray and Y.-Z.~Qian, \emph{{Enhanced Muonization by Active-Sterile Neutrino Mixing in Protoneutron Stars}},  \href{https://arxiv.org/abs/2404.14485}{{\ttfamily 2404.14485}}.

\bibitem{deGouvea:2015euy}
A.~de~Gouv\^ea and A.~Kobach, \emph{{Global Constraints on a Heavy Neutrino}}, \href{https://doi.org/10.1103/PhysRevD.93.033005}{\emph{Phys. Rev. D} {\bfseries 93} (2016) 033005} [\href{https://arxiv.org/abs/1511.00683}{{\ttfamily 1511.00683}}].

\bibitem{Beacham:2019nyx}
J.~Beacham et~al., \emph{{Physics Beyond Colliders at CERN: Beyond the Standard Model Working Group Report}}, \href{https://doi.org/10.1088/1361-6471/ab4cd2}{\emph{J. Phys. G} {\bfseries 47} (2020) 010501} [\href{https://arxiv.org/abs/1901.09966}{{\ttfamily 1901.09966}}].

\bibitem{T2K:2019jwa}
{\scshape T2K} collaboration, \emph{{Search for heavy neutrinos with the T2K near detector ND280}}, \href{https://doi.org/10.1103/PhysRevD.100.052006}{\emph{Phys. Rev. D} {\bfseries 100} (2019) 052006} [\href{https://arxiv.org/abs/1902.07598}{{\ttfamily 1902.07598}}].

\bibitem{NA62:2020mcv}
{\scshape NA62} collaboration, \emph{{Search for heavy neutral lepton production in $K^+$ decays to positrons}}, \href{https://doi.org/10.1016/j.physletb.2020.135599}{\emph{Phys. Lett. B} {\bfseries 807} (2020) 135599} [\href{https://arxiv.org/abs/2005.09575}{{\ttfamily 2005.09575}}].

\bibitem{Alekhin:2015byh}
S.~Alekhin et~al., \emph{{A facility to Search for Hidden Particles at the CERN SPS: the SHiP physics case}}, \href{https://doi.org/10.1088/0034-4885/79/12/124201}{\emph{Rept. Prog. Phys.} {\bfseries 79} (2016) 124201} [\href{https://arxiv.org/abs/1504.04855}{{\ttfamily 1504.04855}}].

\bibitem{Boyarsky:2009ix}
A.~Boyarsky, O.~Ruchayskiy and M.~Shaposhnikov, \emph{{The Role of sterile neutrinos in cosmology and astrophysics}}, \href{https://doi.org/10.1146/annurev.nucl.010909.083654}{\emph{Ann. Rev. Nucl. Part. Sci.} {\bfseries 59} (2009) 191} [\href{https://arxiv.org/abs/0901.0011}{{\ttfamily 0901.0011}}].

\bibitem{Ruchayskiy:2012si}
O.~Ruchayskiy and A.~Ivashko, \emph{{Restrictions on the lifetime of sterile neutrinos from primordial nucleosynthesis}}, \href{https://doi.org/10.1088/1475-7516/2012/10/014}{\emph{JCAP} {\bfseries 10} (2012) 014} [\href{https://arxiv.org/abs/1202.2841}{{\ttfamily 1202.2841}}].

\bibitem{Sabti:2020yrt}
N.~Sabti, A.~Magalich and A.~Filimonova, \emph{{An Extended Analysis of Heavy Neutral Leptons during Big Bang Nucleosynthesis}}, \href{https://doi.org/10.1088/1475-7516/2020/11/056}{\emph{JCAP} {\bfseries 11} (2020) 056} [\href{https://arxiv.org/abs/2006.07387}{{\ttfamily 2006.07387}}].

\bibitem{Mastrototaro:2021wzl}
L.~Mastrototaro, P.D.~Serpico, A.~Mirizzi and N.~Saviano, \emph{{Massive sterile neutrinos in the early Universe: From thermal decoupling to cosmological constraints}}, \href{https://doi.org/10.1103/PhysRevD.104.016026}{\emph{Phys. Rev. D} {\bfseries 104} (2021) 016026} [\href{https://arxiv.org/abs/2104.11752}{{\ttfamily 2104.11752}}].

\bibitem{Boyarsky:2020dzc}
A.~Boyarsky, M.~Ovchynnikov, O.~Ruchayskiy and V.~Syvolap, \emph{{Improved big bang nucleosynthesis constraints on heavy neutral leptons}}, \href{https://doi.org/10.1103/PhysRevD.104.023517}{\emph{Phys. Rev. D} {\bfseries 104} (2021) 023517} [\href{https://arxiv.org/abs/2008.00749}{{\ttfamily 2008.00749}}].

\bibitem{Hannestad:1995rs}
S.~Hannestad and J.~Madsen, \emph{{Neutrino decoupling in the early universe}}, \href{https://doi.org/10.1103/PhysRevD.52.1764}{\emph{Phys. Rev. D} {\bfseries 52} (1995) 1764} [\href{https://arxiv.org/abs/astro-ph/9506015}{{\ttfamily astro-ph/9506015}}].

\bibitem{Dolgov:1997mb}
A.D.~Dolgov, S.H.~Hansen and D.V.~Semikoz, \emph{{Nonequilibrium corrections to the spectra of massless neutrinos in the early universe}}, \href{https://doi.org/10.1016/S0550-3213(97)00479-3}{\emph{Nucl. Phys. B} {\bfseries 503} (1997) 426} [\href{https://arxiv.org/abs/hep-ph/9703315}{{\ttfamily hep-ph/9703315}}].

\bibitem{Gorbunov:2007ak}
D.~Gorbunov and M.~Shaposhnikov, \emph{{How to find neutral leptons of the $\nu$MSM?}}, \href{https://doi.org/10.1088/1126-6708/2007/10/015}{\emph{JHEP} {\bfseries 10} (2007) 015} [\href{https://arxiv.org/abs/0705.1729}{{\ttfamily 0705.1729}}].

\bibitem{Atre:2009rg}
A.~Atre, T.~Han, S.~Pascoli and B.~Zhang, \emph{{The Search for Heavy Majorana Neutrinos}}, \href{https://doi.org/10.1088/1126-6708/2009/05/030}{\emph{JHEP} {\bfseries 05} (2009) 030} [\href{https://arxiv.org/abs/0901.3589}{{\ttfamily 0901.3589}}].

\bibitem{Helo:2010cw}
J.C.~Helo, S.~Kovalenko and I.~Schmidt, \emph{{Sterile neutrinos in lepton number and lepton flavor violating decays}}, \href{https://doi.org/10.1016/j.nuclphysb.2011.07.020}{\emph{Nucl. Phys. B} {\bfseries 853} (2011) 80} [\href{https://arxiv.org/abs/1005.1607}{{\ttfamily 1005.1607}}].

\bibitem{Bondarenko:2018ptm}
K.~Bondarenko, A.~Boyarsky, D.~Gorbunov and O.~Ruchayskiy, \emph{{Phenomenology of GeV-scale Heavy Neutral Leptons}}, \href{https://doi.org/10.1007/JHEP11(2018)032}{\emph{JHEP} {\bfseries 11} (2018) 032} [\href{https://arxiv.org/abs/1805.08567}{{\ttfamily 1805.08567}}].

\bibitem{Coloma:2020lgy}
P.~Coloma, E.~Fern\'andez-Mart\'\i{}nez, M.~Gonz\'alez-L\'opez, J.~Hern\'andez-Garc\'\i{}a and Z.~Pavlovic, \emph{{GeV-scale neutrinos: interactions with mesons and DUNE sensitivity}}, \href{https://doi.org/10.1140/epjc/s10052-021-08861-y}{\emph{Eur. Phys. J. C} {\bfseries 81} (2021) 78} [\href{https://arxiv.org/abs/2007.03701}{{\ttfamily 2007.03701}}].

\bibitem{Kersten:2007vk}
J.~Kersten and A.Y.~Smirnov, \emph{{Right-Handed Neutrinos at CERN LHC and the Mechanism of Neutrino Mass Generation}}, \href{https://doi.org/10.1103/PhysRevD.76.073005}{\emph{Phys. Rev. D} {\bfseries 76} (2007) 073005} [\href{https://arxiv.org/abs/0705.3221}{{\ttfamily 0705.3221}}].

\bibitem{Antusch:2017pkq}
S.~Antusch, E.~Cazzato, M.~Drewes, O.~Fischer, B.~Garbrecht, D.~Gueter et~al., \emph{{Probing Leptogenesis at Future Colliders}}, \href{https://doi.org/10.1007/JHEP09(2018)124}{\emph{JHEP} {\bfseries 09} (2018) 124} [\href{https://arxiv.org/abs/1710.03744}{{\ttfamily 1710.03744}}].

\bibitem{Drewes:2021nqr}
M.~Drewes, Y.~Georis and J.~Klari\'c, \emph{{Mapping the Viable Parameter Space for Testable Leptogenesis}}, \href{https://doi.org/10.1103/PhysRevLett.128.051801}{\emph{Phys. Rev. Lett.} {\bfseries 128} (2022) 051801} [\href{https://arxiv.org/abs/2106.16226}{{\ttfamily 2106.16226}}].

\bibitem{Bruenn:1985en}
S.W.~Bruenn, \emph{{Stellar core collapse: Numerical model and infall epoch}}, \href{https://doi.org/10.1086/191056}{\emph{Astrophys. J. Suppl.} {\bfseries 58} (1985) 771}.

\bibitem{Kamada:2015era}
A.~Kamada and H.-B.~Yu, \emph{{Coherent Propagation of PeV Neutrinos and the Dip in the Neutrino Spectrum at IceCube}}, \href{https://doi.org/10.1103/PhysRevD.92.113004}{\emph{Phys. Rev. D} {\bfseries 92} (2015) 113004} [\href{https://arxiv.org/abs/1504.00711}{{\ttfamily 1504.00711}}].

\bibitem{Holdom:1985ag}
B.~Holdom, \emph{{Two U(1)'s and Epsilon Charge Shifts}}, \href{https://doi.org/10.1016/0370-2693(86)91377-8}{\emph{Phys. Lett. B} {\bfseries 166} (1986) 196}.

\bibitem{IMB:1988suc}
{\scshape IMB} collaboration, \emph{{Angular Distribution of Events From Sn1987a}}, \href{https://doi.org/10.1103/PhysRevD.37.3361}{\emph{Phys. Rev. D} {\bfseries 37} (1988) 3361}.

\bibitem{Mirizzi:2006xx}
A.~Mirizzi, G.G.~Raffelt and P.D.~Serpico, \emph{{Earth matter effects in supernova neutrinos: Optimal detector locations}}, \href{https://doi.org/10.1088/1475-7516/2006/05/012}{\emph{JCAP} {\bfseries 05} (2006) 012} [\href{https://arxiv.org/abs/astro-ph/0604300}{{\ttfamily astro-ph/0604300}}].

\bibitem{Jegerlehner:1996kx}
B.~Jegerlehner, F.~Neubig and G.~Raffelt, \emph{{Neutrino oscillations and the supernova SN1987A signal}}, \href{https://doi.org/10.1103/PhysRevD.54.1194}{\emph{Phys. Rev. D} {\bfseries 54} (1996) 1194} [\href{https://arxiv.org/abs/astro-ph/9601111}{{\ttfamily astro-ph/9601111}}].

\bibitem{Super-Kamiokande:2016yck}
{\scshape Super-Kamiokande} collaboration, \emph{{Solar Neutrino Measurements in Super-Kamiokande-IV}}, \href{https://doi.org/10.1103/PhysRevD.94.052010}{\emph{Phys. Rev. D} {\bfseries 94} (2016) 052010} [\href{https://arxiv.org/abs/1606.07538}{{\ttfamily 1606.07538}}].

\bibitem{Mirizzi:2005tg}
A.~Mirizzi and G.G.~Raffelt, \emph{{New analysis of the sn 1987a neutrinos with a flexible spectral shape}}, \href{https://doi.org/10.1103/PhysRevD.72.063001}{\emph{Phys. Rev. D} {\bfseries 72} (2005) 063001} [\href{https://arxiv.org/abs/astro-ph/0508612}{{\ttfamily astro-ph/0508612}}].

\bibitem{Li:2023ulf}
S.W.~Li, J.F.~Beacom, L.F.~Roberts and F.~Capozzi, \emph{{Old Data, New Forensics: The First Second of SN 1987A Neutrino Emission}},  \href{https://arxiv.org/abs/2306.08024}{{\ttfamily 2306.08024}}.

\bibitem{Fiorillo:2023frv}
D.F.G.~Fiorillo, M.~Heinlein, H.-T.~Janka, G.~Raffelt and E.~Vitagliano, \emph{{Supernova Simulations Confront SN 1987A Neutrinos}},  \href{https://arxiv.org/abs/2308.01403}{{\ttfamily 2308.01403}}.

\bibitem{Bolton:2019pcu}
P.D.~Bolton, F.F.~Deppisch and P.S.~Bhupal~Dev, \emph{{Neutrinoless double beta decay versus other probes of heavy sterile neutrinos}}, \href{https://doi.org/10.1007/JHEP03(2020)170}{\emph{JHEP} {\bfseries 03} (2020) 170} [\href{https://arxiv.org/abs/1912.03058}{{\ttfamily 1912.03058}}].

\bibitem{Ema:2023buz}
Y.~Ema, Z.~Liu, K.-F.~Lyu and M.~Pospelov, \emph{{Heavy Neutral Leptons from Stopped Muons and Pions}}, \href{https://doi.org/10.1007/JHEP08(2023)169}{\emph{JHEP} {\bfseries 08} (2023) 169} [\href{https://arxiv.org/abs/2306.07315}{{\ttfamily 2306.07315}}].

\bibitem{Krasnov:2019kdc}
I.~Krasnov, \emph{{DUNE prospects in the search for sterile neutrinos}}, \href{https://doi.org/10.1103/PhysRevD.100.075023}{\emph{Phys. Rev. D} {\bfseries 100} (2019) 075023} [\href{https://arxiv.org/abs/1902.06099}{{\ttfamily 1902.06099}}].

\bibitem{Ballett:2019bgd}
P.~Ballett, T.~Boschi and S.~Pascoli, \emph{{Heavy Neutral Leptons from low-scale seesaws at the DUNE Near Detector}}, \href{https://doi.org/10.1007/JHEP03(2020)111}{\emph{JHEP} {\bfseries 03} (2020) 111} [\href{https://arxiv.org/abs/1905.00284}{{\ttfamily 1905.00284}}].

\bibitem{Curtin:2018mvb}
D.~Curtin et~al., \emph{{Long-Lived Particles at the Energy Frontier: The MATHUSLA Physics Case}}, \href{https://doi.org/10.1088/1361-6633/ab28d6}{\emph{Rept. Prog. Phys.} {\bfseries 82} (2019) 116201} [\href{https://arxiv.org/abs/1806.07396}{{\ttfamily 1806.07396}}].

\bibitem{SHiP:2018xqw}
{\scshape SHiP} collaboration, \emph{{Sensitivity of the SHiP experiment to Heavy Neutral Leptons}}, \href{https://doi.org/10.1007/JHEP04(2019)077}{\emph{JHEP} {\bfseries 04} (2019) 077} [\href{https://arxiv.org/abs/1811.00930}{{\ttfamily 1811.00930}}].

\bibitem{Gorbunov:2020rjx}
D.~Gorbunov, I.~Krasnov, Y.~Kudenko and S.~Suvorov, \emph{{Heavy Neutral Leptons from kaon decays in the SHiP experiment}}, \href{https://doi.org/10.1016/j.physletb.2020.135817}{\emph{Phys. Lett. B} {\bfseries 810} (2020) 135817} [\href{https://arxiv.org/abs/2004.07974}{{\ttfamily 2004.07974}}].

\bibitem{PIONEER:2022alm}
{\scshape PIONEER} collaboration, \emph{{Testing Lepton Flavor Universality and CKM Unitarity with Rare Pion Decays in the PIONEER experiment}},  in \emph{{Snowmass 2021}}, 3, 2022 [\href{https://arxiv.org/abs/2203.05505}{{\ttfamily 2203.05505}}].

\bibitem{Gelmini:2020ekg}
G.B.~Gelmini, M.~Kawasaki, A.~Kusenko, K.~Murai and V.~Takhistov, \emph{{Big Bang Nucleosynthesis constraints on sterile neutrino and lepton asymmetry of the Universe}}, \href{https://doi.org/10.1088/1475-7516/2020/09/051}{\emph{JCAP} {\bfseries 09} (2020) 051} [\href{https://arxiv.org/abs/2005.06721}{{\ttfamily 2005.06721}}].

\bibitem{Gelmini:2004ah}
G.~Gelmini, S.~Palomares-Ruiz and S.~Pascoli, \emph{{Low reheating temperature and the visible sterile neutrino}}, \href{https://doi.org/10.1103/PhysRevLett.93.081302}{\emph{Phys. Rev. Lett.} {\bfseries 93} (2004) 081302} [\href{https://arxiv.org/abs/astro-ph/0403323}{{\ttfamily astro-ph/0403323}}].

\bibitem{Gelmini:2008fq}
G.~Gelmini, E.~Osoba, S.~Palomares-Ruiz and S.~Pascoli, \emph{{MeV sterile neutrinos in low reheating temperature cosmological scenarios}}, \href{https://doi.org/10.1088/1475-7516/2008/10/029}{\emph{JCAP} {\bfseries 10} (2008) 029} [\href{https://arxiv.org/abs/0803.2735}{{\ttfamily 0803.2735}}].

\bibitem{Escudero:2019gvw}
M.~Escudero and S.J.~Witte, \emph{{A CMB search for the neutrino mass mechanism and its relation to the Hubble tension}}, \href{https://doi.org/10.1140/epjc/s10052-020-7854-5}{\emph{Eur. Phys. J. C} {\bfseries 80} (2020) 294} [\href{https://arxiv.org/abs/1909.04044}{{\ttfamily 1909.04044}}].

\bibitem{An:2013yfc}
H.~An, M.~Pospelov and J.~Pradler, \emph{{New stellar constraints on dark photons}}, \href{https://doi.org/10.1016/j.physletb.2013.07.008}{\emph{Phys. Lett. B} {\bfseries 725} (2013) 190} [\href{https://arxiv.org/abs/1302.3884}{{\ttfamily 1302.3884}}].

\bibitem{Hardy:2016kme}
E.~Hardy and R.~Lasenby, \emph{{Stellar cooling bounds on new light particles: plasma mixing effects}}, \href{https://doi.org/10.1007/JHEP02(2017)033}{\emph{JHEP} {\bfseries 02} (2017) 033} [\href{https://arxiv.org/abs/1611.05852}{{\ttfamily 1611.05852}}].

\bibitem{Sandner:2023ptm}
S.~Sandner, M.~Escudero and S.J.~Witte, \emph{{Precision CMB constraints on eV-scale bosons coupled to neutrinos}}, \href{https://doi.org/10.1140/epjc/s10052-023-11864-6}{\emph{Eur. Phys. J. C} {\bfseries 83} (2023) 709} [\href{https://arxiv.org/abs/2305.01692}{{\ttfamily 2305.01692}}].

\bibitem{An:2014twa}
H.~An, M.~Pospelov, J.~Pradler and A.~Ritz, \emph{{Direct Detection Constraints on Dark Photon Dark Matter}}, \href{https://doi.org/10.1016/j.physletb.2015.06.018}{\emph{Phys. Lett. B} {\bfseries 747} (2015) 331} [\href{https://arxiv.org/abs/1412.8378}{{\ttfamily 1412.8378}}].

\bibitem{Hong:2020bxo}
D.K.~Hong, C.S.~Shin and S.~Yun, \emph{{Cooling of young neutron stars and dark gauge bosons}}, \href{https://doi.org/10.1103/PhysRevD.103.123031}{\emph{Phys. Rev. D} {\bfseries 103} (2021) 123031} [\href{https://arxiv.org/abs/2012.05427}{{\ttfamily 2012.05427}}].

\bibitem{Heeck:2014zfa}
J.~Heeck, \emph{{Unbroken B \textendash{} L symmetry}}, \href{https://doi.org/10.1016/j.physletb.2014.10.067}{\emph{Phys. Lett. B} {\bfseries 739} (2014) 256} [\href{https://arxiv.org/abs/1408.6845}{{\ttfamily 1408.6845}}].

\bibitem{Kazanas:2014mca}
D.~Kazanas, R.N.~Mohapatra, S.~Nussinov, V.L.~Teplitz and Y.~Zhang, \emph{{Supernova Bounds on the Dark Photon Using its Electromagnetic Decay}}, \href{https://doi.org/10.1016/j.nuclphysb.2014.11.009}{\emph{Nucl. Phys. B} {\bfseries 890} (2014) 17} [\href{https://arxiv.org/abs/1410.0221}{{\ttfamily 1410.0221}}].

\bibitem{Archidiacono:2013dua}
M.~Archidiacono and S.~Hannestad, \emph{{Updated constraints on non-standard neutrino interactions from Planck}}, \href{https://doi.org/10.1088/1475-7516/2014/07/046}{\emph{JCAP} {\bfseries 07} (2014) 046} [\href{https://arxiv.org/abs/1311.3873}{{\ttfamily 1311.3873}}].

\bibitem{Huang:2017egl}
G.-y.~Huang, T.~Ohlsson and S.~Zhou, \emph{{Observational Constraints on Secret Neutrino Interactions from Big Bang Nucleosynthesis}}, \href{https://doi.org/10.1103/PhysRevD.97.075009}{\emph{Phys. Rev. D} {\bfseries 97} (2018) 075009} [\href{https://arxiv.org/abs/1712.04792}{{\ttfamily 1712.04792}}].

\bibitem{Li:2023kuz}
S.-P.~Li and B.~Yu, \emph{{A cosmological sandwiched window for seesaw with primordial majoron abundance}},  \href{https://arxiv.org/abs/2310.13492}{{\ttfamily 2310.13492}}.

\bibitem{Chang:2024mvg}
S.~Chang, S.~Ganguly, T.H.~Jung, T.-S.~Park and C.S.~Shin, \emph{{Constraining MeV to 10 GeV majoron by Big Bang Nucleosynthesis}},  \href{https://arxiv.org/abs/2401.00687}{{\ttfamily 2401.00687}}.

\end{thebibliography}\endgroup

\end{document}